\documentclass[useAMS,usenatbib,twoside,a4paper,iop,numberedappendix,appendixfloats]{emulateapj} 
\slugcomment{{\sc Accepted for publication in ApJ on September 20, 2013 issue }}
\usepackage{graphicx}
\usepackage{epstopdf}
\usepackage{color}
\usepackage{natbib}
\usepackage{draftcopy}
\usepackage{longtable}
\usepackage{amsmath}
\newcommand{\Msol}{\mbox{$M_{\odot}$}}
\newcommand{\Lsol}{\mbox{$L_{\odot}$}}

\newcommand{\Halpha}{\mbox{H$\alpha$} }
\newcommand{\Hbeta}{\mbox{H$\beta$} }
\def\deg   {{\ifmmode^\circ\else$^\circ$\fi}}

\shorttitle{The Zurich ENvironmental Study of  Galaxies in Groups along the Cosmic Web. I. }
\shortauthors{Carollo C. M.  et al.} 

\def\lta{{\>\rlap{\raise2pt\hbox{$<$}}\lower3pt\hbox{$\sim$}\>}}
\def\gta{{\>\rlap{\raise2pt\hbox{$>$}}\lower3pt\hbox{$\sim$}\>}}

\begin{document}

\title{The Zurich ENvironmental Study \emph{(ZENS)} of Galaxies in Groups along the Cosmic Web.\\ 
I. Which Environment Affects Galaxy Evolution?\footnotemark[\dag]} 

\author{C.~Marcella Carollo\altaffilmark{1,$\star$}, 
Anna Cibinel\altaffilmark{1}, 
Simon~J. Lilly\altaffilmark{1}, 
Francesco Miniati\altaffilmark{1}, 
Peder Norberg\altaffilmark{5}, 
John~D. Silverman\altaffilmark{2}, 
Jacqueline van Gorkom\altaffilmark{3}, 
Ewan Cameron\altaffilmark{1}, 
Alexis Finoguenov\altaffilmark{4}, 
Yingjie Peng\altaffilmark{1},
Antonio Pipino\altaffilmark{1},  
Craig~S. Rudick\altaffilmark{1}
}

\altaffiltext{$\star$}{E-mail: \texttt{marcella@phys.ethz.ch}}
\altaffiltext{1}{Institute for Astronomy, ETH Zurich, CH-8093 Zurich, Switzerland}
\altaffiltext{2}{Kavli Institute for the Physics and Mathematics of the Universe (WPI), Todai Institutes for Advanced Study, The University of Tokyo, 277-8583, Japan}
\altaffiltext{3}{Department of Astronomy, Columbia University, New York, NY 10027, USA}
\altaffiltext{4}{Max-Planck-Institut f\"ur extraterrestrische Physik, D-84571 Garching, Germany}
\altaffiltext{5}{Institute for Computational Cosmology, Department of Physics, Durham University, South Road, Durham DH1 3LE, UK}
\altaffiltext{$\dagger$}{Based on observations collected at the European Southern Observatory, La Silla Chile. Program ID 177.A-0680}

\begin{abstract}
The {\it Zurich Environmental Study (ZENS)} is based on a sample of  $\sim1500$   galaxy members of 141   groups in the mass range $\sim10^{12.5-14.5} M_\odot$ within the narrow redshift range $0.05<z<0.0585$. \emph{ZENS} adopts novel approaches, here described,   to quantify four different galactic environments, namely: $(1)$ the mass of the  host group halo;  $(2)$ the projected halo-centric distance; $(3)$ the rank of galaxies as central  or satellites within their group halos; and $(4)$ the filamentary large-scale structure (LSS) density.   No self-consistent identification of a central galaxy is found in  $\sim40\%$ of $<10^{13.5} M_\odot$ groups, from which  we estimate  that $\sim15\%$ of groups at these masses are dynamically unrelaxed systems. Central galaxies in relaxed and unrelaxed groups have  in general similar properties, suggesting that centrals are regulated by their mass and not by their environment. Centrals in relaxed groups have however $\sim$30\% larger sizes than in unrelaxed groups, possibly due accretion of small satellites in virialized group halos. At  $M>10^{10} M_\odot$, satellite galaxies in relaxed and unrelaxed groups  have similar size, color and (specific) star formation rate distributions;  at lower galaxy masses,  satellites are marginally redder in  relaxed relative to unrelaxed groups,  suggesting quenching of  star formation in low-mass satellites by physical processes active in relaxed halos.   Finally, relaxed and unrelated groups show similar  stellar mass conversion efficiencies, peaking at halo masses around $10^{12.5} M_\odot$. In the enclosed \emph{ZENS} catalogue we publish all environmental diagnostics as well as the galaxy structural and photometric measurements described in companion \emph{ZENS} papers II and III.
\end{abstract}

\keywords{galaxies: evolution ---
galaxies: formation ---
galaxies: groups: general ---
galaxies: star formation ---
galaxies: stellar content ---
galaxies: structure
}
 
\section{INTRODUCTION}  \label{sec:introduction}

The study of the effect of the environment on the evolution of galaxies is beset by 
a number of difficulties that have made it hard to define a single coherent picture 
and to isolate the main physical processes. It has been clear for many years that both 
the mass and the environment of a galaxy affect its evolution and
 its appearance today.  Since the pioneering work of e.g., \citet{Oemler_1974}, \citet{Dressler_1980}, 
 \citet{Postman_Geller_1984}, many studies have highlighted clear trends between different observational 
 diagnostics of evolution such as stellar absorption line strengths, color or morphology and either galactic mass or environment or both
 \citep[e.g.][]{Carollo_et_al_1993,Balogh_et_al_1999,Goto_et_al_2003,Blanton_et_al_2005,Zehavi_et_al_2002,
 Weinmann_et_al_2006,Weinmann_et_al_2009,Croton_et_al_2005, Park_et_al_2007,Kovac_et_al_2010,Peng_et_al_2010,Cooper_et_al_2010,Peng_et_al_2012,Wetzel_et_al_2012,Calvi_et_al_2012,Woo_et_al_2013}, 
 but the detailed phenomenology, as well as physical understanding, remain unclear.  
 This can be traced to several complicating factors or difficulties.

First, there are  a number of galactic properties that are relevant to define its evolutionary state. 
Galaxy evolution may be traced by changes in the star-formation rates (SFRs) of galaxies 
\citep[e.g.][]{Lilly_et_al_1996, Madau_et_al_1996,
Chary_Elbaz_2001,Rodighiero_et_al_2010,Wetzel_et_al_2012,Woo_et_al_2013}, leading to differences 
in the integrated stellar populations, and therefore in spectral properties and colors of galaxies 
(e.g. \citealt{Carollo_Danziger_1994,Carollo_et_al_1997,Masters_et_al_2010,Bundy_et_al_2010}; 
perhaps modified by the effects of dust; e.g. \citealt{Labbe_et_al_2005,Williams_et_al_2009,Wolf_et_al_2009}).  
Galaxy evolution may also be manifested by changes in the  morphologies of galaxies, 
both in terms of the overall 
structural morphology of bulge-to-disk ratios and the structural properties of each component
 \citep[][among others]{Carollo_et_al_1998,Carollo_1999,Brinchmann_Ellis_2000,Carollo_et_al_2007,Kovac_et_al_2010,Oesch_et_al_2010,Feldmann_et_al_2011,Skibba_et_al_2012,Calvi_et_al_2012,Cooper_et_al_2012,Raichoor_et_al_2012}
and also in the appearance  of features such as spiral arms or bars.  Color and morphology clearly broadly correlate within 
the nearby galaxy population, but with a significant and poorly understood scatter  \citep{Strateva_et_al_2001}. 
 Morphology and color may reflect different aspects of a single   evolutionary sequence, or may reflect the outcome of quite different physical   processes that may conceivably occur either synchronously or asynchronously.  Many previous studies have focused on just one or other of this color-morphology duality.  A comprehensive picture is likely to require the simultaneous treatment of all such physically relevant properties.

Second, with both mass and environment, it is not clear exactly $which$ mass or environment is likely to be the most relevant for ÒcentralsÓ, i.e., galaxies which appear to dominate their halos, and ÒsatellitesÓ, i.e. galaxies which orbit another more massive galaxy within a single dark matter halo \citep[e.g.][]{Cooper_et_al_2005,DeLucia_et_al_2012,Haas_et_al_2012,Muldrew_et_al_2012}. Observationally, the existing stellar mass of a galaxy is the most easily accessible, but the physical driver of the evolution could be the mass of the dark matter halo of a galaxy, or, in the case of satellite galaxies, the mass of the dark matter halo in which the galaxy resides, leading to an environment-like measure of mass. 
Similarly, the environment that could influence the evolution of a galaxy could reflect either very local effects, e.g. the location of a galaxy in a dark matter halo, or the interaction with nearby neighbors through the mass of the dark matter halo (as above), or  the broader environment beyond the halo, as defined by the cosmic web of filaments and voids.  Clearly some of the definitions of environment are closely linked to the mass of a galaxy, especially for galaxies which dominate their dark matter haloes.  Even for galaxy stellar mass we could imagine some direct crosstalk between it and environment if the stellar mass function of galaxies was itself dependent on environment (\citealt{Bundy_et_al_2006}; \citealt{Baldry_et_al_2006}; \citealt{Bolzonella_et_al_2010};   \citealt{Kovac_et_al_2010}), necessitating the careful isolation of these two variables.

A recent analysis in the Sloan Digital Sky Survey (SDSS; \cite{York_et_al_2000}) of 
the three-way relationships between color, stellar mass and environment, the latter defined simply in terms of a 5th-nearest galaxy-neighbor density, reveals some interesting simplicities within the galaxy population \citep{Peng_et_al_2010}.  Not least, the effects of environment and stellar mass on the fraction of galaxies that are observed to be red (the Òred fractionÓ) are straightforwardly separable in the sense that the chance that a given galaxy is red is the product of two functions, one of mass independent of environment, and the other of environment, independent of mass. 
 This led Peng et al.\ to identify two separate physical processes, termed Òmass-quenchingÓ and Òenvironment-quenchingÓ.  A conclusion of this analysis was that for galaxy stellar masses below $\sim10^{10}M_\odot$ the effects of the environment dominate, while above $\sim10^{11}M_\odot$  the galaxy population is dominated by the effects of merging, which again is environmentally determined. The differential effects of galactic stellar mass and environment can be most clearly seen in the $\sim10^{10-11}M_\odot$ galaxy population. \citet{Peng_et_al_2012} extended their original formalism to the central-satellite dichotomy of galaxies, using a large group catalogue \citep{Yang_et_al_2005, Yang_et_al_2007}.  Although the characteristics of mass- and environment-quenching were identified, their physical origin remains uncertain. 
 
 Also unclear remains whether morphological transformations are causally connected with, and whether they anticipate or lag behind, the spectrophotometric transformations which shift blue, star forming galaxies onto the red sequence of bulge-dominated systems \citep[e.g.][]{Arnouts_et_al_2007,Faber_et_al_2007,Pozzetti_et_al_2010,Feldmann_et_al_2010,Feldmann_et_al_2011}. Many processes can lead to the disruption of disks and quenching of star formation, e.g., galaxy mergers or tidal interactions \citep[e.g.][and references therein]{Park_et_al_2007}, ram pressure stripping of cold gas \citep[][but see also Rasmussen et al. 2008]{Gunn_Gott_1972,Feldmann_et_al_2011} or ÒstrangulationÓ of the galactic system by removal of hot and warm gas, necessary to fuel star formation \citep{Larson_et_al_1980,Balogh_et_al_2000,Font_et_al_2008,Rasmussen_et_al_2012}. In a hierarchical picture, a gaseous disk can be re-accreted around pre-made spheroids at relatively late epochs. This evolutionary path is observed to happen in high-resolution cosmological hydrodynamical simulations \citep{Springel_Hernquist_2005,Feldmann_et_al_2010}.  

The intermediate-mass scales of galaxy groups, which are the most common environments of $\sim L^*$  galaxies in the local Universe (\citealt{Eke_et_al_2004}), have a reputation for being the place where environmental drivers of galaxy evolution should be at their peak efficiency. With an in-spiral timescale of dynamical friction that varies in proportion to $\sigma^3/\rho$, with $\sigma$ and $\rho$ the dark matter halo velocity dispersion and density, respectively, galaxy tidal interactions and mergers should take place on a cosmological short timescale in group potentials with relatively low velocity dispersion, unlike the most massive galaxy clusters where the velocity dispersion is much higher. 
Also, with ram pressure efficiency varying as $\rho_{igm}v^2$, with $\rho_{igm}$ and $v$ respectively the density of the intergalactic/intragroup medium (IGM) and relative velocity of the galaxy toward the IGM, galaxies may well begin to loose their gas already at intermediate environmental densities typical of galaxy groups \citep{Rasmussen_et_al_2006,Rasmussen_et_al_2008}. Resulting internal dynamical instabilities may also contribute to galaxian evolution,  e.g., by fuelling star formation
and supermassive black holes in the centers of galaxies (see e.g. \citealt{DiMatteo_et_al_2007,Hopkins_et_al_2008} for a theoretical perspective, and \citealt{Genzel_et_al_1998,Kewley_et_al_2006,Smith_et_al_2007,Silverman_et_al_2011} for some observational evidences) and establishing feedback loops that affect  whole galaxies  \citep{Croton_et_al_2006}.

These considerations motivate the present study, termed  \emph{ZENS} (the Zurich Environmental Study), where we use a statistically complete sample of  1627  galaxies brighter than $b_J=19.45$, known to be members of 141 nearby groups spanning the  mass range between $\sim10^{12.5} M_\odot$ and $\sim10^{14.5} M_\odot$. The \emph{ZENS} sample is complete at stellar masses above $10^{10}\Msol$ for passively evolving  galaxies with old stellar populations, and above $10^{9.2}\Msol$ for star forming galaxies. In \emph{ZENS} we aim at simultaneously $(1)$ characterizing the present evolutionary state of galaxies in as broad a way as possible, using both diagnostics based on stellar populations and structural morphology, and $(2)$ studying as broad a range of environments as possible and characterizing the environments in a number of ways that sample different physical scales, and include a careful distinction between central and satellite galaxies. Specifically, in our study we directly compare, at fixed galaxy stellar mass, the dependence of key galactic populations diagnostics on the large-scale environmental (over)density ($\delta_{LSS}$), on the mass of the host group halo ($M_{GROUP}$), on the location of galaxies within their group halos (the latter expressed in terms of projected distance from the halo center,  $R/R_{200}$, with $R_{200}$ the characteristic size of the group) Ð maintaining a central-satellite distinction when possible and relevant. 

The \emph{ZENS} sample is extracted from the 2-degree Field Galaxy Redshift Survey (2dFGRS; \citealt{Colless_et_al_2001}), which contains nearly 225,000 redshifts for galaxies with $14 < b_J < 19.45$ and a median redshift $z\sim0.11$, with a redshift completeness of $85\pm 5\%$. In combination with a dynamic range of five magnitudes at each redshift, the 2dFGRS is the ideal basis for constructing a homogeneous catalogue of nearby galaxies in a wide range of environments. We have followed up the \emph{ZENS} sample with $B$ and $I$ deep WFI imaging at the ESO/2.2m to derive, for all galaxies in the sample, detailed properties of substructure such as bulges, disks, bars and tidal tails. The wealth of data on the \emph{ZENS} groups enables us to define very carefully the nature of the group, including its likely dynamical state (relaxed or unrelaxed), to do a careful group-by-group identification of the most likely dominant member, to derive accurate photometric and structural measurements for galactic subcomponents (disks, bulges and bars) Ð all analyses unaffected by distance, size, magnitude, mass, type and other biases, which often complicate the interpretation of comparisons of independent studies published in the literature.

In this first paper in the \emph{ZENS} series:

$(i)$ We describe the \emph{ZENS} design and database (Section \ref{design});

$(ii)$  We present  our definitions and calculations of the four  environmental parameters $\delta_{LSS}$, $M_{GROUP}$,  $R/R_{200}$ plus the central-satellite distinction (Section \ref{sec:environs}). Specifically, in this section we detail the  approaches that we adopt to identify central and satellite galaxies and  thus the centers of the groups, and to measure a large-scale structure (over)density proxy which, at relatively low group masses, provides a measurement which, in contrast with the often used Nth-neighbor-galaxies estimators, is  independent of the richness and mass of the host group halos.  We furthermore quantify how  random and systematic errors in  the computation of each environmental parameter affect the studied trends of  galaxy properties with such environment;

$(iii)$ We publish the   \emph{ZENS}   catalogue  (Section \ref{sec:ZENS_catalogue}) which lists, for every galaxy in the sample, the environmental parameters derived in this paper, as well as structural (from \citealt{Cibinel_el_al_2013a}, Paper II) and  spectrophotometric  measurements (from \citealt{Cibinel_el_al_2013b}, Paper III). The structural measurements  are corrected for magnitude-, size-, concentration-, ellipticity-, and PSF-dependent biases; 

$(iv)$ We discuss our  classification of groups in dynamically `relaxed' and `unrelaxed' systems (Section \ref{sec:relaxedUnrelaxed}), and briefly investigate whether their galaxy members, both central and satellites, differ in fundamental structural (size), star formation (specific star formation rate, sSFR) and surface density of star formation rate ($\Sigma_{SFR}$) and optical $(B-I)$ properties (see also Appendix \ref{newtable}). Finally,
 
$(v)$ we summarize our main points in Section \ref{sec:endfinally}. 

In Appendices \ref{App:2dFGRS_limits}, \ref{App:missedgals}, \ref{App:densityTests}  and \ref{App:readmecat} we present details on, respectively,  $(a)$ the impact on our study of the 2dFGRS magnitude limits in the \emph{ZENS} fields, $(b)$ the impact of `missed' galaxies, either by the 2dFGRS, or by the new $B$ and $I$ ESO 2.2m/WFI imaging for the \emph{ZENS} sample,  $(c)$  2PIGG incompleteness  in group-membership,  and $(d)$ additional tests on the robustness of  our fiducial LSS density estimates and the comparison with traditional Nth-neighbor-galaxies estimators, and, finally, $(e)$ the $Readme$ file of the published \emph{ZENS} catalogue.

	For the relevant cosmological parameters we assume the following values: $\Omega_m=0.3$, $\Omega_{\Lambda}=0.7$ and $h=0.7$. Unless otherwise stated, group masses and luminosities are given in units of $\Msol$ and $\Lsol$, i.e., we incorporate the value $h=0.7$ in the presentation of our results. All magnitudes are in the AB system. These choices are also adopted in \citealt{Cibinel_el_al_2013a,Cibinel_el_al_2013b} which present respectively the structural and photometric measurements included in the catalogue associated with this paper.

\section{THE ZURICH ENVIRONMENTAL STUDY (ZENS)}\label{design}
\subsection{Design and Sample Specifications}\label{sec:designData}

The entire \emph{ZENS}  sample of 141 galaxy groups was selected from the 2dFGRS Percolation-Inferred Galaxy Group catalogue \citep[2PIGG,][]{Eke_et_al_2004}, which is based on a friends-of-friends (FOF)  \citep{Huchra_Geller_1982} percolation algorithm thoroughly tested on realistic mock galaxy catalogues generated from cosmological N-body simulations.  We refer to \citet{Eke_et_al_2004} for the details of the group finding algorithm and the procedures adopted for the identification of the groups. The 2PIGG catalogue covers the 1500 square degrees of the 2dFGRS and provides one of the largest homogeneous samples of galaxy groups currently available, with around 7000 groups with $\geq 4$ catalogued members and spanning a wide range in luminosity, from $\sim10^{10} L_{\odot}$ up to $\sim10^{12} L_{\odot}$, and dynamical mass from a few $10^{12} M_{\odot}$ up to clusters of mass $10^{15} M_{\odot}$ (\citealt{Eke_et_al_2004b}). The catalogue is selected from a volume of $\sim250,000$ (Mpc h$^{-1})^3$, and it is so large that one not only has information on the groups themselves from the 2dFGRS data (e.g., velocity dispersions, spatial positions of members, mass, density, compactness etc.), but also on their proximity to large clusters, filaments, and voids of the large-scale structure web. The 2PIGG catalogue is representative of the Universe as a whole, and contains a large number of groups that are close enough to allow detailed studies of the galaxy members. It is thus ideal for undertaking the study of the nearby galaxy properties as a function of the environment, and in particular, for directly comparing how galaxy properties and key galaxy population diagnostics depend respectively on group mass, on the location of galaxies within their host groups, and on the location of the host groups relative to the large scale filamentary structure (i.e. on the local density of the cosmic web).  The 2dFGRS  fields  are located well above the Milky Way disk, minimizing the effect of extinction from Galactic dust (typically 0.1 mag in the B-band). 

The \emph{ZENS} groups were randomly extracted from the complete sample of 185 2PIGG groups  (excluding few groups falling in very incomplete fields of the survey)  to be in the narrow redshift bin  $0.05<z<0.0585$ and to have at least 5 spectroscopically-confirmed galaxy members in the 2dFGRS. 
Note that,  by construction, \emph{ZENS} excludes both  ÔfieldÕ galaxies or groups with less that five galaxy members. The motivation for this selection is to increase the probability that the associated members are truly linked within a common halo. Within these selection boundaries, the \emph{ZENS} sample provides a statistically complete and representative census of the nearby galaxy population inhabiting the group environment.  

\begin{figure}
\begin{center}
\includegraphics[width=92mm]{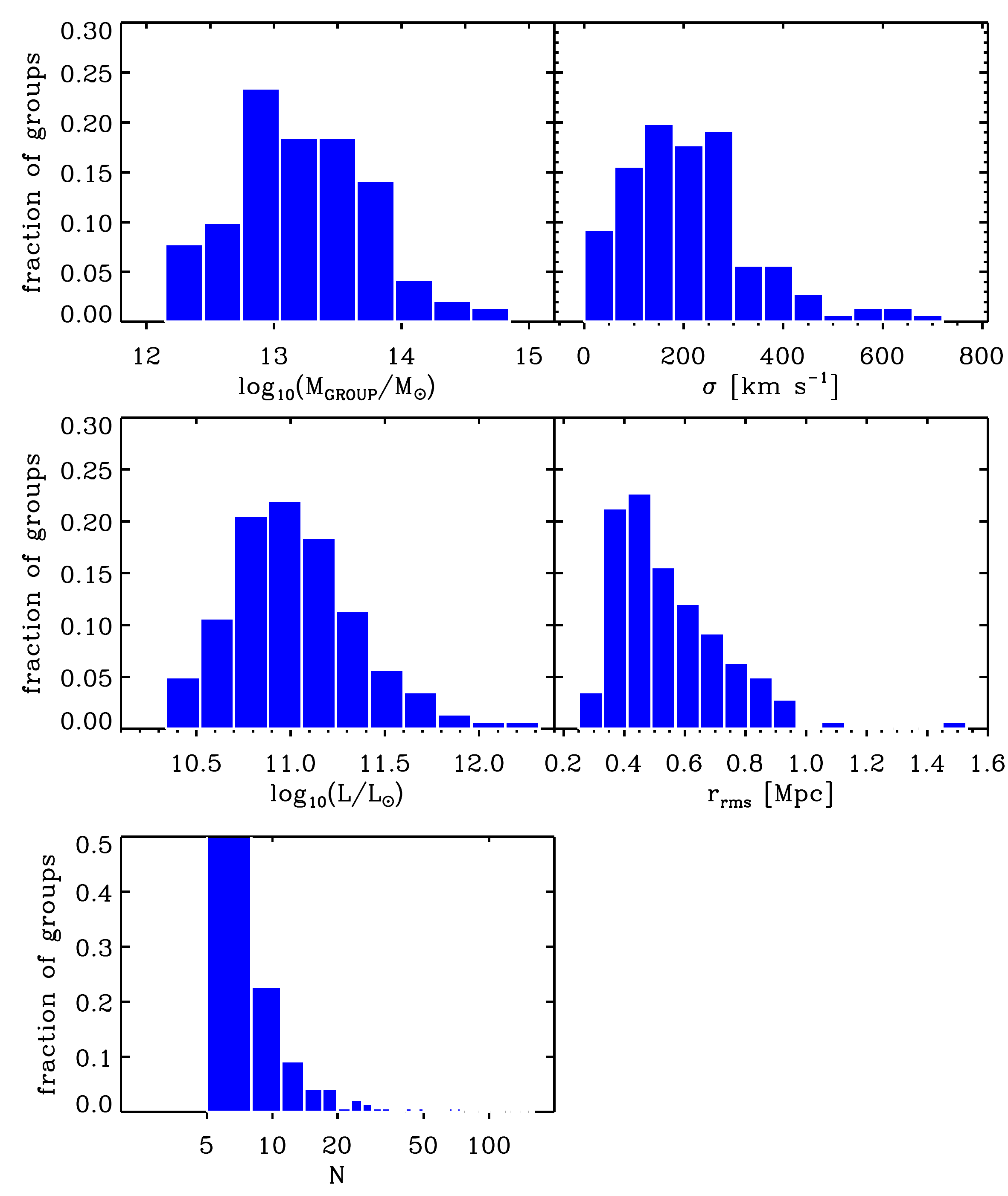}
\end{center}
\caption{\label{fig:groups_prop}
Distributions of main properties for the 141 \emph{ZENS} groups.  From left to right and top to bottom  we plot the distributions of \emph{ZENS} group masses (derived as described in Section \ref{sec:GroupMasses}), velocity dispersions and luminosities, and also the distributions of projected r.m.s. galaxy separations within the groups and of number of group member galaxies.}
\end{figure}

The very narrow redshift range of the \emph{ZENS} sample was chosen to optimize several issues: $(i)$ The 2dFGRS magnitude limits translate at this redshift to luminosities between $[M^*-2]$ to $[M^*+3]$ (\citealt{Norberg_et_al_2002}), meaning that the existing redshift catalogue already samples all of the luminosity function of massive galaxies and  straddles well the break or bimodality in galaxy properties around $M^*$ (\citealt{Kauffmann_et_al_2003}); $(ii)$ This redshift range is located just below the peak in $N(z)$ in 2PIGG, and thus ideally samples the targeted range of group mass $\sim10^{12.5-14.5} M_{\odot}$. $(iii)$ Likewise, the groups fully cover the entire range of large-scale structure environments, with some groups residing in very dense regions and others residing in much lower density environments, allowing us to study the effects of the large-scale structure on group and galaxy evolution. $(iv)$ At this redshift the group selection is robust and less affected by the peculiar velocities of the galaxies than is the case at lower redshifts. $(vi)$ Finally, deep, ground-based imaging with typical seeing $\sim1^{\prime\prime}$ is well suited for the determination of morphologies, substructure units such as bars, bulges and disks, and presence and properties of faint structures. It is also directly relatable to Hubble Space Telescope $\ll1^{\prime\prime}$-resolution images of the $z > 0.5$ Universe (with a relative angular diameter distance of a factor $\sim$8), and therefore provides an ideal benchmark for a direct comparison with HST images of galaxies in high-z groups (e.g. \citealt{Knobel_et_al_2009,Kovac_et_al_2010,Gerke_et_al_2012}).
	A summary of the  properties of the ZENS groups is given in Table \ref{tab:ZENs_prop} and Figure \ref{fig:groups_prop}.

\subsection{Impact of the 2dFGRS selection function on ZENS}\label{sec:2dFGRScompl}

\subsubsection{Impact of the 2dFGRS redshift incompleteness and of field-to-field scatter in the 2dFGRS magnitude limits}

The depth and completeness of the 2dFGRS are not uniform over the sky for a number of reasons \citep{Colless_et_al_2001}: 
the 2dFGRS parent catalogue (APM survey, \citealt{Maddox_et_al_1990}) was recalibrated and the extinction corrections were revised after the survey limit of $b_j=19.45$ was originally set; moreover the number of successful or repeated observations varies with position on the sky. We quantify 
in Appendix \ref{App:2dFGRS_limits} the impact of the original 2dFGRS  magnitude limits in our targeted fields. 
 These limits  translate into a minimum mass at which \emph{ZENS}  
 is complete; this corresponds to $10^{10}\Msol$ for ``quenched", red-and-dead galaxies with 
 old stellar populations, and to $10^{9.2}\Msol$ for galaxies with star forming SEDs. 
 A detailed description of the derivation of galaxy stellar masses and of the mass completeness limits of \emph{ZENS} is given in Paper III.

We also note that, unless explicitly stated otherwise, all statistical analysis that we present in this and in the following \emph{ZENS} papers  refer to quantities corrected for spectroscopic incompleteness of the 2dFGRS.

\subsubsection{Sample completeness tested on the SDSS}\label{sec:SDSSComparison}

We  investigated the resulting incompleteness  in the \emph{ZENS} group sample, originating from the catalogue limitations described above. To this end we searched the SDSS DR7 spectroscopic catalogue 
\citep{Abazajian_et_al_2009} for galaxies within up to $\pm30\%$ of the redshift distribution of a given \emph{ZENS} group, not present however in the parent 2dFGRS catalogue. The details of this comparison are given in Appendix \ref{App:SDSSMatch}.
 There are many aspects of our analyses that could in principle be
  affected by any such incompleteness, i.e., 
  the estimates for the group masses, the determination of the central galaxy and thus of the group centers,
  and consequently of radial trends with group-centric distance. 
The results of our tests, presented in the Appendix, 
   indicate however that our main results are not severely affected by significant
    biases due to incompleteness in the parent 2dFGRS catalogue.

\subsection{New B and I data with the ESO/2.2m Wide Field Imaging Camera}

\emph{ZENS} capitalizes on the wealth of data and measurements 
already available from the original 2dFGRS analyses and other surveys 
(e.g. GALEX, 2MASS and SDSS).
In particular, for all ZENS galaxies  $b_j$ and $r_F$ photometry is available from the  2dFGRS catalogue.
 However, pixel data for the original APM scans of the photographic plates is not provided. The digitalized version of the SuperCOSMOS Survey plates \citep{Hambly_et_al_2001}, which has also been used in the definition of the 2dFGRS photometric catalogue, are available on-line but with no photometric calibration information.
Furthermore, the resolution of these images ($\sim 2-2.5^{\prime \prime}$) is not suited for detailed structural analyses of typical galaxies in the sample.

In order to obtain accurate measurements 
for the structural \citep{Cibinel_el_al_2013a} and stellar population properties \citep{Cibinel_el_al_2013b}  of 
galactic subcomponents, we thus  acquired new  deep $B$  (\emph{$BB\#B/123\_ESO878$}) and $I$ (\emph{$BB\#I/203\_ESO879$}) images with the WFI at the MPG/ESO 2.2m telescope. 
 The WFI observations reach the limiting magnitudes, defined as the magnitude of an uniform area of 2$^{\prime \prime}$ having a signal 5 times higher than the typical noise,  of $B_{AB}=25$ mag and $I_{AB}=23.4$ mag, respectively (to compare with the corresponding depth of the $b_{J,AB}$ and $r_{F,AB}$ plates, which are $\sim 22.5$ mag and $\sim21.7$ mag). Note however that  our new deeper photometry  was not utilized to extend the group membership, in order to keep consistency with the original 2dFGRS/2PIGG catalogues for which the spectroscopic information is available. 
 
 The data were taken in several observing
 runs over the period 2005-2009. Following a pilot-project time allocation in 2005, most of the observations
  were carried out as service-mode observations in the context of the ESO Large Program 177.A-0680.
   For a combination of weather and technical issues, the service-mode observations were distributed over several runs 
    during the period 2006-2008. The last two observing runs were carried out in visitor-mode at the end of 2008 and 2009, 
    achieving a final sample of 141 groups randomly extracted from the original complete sample of 185 targets.

These 141 ZENS groups host a total of 1627 catalogued galaxy members brighter than the magnitude limit of the 2dFGRS survey, 1484 of which within our WFI pointings (see Appendix \ref{App:missedgals}; note that the vast majority of galaxies outside our WFI pointings have stellar masses below our completness limits discussed above, and would thus not be included in the majority of our analyses).
 For these 1484 galaxies\footnote{Note that there are only 1455 galaxy entries in the 2dFGRS catalogue corresponding to our 141 2PIGG groups. However, 29 of these entries correspond to galaxy pairs, for which we measure all quantities individually.} we have derived accurate structural measurements (sizes, bulge-disk decompositions, bar sizes and strengths, 
 non-parametric diagnostics such as concentration, Gini, asymmetry etc.\, as well as a quantitative, 
 robust morphological classification which corrects for seeing, inclination and dust effects; see Paper II) 
 and photometric measurements (e.g., colors, specific and total SFRs, stellar masses,
  for the whole galaxies, as well as for bulges and disks, including corrections for inclination, dust and fiber-area effects; 
  see Paper III). 
  
In  Paper II we provide details on the observing runs, the raw data properties, the data reduction procedures and the photometric calibration for WFI data of the \emph{ZENS} groups.

\subsection{Fossil groups in the ZENS sample} \label{sec:FossilGroups}

 The \emph{ZENS} sample includes groups whose luminosity budget is dominated by a single bright 
central galaxy, highlighting peculiar halo occupation properties, and thus formation or evolution histories.
 A widely used definition for such ``fossil" groups is that they show an absolute magnitude
gap between the most luminous galaxy and the second brightest member of $\Delta m_{12} > 2$ mag 
in the R band \citep{Jones_et_al_2003}. 

We adopt this optical criterion using the $r_F$ magnitudes which are available for the 2dFGRS galaxies from the SuperCosmos Survey \citep{Hambly_et_al_2001}.  
 To derive the $k$-correction from the galaxy $C=b_j-r_F$ color provided by the 2dFGRS and thus compute $r$-band absolute magnitudes, we use equation 3 of \citealt{Cole_et_al_2005}, $ k_{r_F}=(-0.08+1.45C)\frac{z}{z+1}+(-2.88-0.48C)\left(\frac{z}{z+1}\right)^2$. 
About 7$\%$ of  \emph{ZENS} groups (i.e., a total of 10 groups) satisfy the  fossil selection criterion above. 
These fossil groups are marked with an asterisk symbol in Table \ref{tab:ZENs_prop}.

Only in 3 out of the 10 \emph{ZENS} fossil groups the dominant galaxy is 
a giant E/S0 galaxy. These groups'  masses span a wide range,  from $\sim6\times 10^{12}\Msol$ up to  $\sim1.5\times 10^{14}\Msol$, similar to  fossil groups  which are found to host an early-type central galaxy  in other studies  \citep[e.g.][]{Ponman_et_al_1994,Romer_et_al_2000,Khosroshahi_et_al_2004}.
The mass range of the remaining seven fossil groups in our sample
which host morphological late-type centrals overlaps at the low end with the  range above, extending down to $\sim4\times 10^{12}\Msol$, but remains confined to generally lower masses $<2\times 10^{13}\Msol$. Interestingly, in 7 of  the 10 fossil-like groups, including two with  E/S0 centrals,  the 
central galaxy is either undergoing a merger, or is in a close pair with a 
satellite, or shows a disturbed morphology.
Only three \emph{ZENS} fossil groups, one of which hosting a central elliptical,  are 'unrelaxed' systems according to the definition which we described in Sections \ref{sec:centralsDef} and \ref{sec:relaxedUnrelaxed}, and none show sub-clustering according to the Dressler and Schectman test \citep{Dressler_Shectman_1988} that we present in Section \ref{subclass}.

The small fraction of compact groups that we find in our sample echoes other previous studies showing that such groups are, at any mass scale, a small  fraction of the population (e.g., \citealt{van_den_bosch_2007}; see however \citealt{yang2008} for larger estimates towards lower halo masses).
Unless explicitly stated, we will therefore include these groups in our \emph{ZENS} analysis.

\subsection{The strength of  ZENS}

Relative to previous work, the \emph{ZENS} database offers additional power in several aspects for studying environmental effects on galaxy properties. Previous detailed analyses have often focused on biased group samples, e.g., X-ray selected, `compact' or `fossil' groups \citep[e.g.][]{Lee_et_al_2004,McConnachie_et_al_2009,Harrison_et_al_2012}. While understanding these systems is important, a comprehensive study of the role of environment  on different scale necessitates a less biased selection of the  sample and a definition of a `group' that is as general as possible. As discussed in Section \ref{sec:designData},  \emph{ZENS} is fully representative of the local population of galaxy groups. 

Other studies have   adopted similarly general identification and selection criteria as those employed in 2PIGG, from which ZENS is extracted, to produce large group catalogues for the major surveys \citep[e.g.,][]{Merchan_Zandivarez_2002,Yang_et_al_2005,Berlind_et_al_2006,Yang_et_al_2007,Tago_et_al_2008,Tago_et_al_2010,Calvi_et_al_2011,Robotham_et_al_2011,Tempel_et_al_2012}.  Relative to these efforts, \emph{ZENS} trades off sample size to measurement accuracy. The relatively smaller size of the \emph{ZENS} sample enables us to analyze and measure properties for each of the galaxies individually, rather than rely on automatic algorithms which introduce a substantial `impurity' in the measurements and thus classifications of larger galaxy samples.  Furthermore, in \emph{ZENS} we have attempted to identify problems inherent in standard definitions of the different environments, and to implement  solutions that minimize or at least flag such potential causes for signal contamination. Finally, all environmental and galactic estimates performed on the \emph{ZENS} sample have been accurately calibrated against several intrinsic and observational biases, as discussed below (see also Papers II and III for details on the derivation of the structural and morphological galaxy parameters, respectively). The sum of the above  minimizes or, when possible, eliminates uncertainties, often of order $\sim30\%-40\%$, which affect  statistically-handled measurements in the larger samples, giving  \emph{ZENS} a unique niche to study aspects of galaxy evolution in group halos which are complementary to those that are enabled by the larger but less detailed samples.

\section{Four Environments in Comparison with One Another}\label{sec:environs}

With the goal of identifying {\it which} environment affects galactic evolution at different galaxy mass scales, we quantify four environmental diagnostics that will enable us to search for differential trends with these environments in different galactic populations. The four environments are, respectively: $(1)$  the mass of the host group halo, $(2)$ the distance of a galaxy from the center of its group halo, $(3)$ the average LSS density at the position of the host group, determined by the underlying filamentary structure of the cosmic web, and $(4)$ the central versus satellite dichotomy, considered here to be also an  Òenvironmental conditionÓ that  galaxies experience in their life within a bound common halo.

\subsection{Environment number one: The mass of the Host Group Halo} \label{sec:GroupMasses}

The 2PIGG catalogue from which we have extracted the \emph{ZENS} sample lists the velocity dispersions $\sigma$ returned by the friends-of-friends algorithm which was used to construct the catalogue, and the radii of the groups (reported in Table \ref{tab:ZENs_prop}), defined as the weighted r.m.s. of the projected separations between the nominal 2PIGG center and the remaining group members. \cite{Eke_et_al_2004}  and \cite{Eke_et_al_2004b} discuss in detail the tests performed to ascertain the robustness of these estimates, which were optimized to best reproduce the global properties of the 2dFGRS mock catalogues \citep{Eke_et_al_2004}. Dynamical halo masses computed as $M_{dyn}=5\frac{\sigma^2r_{rms}}{G}$ are however affected by large uncertainties, especially for groups with a relatively low number of members (redshift errors  in the 2dFGRS of are typically $\sim 70$ km s$^{-1}$ at $z\sim0.05$).  In contrast, group total (stellar) luminosities can be measured with a  higher accuracy even in poor groups. Using mock catalogues, \citet{Eke_et_al_2004b}  calibrated the observed group total luminosities into total group masses, providing robust estimates for the halo mass-to-light ratios ($\Upsilon_{b_j}$) needed to convert the ${b_j}$ luminosities into total halo masses $M_{GROUP}$. We adopt such  ${b_j}$ luminosity-based halo masses as our fiducial estimates for the matter content of the \emph{ZENS} groups\footnote{Note that, although an expression for  $\log_{10} \Upsilon_{r_F}$ is also provided by Eke et al.\ (their Eq. 4.5), we  chose to use $b_j$ luminosities for our estimates, since the overall shape of the mass-to-light ratio vs. luminosity relations is very similar in both  passbands, as discussed by \citet{Eke_et_al_2004b},  and the  $\log_{10} \Upsilon_{r_F}$ estimates  are anyhow based on the total $b_j$  luminosity.}. 

Specifically, following the prescription of \citet{Eke_et_al_2004b}, we computed the observed group 
luminosity as the weighted sum of the luminosities of the individual galaxy members 
 $L_{i,b_j}$, i.e., $L_{GROUP,OBS}=\sum_i^N w_i L_{i,b_j}$, with $N$ the number of member 
 galaxies in the group, $w_i$ the weights used in the construction of the 2PIGG catalogue that account for the 2dFGRS redshift incompleteness,
and  $b_j$  the original apparent (Vega) 2dFGRS magnitudes.
 The latter were converted to absolute magnitudes by applying the 
 mean $k+e$ correction as given in equation 2.4 of \citet{Eke_et_al_2004b}, 
 $k+e=\frac{z+6z^2}{1+8.9z^{5/2}}$.
  This observed luminosity was corrected into  total luminosity ($L_{GROUP}$)
   by integrating a Schechter function to zero luminosity, namely by dividing $L_{GROUP,OBS}$ 
   by the incomplete Gamma function $\Gamma(\alpha+2,L_{min}/L_{*})/\Gamma(\alpha+2)$. 
   In the above formula $(\alpha,L_*)$ are the slope and cut-off luminosity of the Schechter 
   function, and $L_{min}$  is the luminosity corresponding to the magnitude limit 
   of the 2dFGRS survey at the considered position in the sky.  The correction was done by keeping the slope and $L_*$ fixed for all  groups and assuming the values $\alpha=-1.18$ and $M_*=-19.725$,
obtained by \citet{Eke_et_al_2004b} from a global Schechter function for the 2dFGRS galaxies. Our fiducial total halo masses $M_{GROUP}$ were finally obtained using eq. 4.4 in \cite{Eke_et_al_2004b} for the mass-to-light ratio: $\log_{10} \Upsilon_{b_j}=2.28+0.4\tanh[1.9(\log L_{GROUP}-10.6)]$.

As expected, the comparison of the so obtained masses for the \emph{ZENS} groups with the dynamical  $M_{dyn}$ estimates defined above shows that their difference decreases  with increasing $M_{GROUP}$,  from $\Delta( \log M)\equiv \log_{10} M_{dyn}$-$\log_{10}M_{GROUP}$=0.40 $\pm$0.12 at $M_{GROUP}<10^{12.7}\Msol$ down to  $\Delta (\log M)$=-0.01 $\pm$0.06 at $M_{GROUP}>10^{13.5}\Msol$. 
The median difference for the total ZENS sample is $\Delta (\log M)=0.12\pm0.05$.
In our  future \emph{ZENS} analyses we will systematically test and report whether any of our results will significantly change if the dynamical mass estimates were to be used  instead of our adopted definition of $M_{GROUP}$; this is not the case for any of the analyses  of Paper I, II and III.

\subsubsection{Sources and Effect of Errors on our Fiducial Group Masses}

The conversion of group luminosity into dark matter halo mass outlined above is affected by several factors, in addition to errors in the relevant galactic measurements (such as redshifts and luminosities). The most important additional contributions to the uncertainty in the conversion come from $(i)$ errors in group membership, either by ``missing" group members above the galaxy luminosity completeness cut in the survey, or by including interloper galaxies which are not physically associated with the given group; $(ii)$ peculiarities in the groups' luminosity functions; and $(iii)$  the intrinsic uncertainty/scatter in group mass-to-light ratio relation, partly due to uncertainties in the physics underlying this conversion. These sources of error on the group masses are not easy to eliminate, and thus we try to assess their impact on our analyses.

\paragraph{Erroneous group membership assignments: Interlopers and `missing'  galaxies} 
\label{sec:MembershipTest}
How sensitive are our halo mass estimates to the spurious inclusion Ð or exclusion Ð of member galaxies
 from a given group? The most obvious source of such kind of error is a non-optimal performance of the
  friends-of-friends algorithm used to generate the 2PIGG catalogue.  
     
We start by assessing the impact of interlopers. At the typical redshift and mass scale of the \emph{ZENS} groups, the 
level of contamination from such interlopers ranges between $20 - 40\%$, depending 
on the  mass of the group; this is clear from Figure 2 of \citealt{Eke_et_al_2004}, which shows the fraction of interlopers as a function of halo mass for the parent 2PIGG sample. Note that similar levels of interloper fractions are found in other studies (see e.g., the \citealt{Yang_et_al_2005} compilation, further discussed in Section \ref{sec:YangComp}). 
Interlopers are thus a substantial source of error in the estimate of the group masses, and may also affect   estimates of   global properties of galaxies in groups. Since we have no a priori knowledge of which galaxies could be ÒfalseÓ members, we
 need to approach this problem in a statistical manner. 
 
\begin{figure}
\begin{center}
\includegraphics[width=65mm,angle=90]{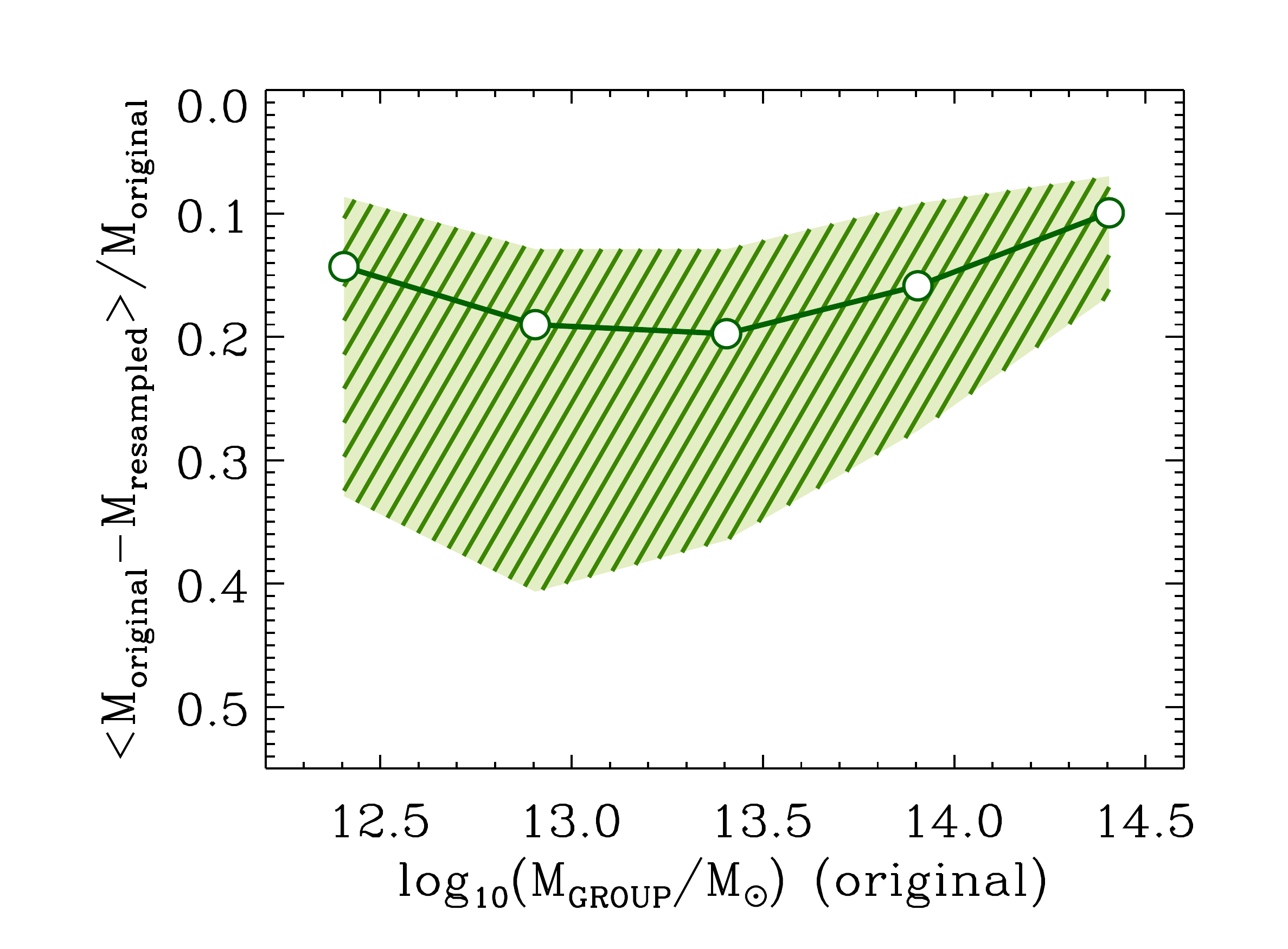}
\end{center}
\caption{\label{Fig:MassError}Quantification of the effect of interloper galaxies on  group halo mass estimates. The global  effect is an overestimate of typically $\sim20\%$, up to $\sim40\%$ at low masses. The solid line shows the  median of the distributions of differences, normalized to the fiducial mass of a group as determined in Section \ref{sec:GroupMasses}, between the fiducial mass (`original') and the mass in a given bootstrap realization (`resampled') obtained  by removing, for each group, between 20 - 40$\%$ of the member galaxies (also including potential extra galaxies, as discussed in Section \ref{sec:MembershipTest}). The  points are the corresponding averages within bins of group mass of width 0.5 dex; the shaded area shows the typical scatter around the median value. }
\end{figure}

 To specifically estimate the impact of interlopers on the group mass estimates, we  removed from each group,
 in 3000 bootstrap realizations, a random $20 - 40\%$ of the member galaxies 
(both including, and, in a separate set of tests, excluding the  possible extra candidates  discussed below in this Section). For each realization we recalculated the mass of the groups from the luminosity function, as outlined in 
 Section \ref{sec:GroupMasses}.
  Note that for groups with low numbers of galaxies this is 
 equivalent to spanning all the possible combination of rejected galaxies.  
 For  groups with $N_{gal}>15$ this is not the case anymore; however, 
 given the large number of bootstrap sample we employ in the analysis, 
 the derived distributions will be representative of the complete mass range.
This bootstrap approach provides us, for each group, with a distribution of masses for each  configuration, and a median of this distribution.
 The results are shown in Figure \ref{Fig:MassError}. 
 The symbols in the plot are the medians, in different bins of group mass, of the distributions of differences,
  normalized to the fiducial mass of a group (as determined in Section \ref{sec:GroupMasses}), between the fiducial mass of that group (`original') and that in a given bootstrap realization (`resampled'). The shaded area shows the 1$\sigma$ scatter around the plotted medians.

 The uncertainty on the mass estimates driven by the presence of the interlopers is between $15\%$ at the highest masses, and $40\%$ at and below $\sim10^{13}$M$_\odot$; this trend with halo mass is a direct consequence of the trend with halo mass of the fraction of interlopers already assessed by Eke et al.
 We also note that the uncertainty on the group mass reported in Figure  \ref{Fig:MassError} is in the direction of an overestimate relative to the true value. This is the result of interlopers adding to the real group members, and, by contrast, of a  negligible number of galaxies having been 'missed' by the FoF algorithm that has been used to generate the parent 2PIGG sample.
 
 To quantitatively assess  this latter issue, we searched 
  the parent 2dFGRS catalogue for galaxies which in principle could have been physically associated
   with a 2PIGG group, but were not included in that group by the 2PIGG algorithm. In particular, 
   using similar criteria to those used to search for missed galaxies in the SDSS (Section \ref{sec:SDSSComparison} and Appendix \ref{App:SDSSMatch}), 
   we searched the 2dFGRS catalogue for galaxies within $\pm30\%$ of the redshift window
    of our \emph{ZENS} groups, lying within a circular projected areas, centered on the central galaxy, of radius equal to 1.5 times the r.m.s. radius of the group, and which were not associated with the given group by the  2PIGG algorithm. For those groups for which our centers differed from the original 2PIGG centers, the choice of using our own definition for the centers also enables us to simultaneously test the impact of this definition on the resulting group membership.
   
	Following these criteria we found a total of 52 `extra' 2dFGRS candidate galaxy members
    for 24 of the  \emph{ZENS} groups. We present in Appendix \ref{App:ExtraCandidatesProp} details on
     the spatial and velocity distributions of these new candidates in relation to the galaxies
      which are identified as group members in the 2PIGG catalogue, as well as the distributions of 
      fiducial halo masses for the groups which may miss these extra candidates. 
   
These potential extra members are found for relatively massive \emph{ZENS} 
   groups with $M_{GROUP}\gta 10^{13} M_\odot$. Including these potential galaxy 
   members in the computation of the group masses has therefore a small effect at these mass scales, 
   typically within 0.1 dex. Only for two groups the difference between the fiducial group mass and 
   the recalculated mass is larger than this value (of order 0.2-0.3 dex). 
   We therefore conclude that missing 2dFGRS galaxies in the 2PIGG associations is not a dominant source of error in the computation of the fiducial  \emph{ZENS} group masses.

We conclude that interlopers are a $\sim20\%$ (at high group masses) up to $\sim40\%$ (at low group masses) source of overestimation  of   halo mass.
Note also that interlopers may affect the identification of the central galaxies and thus the determination of the dynamical state of a group; again, the impact is expected to be larger at lower group masses, since the latter suffer from a higher fraction of interlopers (see discussions in   Sections \ref{sec:cenErrors} and \ref{sec:relaxedUnrelaxed}).
Finally, as mentioned above, interlopers may also contaminate estimates of  properties of galaxies in groups.  To monitor their impact on our ZENS analyses, we will systematically  take into account the uncertainty on possible trends with halo mass that they introduce, as outlined in Section \ref{sec:nlab}, and will establish through statistical simulations the impact of their contamination on the global properties of group member galaxies.

\paragraph{Comparison with independent group catalogues for the 2dFGRS}\label{sec:YangComp}

Another question we ask is of how sensitive are the halo mass estimates, as inferred in Section \ref{sec:GroupMasses}, to the details of the algorithm adopted for identification of the groups. To learn about this issue in a `post-processing' approach, we cross-matched the \emph{ZENS} sample, extracted from the 2PIGG catalogue, 
with the independent 2dFGRS group catalogue of  \citet{Yang_et_al_2005}.   
In this catalogue the basic identification of potential groups also follows a friends-of-friends algorithm,
but the final group membership -- hence properties -- are iteratively refined, starting from the initial FOF estimates, by assuming that groups at any redshift are described by a NFW density profile \citep{Navarro_Frenk_White_1997} and that the phase space distribution of galaxies is similar to that of dark matter particles. 
The membership assignment is made by comparing the local density contrast, calculated following the above halo occupation recipe,  with a background level which is comparable to the density in the outskirts of a halo and that is chosen such to minimize the contamination while maximizing the completeness of the groups.

Furthermore, the Yang et al.\ 2dFGRS catalogue imposes a minimum redshift completeness of $80\%$.  The parent 2dFGRS catalogue adopted by Yang et al.\ hence excludes galaxies (the sample used is 25$\%$ smaller than the one in 2PIGG), which means that  128  of our 141 \emph{ZENS} sample groups can
  be compared with the group catalogue of these authors.  Note that  not all galaxies in a \emph{ZENS} group are necessarily  associated to a group in the Yang et al.\ catalogue. In 21 of the  128 groups in common, only one galaxy is found in the Yang et al.\, catalogue.
  
 \begin{figure}
\begin{center}
\includegraphics[width=65mm,angle=90]{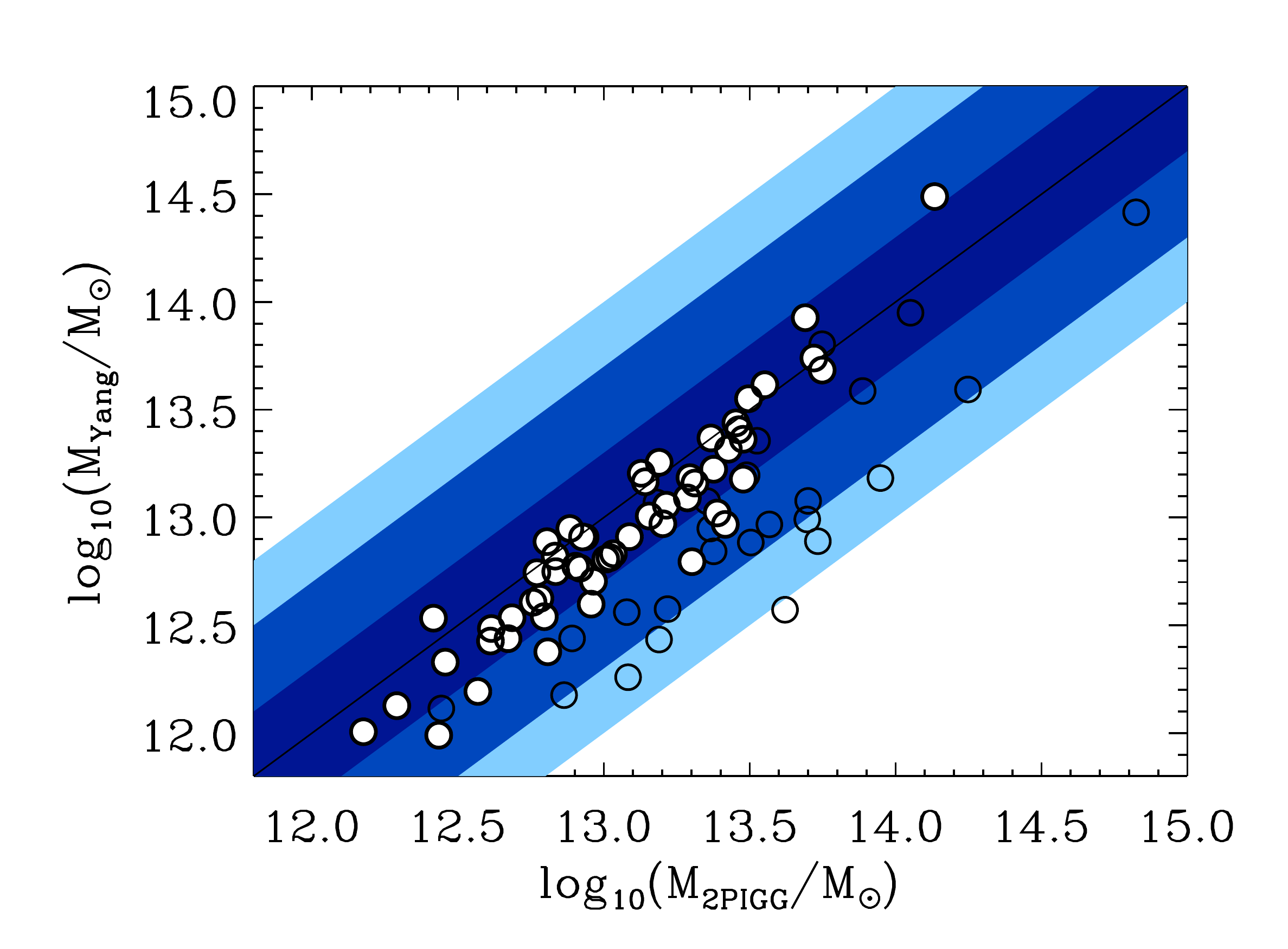}
\end{center}
\caption{\label{Fig:YangComp}Comparison between the \emph{ZENS} group masses as 
originally defined by the 2PIGG catalogue and the mass of the cross-matched groups 
in the \citet{Yang_et_al_2005} catalogue. 
The latter are defined as the most massive groups in this catalogue which are associated
 to each \emph{ZENS} group. The shaded areas show  a factor of two (dark-color strip), 
 five (intermediate) and ten (light-color strip) difference in mass.
Groups which are fragmented into more than one group in the Yang et al. 
 catalogue are shown with empty points. Note that not all galaxies in \emph{ZENS} 
 group are necessarily associated to groups in the Yang et al. catalogue.
\emph{ZENS} groups which are associated to a single galaxy in the Yang et al. group catalogue
 or are fragmented into sub-groups containing less than $40\%$ of the 2PIGG members 
 are not shown in this figure.}
\end{figure}

In practice, for any given group in the \emph{ZENS} sample, we searched in the Yang et al.\ catalogue 
to which of its groups the \emph{ZENS} member galaxies were assigned, and
 we associated the most massive of the Yang et al.\ groups so identified to the given 
 \emph{ZENS} group.
We show in Figure \ref{Fig:YangComp}  the
comparison between the corresponding group masses in the two catalogues, for \emph{ZENS} groups which are associated with  Yan et al.\ groups which contain  at least $40\%$ of the original galaxy members.

\begin{figure*}
\begin{center}
\includegraphics[width=\textwidth,height=0.9\textwidth]{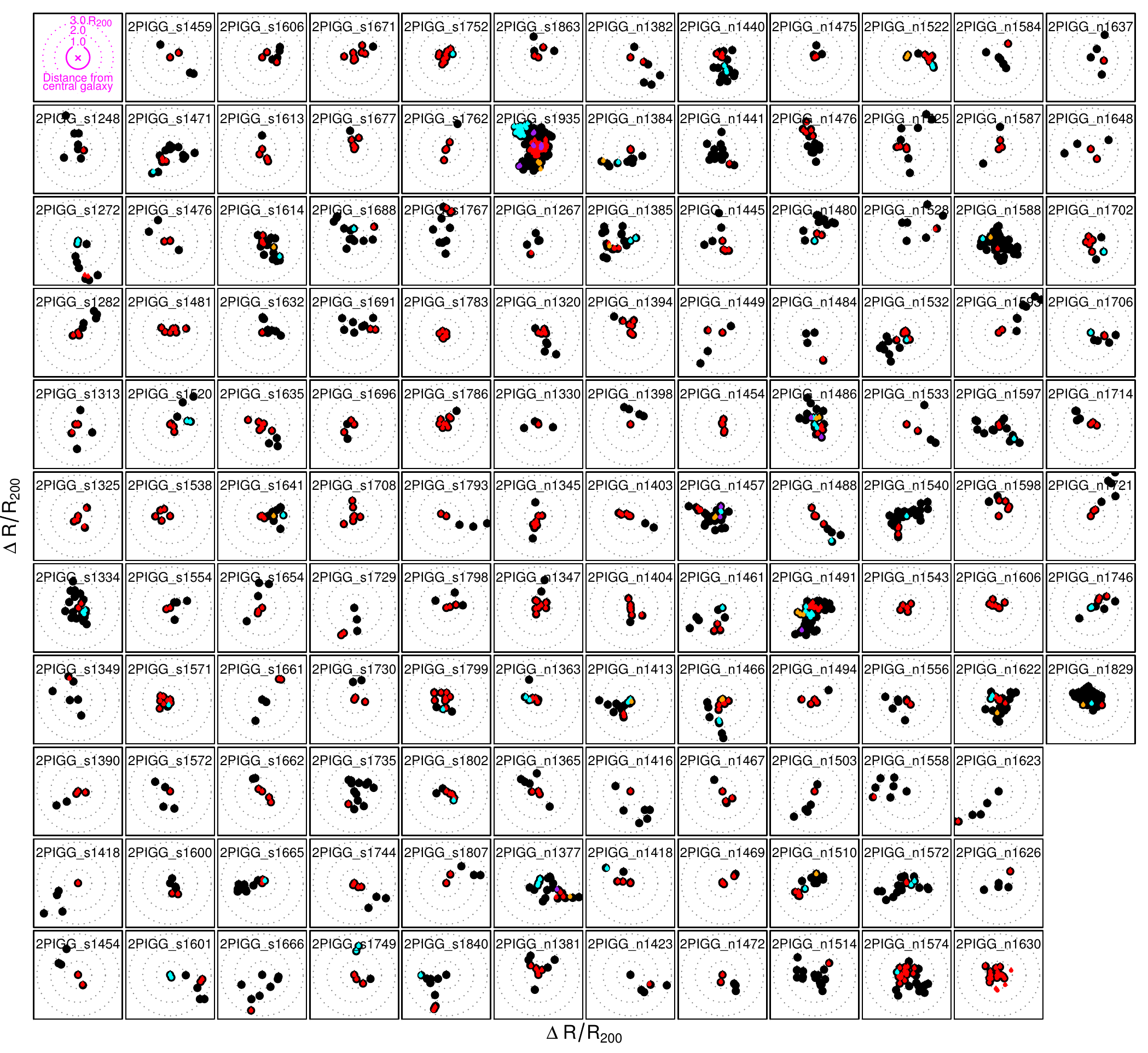}
\end{center}
\caption{\label{fig:YangFragmentation} For each of the 128 \emph{ZENS} groups which have at least one matching group in the Yang et al.\, group catalogue,  shown is the spatial distribution, with respect to the 
identified central galaxy, of the nominal 2PIGG members (black dots). Highlighted  with   different colors are galaxies that are identified as members of different groups in the Yang et al.\,  catalogue (same color identifies galaxies linked into the same group in the Yang et al.\, catalogue; black symbols  are galaxies that do not belong to any group in the Yang et al.\ compilation.}
\end{figure*}

 The masses of the Yang et al.\ groups matching the \emph{ZENS} groups are typically smaller than the
 fiducial \emph{ZENS} masses as estimated in Section \ref{sec:GroupMasses}.
 An inspection of the two cross-matched catalogues shows that, in most cases, 
 the \emph{ZENS} groups are fragmented into smaller subgroups in the Yang et al.\ catalogue.  
 This is shown in Figure \ref{fig:YangFragmentation},
  which plots, for each of the 128 groups that appear  in both catalogues,
 the position of the nominal 2PIGG member galaxies relative  to their `central galaxy' (see Section \ref{sec:centralsDef}),
with highlighted in different colors galaxies associated with different groups in  the Yang et al.\,  catalog. 
Note that some of the 2PIGG/ZENS groups are fragmented in this catalogue  into single galaxies.
Factors that contribute to these differences include the attempt to  take into account, in the Yang et al.\, catalogue, of  the effects of interlopers discussed above, and also missing galaxies in the input 2dFGRS catalogue that these authors  adopt, as discussed above.  
 
\begin{figure*}
\begin{center}
\includegraphics[width=\textwidth,height=0.9\textwidth]{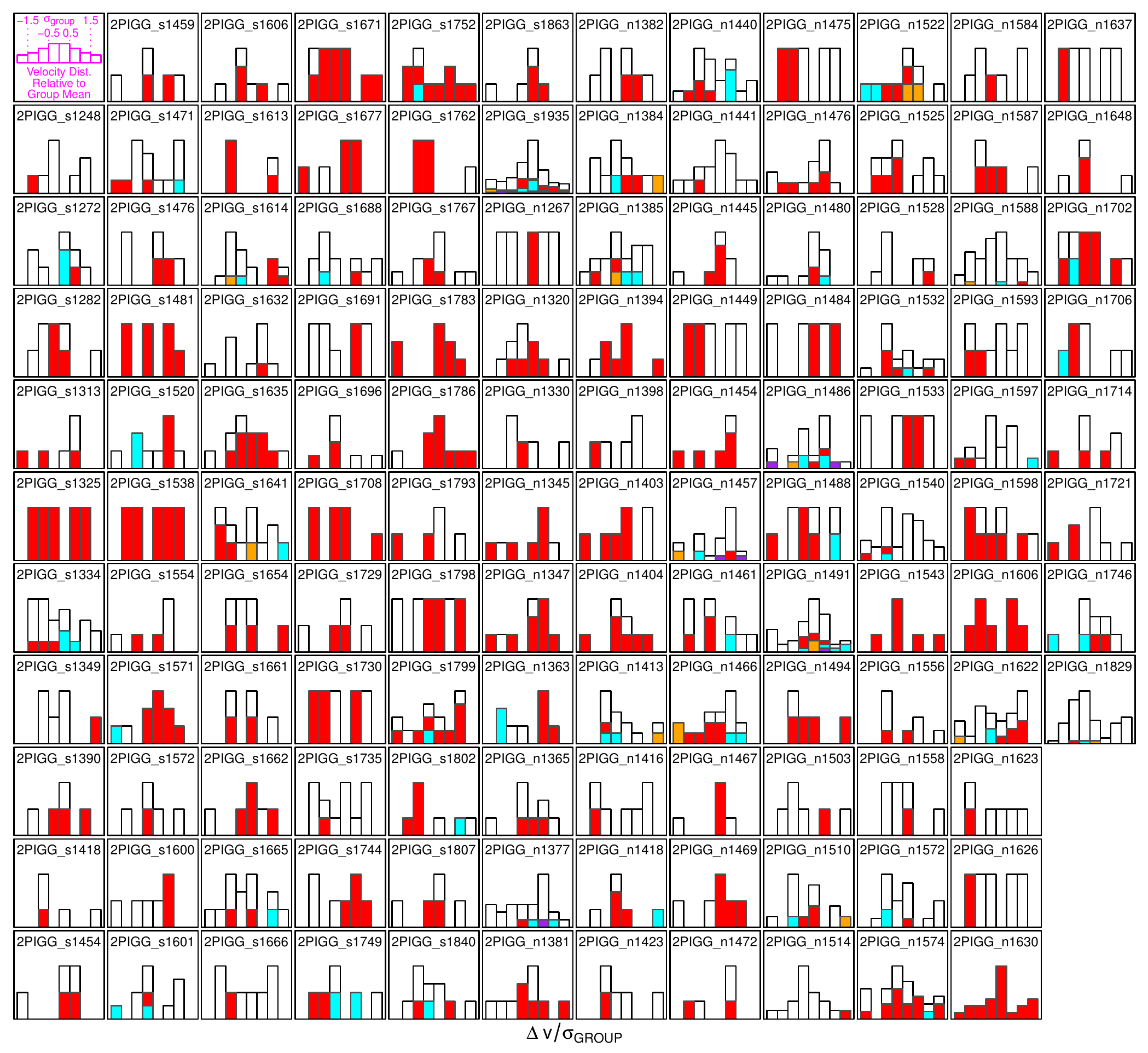}
\end{center}
\caption{\label{fig:YangFragmentation2} As in Figure \ref{fig:YangFragmentation}, but for   the distribution of velocities of 2PIGG group member galaxies relative to their group redshift   (black histograms). Superimposed in a different color, using the identical color scheme as in Figure \ref{fig:YangFragmentation}, are the corresponding histograms for galaxies that are identified as members of different groups in the Yang et al.\,  catalogue (same color for galaxies that are linked into the same group in the Yang et al.\, catalogue).}
\end{figure*}

Indeed in tests performed by \citet{Yang_et_al_2005} these authors show that their grouping algorithm performs slightly better in terms of reducing the interlopers fraction with respect to the standard FOF algorithm used in 2PIGG, especially for the case of flux limited samples (see their Figure 7). Overall,
 their fraction of interloper galaxies remains however at the $\sim 20\%$ level, i.e.,  coarsely  comparable to the $\sim20-30\%$ fraction  estimated by \citet{Eke_et_al_2004} on their mock 2dFGRS catalogues for groups with $M\lesssim 10^{14}\Msol$ (see their Figure 2). At the same time, the fragmentation of the 2PIGG groups in the Yang et al.\ compilation not necessarily leads to a cleaner definition of the bound structures. This is illustrated  in Figure \ref{fig:YangFragmentation2}, where we plot the velocities of the galaxies in the  Yang et al.\, catalogue with respect to the group mean redshift and the velocity of the nominal 2PIGG group members. Galaxies that are assigned to   distinct groups  in the Yang et al.\ catalogue have often positions and velocities  within the extremes in these quantities shown by the galaxies that are assigned to a single group by the 2PIGG algorithm. 
 
The different group identification methods and treatments of interloper galaxies used  in the two studies may  furthermore affect the division into dynamically `relaxed' and `unrelaxed' groups, as well as the central vs. satellite distinction,  that we will describe respectively in Section \ref{sec:relaxedUnrelaxed} and Section \ref{sec:centralsDef}. 
Assuming that, ideally,  fully virialized groups should be equally well identified in both catalogues, we can use  mismatches in the identification of the state of a  group  (`relaxed' and `unrelaxed', see Section \ref{sec:relaxedUnrelaxed})  to assess how much this latter classification, as well as the identification of  the central galaxies, is affected by the different algorithmic choices. To make a sensible comparison, we use  \emph{ZENS} groups for which at least 50$\%$ of the 2PIGG members fall in the Yang et al. galaxy selection, for a total of 121 -- respectively 67  and 54 of our `relaxed' and `unrelaxed' -- groups, i.e., 86\% of the \emph{ZENS} catalogue.

Reassuringly, the ZENS `relaxed' groups are well-identifiable structures with similar properties also in the Yang et al.\ catalog. A fraction $\sim60\%$  of our relaxed groups are matched by a single Yang et al.\  group  with $>50\%$ of the original galaxy  members, and in all but four ($89\%$) of these groups the identification of the central galaxy is confirmed  also in the Yang et al.\ counterparts.
On the other hand,  for `unrelaxed' groups the fraction of 2PIGG  systems which are matched by a single Yang et al.\  group  with$>50\%$ of the original galaxy  members  decreases to  $\sim25\%$.  This supports the picture where  groups that are classified as relaxed are  genuinely bound structures, coherently detected by both algorithms, while groups  classified as unrelaxed are likely a more heterogeneous class, including  both  structures which are genuinely dynamically-young, as well as systems whose identification and characterization is affected by observational limitations. 
This suggests that  only about a quarter of the nominally unrelaxed  groups, i.e., about $10-15\%$ of  all \emph{ZENS} groups,
are genuinely dynamically-young systems in an early stage of assembly. This estimate   is also supported by  comparisons with the SDSS catalogue,  which we have performed to quantify the impact of spectroscopic incompleteness in the 2dFGRS on the relaxed/unrelaxed classification, and also by  a  group sub-clustering analysis (see Section \ref{sec:relaxedUnrelaxed}).   

Summarizing, it is important to keep in mind that  different parent sample selections and grouping methods, e.g., those adopted in the Yang et al.\ and in the 2PIGG  catalogues, can undoubtedly results in individual  cross-matched groups with different  properties (in particular, as already discussed,   systematically lower number of member galaxies  and thus  systematically lower group masses in \citet{Yang_et_al_2005} than  in 2PIGG). Nevertheless,  $(i)$ for $75\%$ of the \emph{ZENS} groups,  the difference in their mass estimates remains smaller than a factor of two. 
 This uncertainty is comparable with the error in the total group mass that we estimated in Section \ref{sec:Lum2DM}, and does not affect  our study of galaxy properties as a function of (also) group mass over about two order of magnitudes in the latter; $(ii)$ the determination of the (apparent) `dynamical state' of the groups remains in the majority of cases stable, independent of the specific choices for the group identification algorithm, and $(iii)$  the identification of the central galaxies in relaxed groups is very robust.

\paragraph{How well do group halo masses inferred from integration of luminosity functions approximate the true halo masses?} \label{sec:Lum2DM}

Existing physical trends in galaxy properties with group mass may be smeared across different group mass bins and even washed out by the relatively large random and systematic errors in the estimates of the group masses. To quantify the uncertainties introduced by the specific algorithm of Section \ref{sec:GroupMasses}, that we adopted for computing the group masses, we used the Millennium simulation \citep{lemson} with the semi-analytic model of De Lucia \& Blaizot (2007). The model is not specifically designed to match the 2dFGRS properties and selection function; nonetheless, it enables us to gain useful insight on the limitations of our analysis. 

Details of the model are given in the original reference. The dark matter component is taken from the Millennium simulation (Springel et al.\, 2005b); details of the baryon physics are 
added in the fashion that is customary in semi-analytic modeling of galaxy evolution. Recipes for gas cooling, star-formation, AGN and supernovae feedback are included. The typical baryonic resolution of the adopted models corresponds to a galaxy mass of $\sim3\times10^9 \Msol$, comparable with the limiting mass of completeness for \emph{ZENS} galaxies.

For each halo  of mass  above $10^{12.2} M_\odot$ in the  volume we computed the total 2dFGRS $b_j$ luminosity of the host galaxies above the survey limit of $b_j=19.45$. A thousand random realizations were obtained, sampling each halo with 80\% of the galaxies, to simulate the spectroscopic completeness of our sample. The `inferred'  halo mass  was obtained adopting the same approach that we used for the \emph{ZENS} groups (see Section \ref{sec:GroupMasses}).
	
\begin{figure}
\begin{center}
\includegraphics[width=60mm,angle=90]{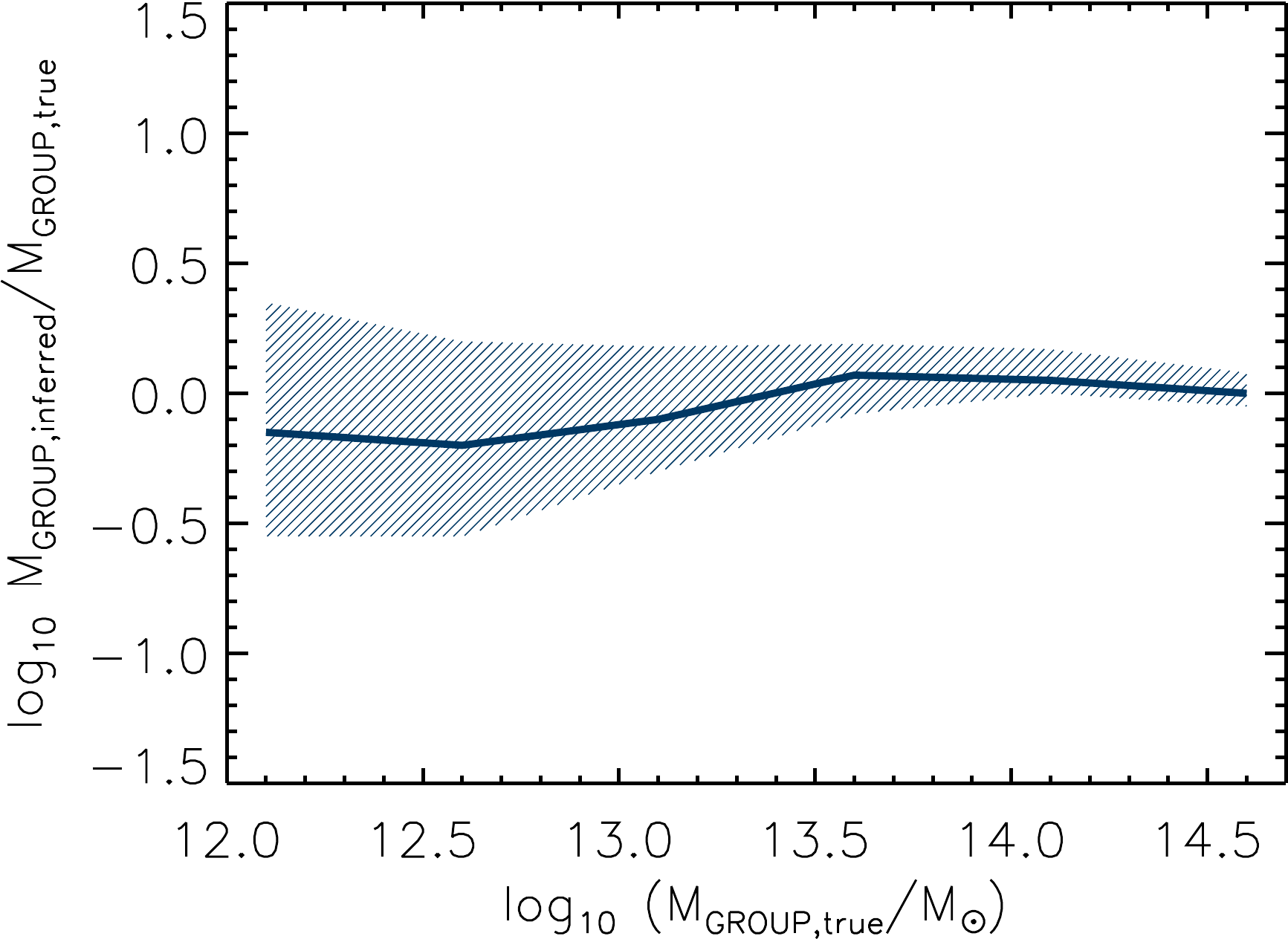}
\end{center}
\caption{\label{fig:lum2mass} Relation between inferred halo mass, estimated from the total group luminosity as
discussed in Section \ref{sec:GroupMasses}, and input (`true') dark matter halo mass from the DeLucia
\& Blaizot (2007) semi-analytic model applied to the Millennium simulation \citep{lemson}.
The shaded area show the scatter around the median relation (solid line) obtained in a thousand
random realizations.
}
\end{figure}

The comparison between `true' dark matter halo masses, and halo masses 
inferred from the extrapolation of the group luminosity function, is shown in Figure \ref{fig:lum2mass}. 
The  shaded area in Figure \ref{fig:lum2mass} shows the scatter around the median relation in the thousand  realizations. 
The average trend tracks the 1:1 relation above $M_{GROUP}\sim 10^{13.5} M_\odot$ with a modest scatter $<10\%$. At lower group masses, the data tend to underestimate the true values  on average  at the  $\sim20\%$ level, with a scatter up to $40\%-50\%$. 
Not surprisingly, the scatter in the relation decreases with increasing mass of the group. 

These uncertainties are comparable with the errors in the inferred group total luminosities 
 as a function of number of member galaxies shown in Figure 3 of \citet{Eke_et_al_2004b}.

\paragraph{Summary: Impact of the uncertainties on group masses on trends of galaxy properties with halo mass }\label{sec:nlab}

The tests above indicate that our fiducial group mass estimates 
suffer from a global uncertainty of about 0.3 dex, ranging from 0.2 dex at high group mass up to 0.4 dex at low group mass, which thus we consider to be the 
typical error on these estimates. We therefore ask what maximum trends in galaxy properties
 with group mass could remain undetected in our sample, due to this level of uncertainty in the 
 measurements of group masses. In other words, how strong a dependence of a given quantity 
 on halo mass could disappear in our data, due to the uncertainty in our practical realization of the group masses? 
 We address this question by computing how much the observed slope of a measured trend can change
  with respect to the `true' one, also given the statistical size of the total  \emph{ZENS}  sample.
  
\begin{figure}
\includegraphics[width=85mm]{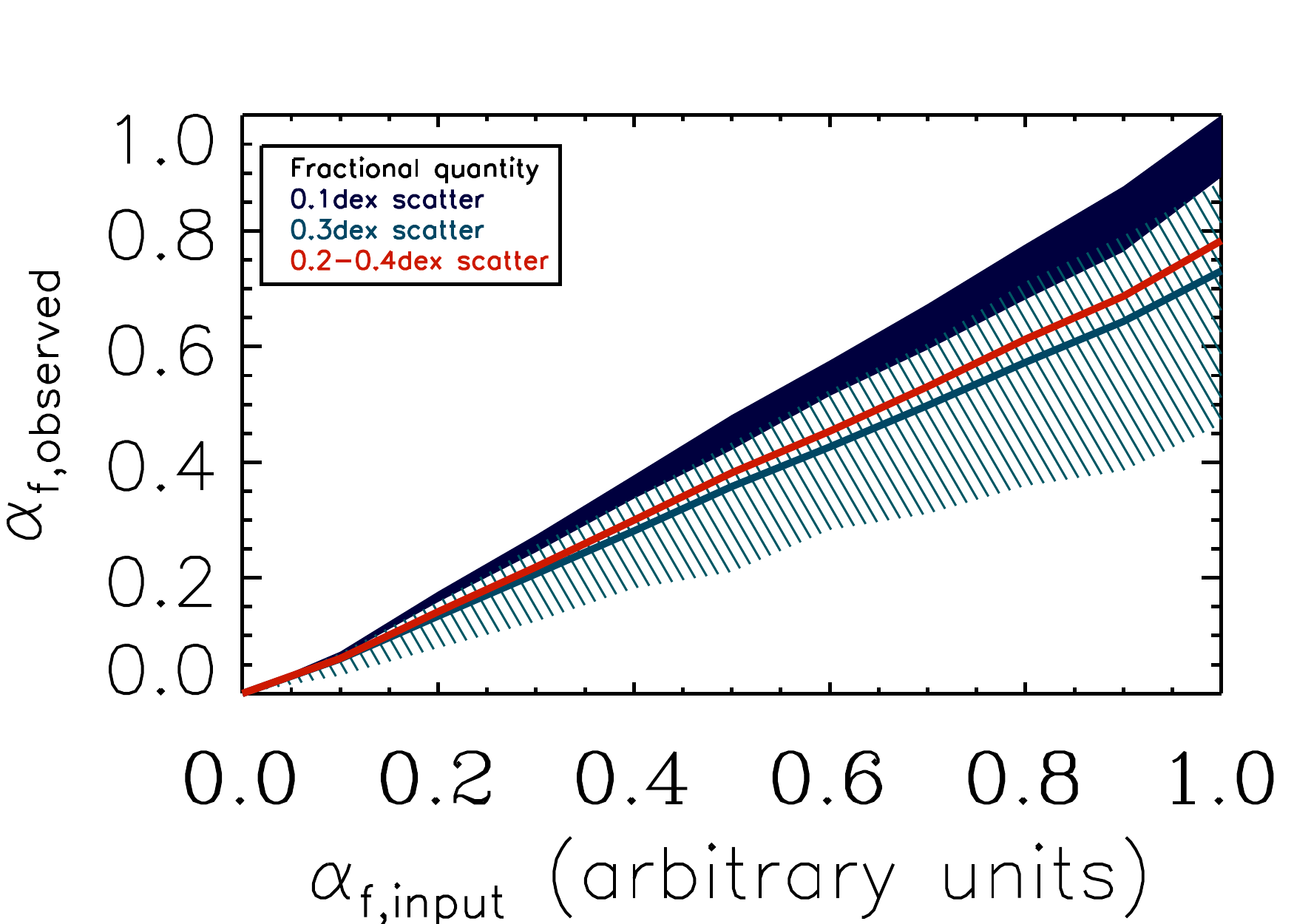} 
\caption{\label{fig:slopeChange} Effect of  errors in the group masses on the  
measured slope $\alpha_{f,observed}$ of the relationship with group mass of a given fractional property (e.g., fraction of  quenched galaxies, etc.\,).
 Plotted are the effects of a 0.1 dex (dark area) and 0.3 dex (dashed area), assuming Gaussian errors for the group masses. The median relation for the typical case
  of a 0.3 dex error is highlighted as a solid line.
The red solid line is the median  $\alpha_{observed}$-$\alpha_{input}$ relation calculated assuming a 0.4 dex uncertainty in group mass at low  group masses below $10^{13} M_\odot$,  and an uncertainty decreasing to 0.2 dex  at the highest group masses. Either considering an average  0.3 dex scatter or an increase in the error with decreasing halo mass returns similar results, namely $(i)$ a correction factor of order $\sim1.3$ to recover the intrinsic slope $\alpha_{f,input}$ of a trend with group mass, from the slope $\alpha_{f,observed}$ that is measured using the full \emph{ZENS} dataset; $(ii)$ an uncertainty on  a flat relationship, i.e., for $\alpha_{f,observed}=0$   of $\sim0.05$.}
\end{figure}

\begin{figure}
\includegraphics[width=85mm]{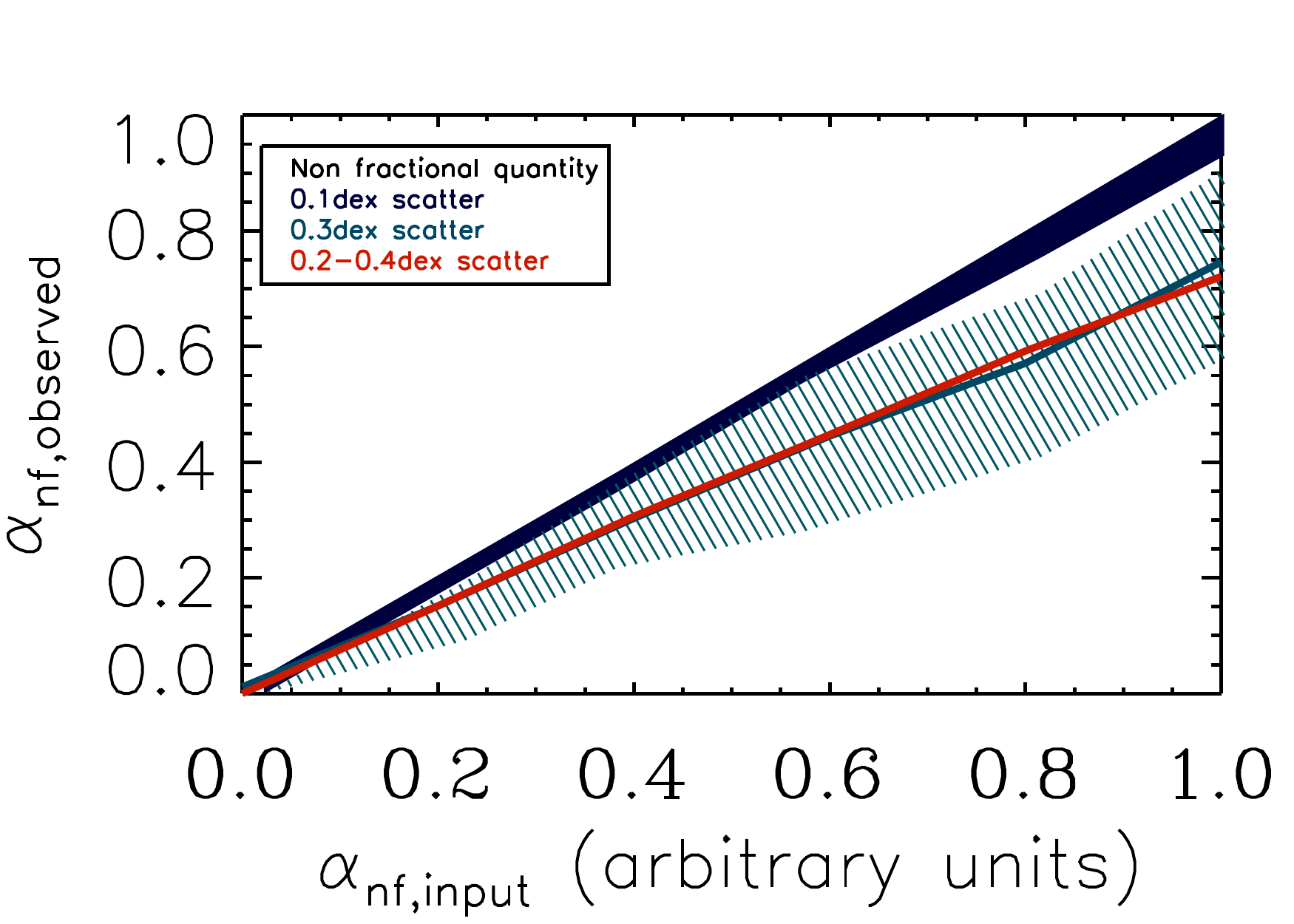}
\caption{\label{fig:slopeChange2}As in Figure \ref{fig:slopeChange}, 
but for the case in which the studied property is not a fractional quantity. In this case, 
the errors on the group masses flatten the intrinsic slope of the relationship of this quantity with halo mass of a factor of order $\sim1.4$. The uncertainty on  a flat relationship, i.e., $\alpha_{nf,observed}=0$   is $\sim0.1$. Symbols and colors are as in Figure \ref{fig:slopeChange}.
 }
\end{figure}

	We consider the two cases in which the observed property is either a fractional
   quantity (for example, the fraction of  `quenched' satellite galaxies of a given morphology  in any of the environmental conditions that we study in \emph{ZENS}, that we present in Carollo et al.\ 2013b, Paper IV, in preparation), 
   or a non-fractional quantity (for example, the colors of the bulge  and disk components of these satellite galaxies, also in preparation) -- see  Figures \ref{fig:slopeChange} and \ref{fig:slopeChange2}.  
   We split the simulated sample in two bins of group mass separated at $10^{13.5} \Msol$. 
The  $10^{13.5} \Msol$  value is chosen because $(i)$ it roughly divides our ZENS galaxy sample in two sub-samples of  comparable sizes, and it also is close to the median group mass in the ZENS sample (see Figure 1), and $(ii)$ it roughly represents the separation between the typical 'group' and 'cluster'   environment. We stress however that the results of the tests reported below are largely independent of the precise separating group-mass value that is used to define a low and a high group mass bin in our sample.

	  We impose as an initial condition for our test a similar  number of galaxies in the low- and high- group mass bins as we have in our \emph{ZENS} sample.
For the case of a fractional quantity, we  assume that each group has a total 
    number of galaxies $n_i$ of which $k_i$ have a given property, 
    producing a fraction $f_{i}$ of galaxies in that group with this characteristic.  
       As a starting point we assume that the mass of the group is known without errors, and that all groups in the low (high) group mass  bin have initially
     the same $f_{i,Low}$ ($f_{i,High}=f_{i,Low}+\delta_f$, with $\delta_f$ the `intrinsic' difference in the two group mass bins).
      This determines the {\it slope $\alpha_f = \delta_f/\Delta \log_{10} M$} of the relationship between the fractional quantity under study and the group mass. For simplicity, in the analysis, we set $\Delta \log_{10} M = 1$, so the values of $\alpha_f$ vary between zero and one. 
We  calculate the global input fractions $f_{Low}$ and $f_{High}$ of all galaxies in the low and high mass bins, which gives the assumed  `input' (i.e., true) slope of the relation.  A thousand realizations of the sample   are then obtained by perturbing the initial group masses with logarithmic random Gaussian errors, 
  for several initial $f_{i,Low}$ spanning the range [0,0.9]. Due to the errors, groups in the high group
   mass bin will move into the low group mass bin and vice versa. 
  The observed fractions $\tilde{f}_{Low}$ and $\tilde{f}_{High}$  are computed for 
each realization of the error-perturbed sample, and the resulting new slopes are estimated.

 In  Figure \ref{fig:slopeChange} we show the comparison between the input (true) slope $\alpha_{f,input}$ with the output, error-affected slopes $\alpha_{f,observed}$. This enables us to  assess  the impact of the errors on group masses on our capability of measuring the input, true slope of any trend with group mass that we will seek to measure. In the figure, the two shaded areas  bracket the minimum and maximum resulting slopes resulting from 0.1 dex (dark color)  and 0.3 dex (dashed area) perturbation
amplitudes, respectively, around a given value of the input slope.  
  The solid line is the median relation for the typical 0.3 dex uncertainty in ZENS group masses;   the red line shows the result obtained with varying the uncertainty with  halo mass between 0.2 dex (at high group masses) and 0.4 dex (at low group masses), as determined above.

  	As expected, the effect of the errors on the group masses is to lead to an underestimation of the slope of any relationship of fractional galaxy properties with this environmental quantity.   Nevertheless, the \emph{ZENS} data  enable us to detect even   moderate trends with group mass. A median `correction factor' can be estimated from this test, i.e.,  $\alpha_{f,input} \sim 1.3 \times \alpha_{f,observed}$, with an uncertainty of order $\sim0.05$ for a flat relationship with $\alpha_{f,observed}\sim0$. 
   
 To establish the error on trends with group mass for  non-fractional quantities, we assumed that all groups in the low group-mass bins
      have an initial distribution of such quantity with a certain mean value, equal to $<q_{nf,Low}>$,
      and a  standard deviation equal to $\sigma_{nf}=0.2$. The groups in the high group mass bin have
        an initial distribution of values with the same dispersion as the low mass groups,
       but centered at $<q_{nf,High}>=\delta_{nf} + <q_{nf,Low}> $, with  $\alpha_{nf} =  \delta_nf/\Delta \log_{10} M$ the slope of the relationship between the non-fractional quantity under study and group halo mass. We exemplify this case  in Figure \ref{fig:slopeChange2} by setting $<q_{nf,Low}>=0$
and exploring an  range of $<q_{nf,High}>$ from 0 to 1 in arbitrary units. 
Again we set $\Delta \log_{10} M = 1$, so that the slope will vary between zero and one as well.
Starting from the given initial condition,
 we perturb the group masses with 0.1 and 0.3 dex Gaussian errors, as before,   and also with an increasing error with decreasing group mass from 0.2 dex at the highest masses to 0.4 dex at masses below $10^{13} M_\odot$, and
  recalculate the average values, deviations and difference of the measurements
  between the low and high groups mass bins, as above.
 Figure \ref{fig:slopeChange2} shows that the observed slope $\alpha_{nf,observed}$  is typically $\sim1.4$ times flatter than the true slope $\alpha_{nf,input}$ (both considering an average  0.3 dex scatter and an increase in the error with decreasing halo mass, as both these cases return again similar results).
Also, an observed $\alpha_{nf,observed} \sim 0$ may hide an intrinsic slope of  $\sim10\%$.

These tests give us a benchmark for interpreting correctly the trends with group mass    that we explore in \emph{ZENS}.

\subsection{Environments number two and number three: 
Galaxy distance from the group center and the central/satellite dichotomy}

Galaxies may suffer environmental effects as they enter or  orbit 
the dark matter halos of their host groups. Also, central and satellite
 galaxies are expected to experience different  physical conditions through their evolution with cosmic time \citep{Somerville_Primack_1999,Simha_et_al_2009}. 
 Ideally we  would want to know the precise 
 location and velocity vector of each galaxy within the three-dimensional
  potential  of its host group halo, relative to the characteristic ÒsizeÓ of 
  the halo, e.g., here taken to be the $R_{200}$ radius at which the density in the halo is a factor
   200 higher than the density of the Universe at the redshift of the structure. 
  A valuable proxy for this is given by the projected distance
of the galaxy from the assumed center of the group: this is the quantity that we use in our
 studies to explore the dependence of galaxy properties on the location of galaxies within 
their host groups. 

While simulations easily assign the rank of central or satellite to a galaxy in a common halo, with real data determining which is the central galaxy and which its satellite galaxies and, related to this, establishing the group centers, is not without
  challenges. We therefore adopt an operational definition, and subsequently  establish the impact on the final results of our  choice. 

To separate central galaxies, assumed to sit on the center of the groups, from satellite galaxies, assumed to orbit the central galaxies within the group potentials,  previous literature has typically adopted galaxy luminosity \citep[e.g.][]{Weinmann_et_al_2006, Skibba_2009,Hansen_et_al_2009} 
or stellar mass (e.g. \citealt{vanDenBosch_et_al_2008,Guo_et_al_2009,Kimm_et_al_2009}, the latter in some cases in addition to luminosity, e.g., \citealt{Peng_et_al_2012}). For our sample, the 2PIGG catalogue \citep{Eke_et_al_2004} provides a group center determined with an iterative approach: the weighted mean position of the member galaxies is calculated, and the most distant galaxy subsequently rejected until only two galaxies are left, at which point the center of the given group is associated with the galaxy with either the larger `weight' that models the local incompleteness of the 2dFGRS data, or, for identical weights, the larger flux. 

In our work we scrutinized three different definitions for the center of a group, i.e., the 2PIGG centers, 
the geometric center of stellar mass, and the (center of the) galaxy with the highest stellar mass, respectively. Ultimately we opted for the latter definition as our fiducial estimate for the group center and central galaxy of a group; 
however, we imposed that the resulting central galaxy should satisfy a consistency requirement for it   being also at the spatial and velocity center of the group. Furthermore, in determining which galaxy had the highest mass, we considered not only the mass estimates provided by the `best-fit' templates to the observed galactic SEDs, but also the errors in these estimates.  We discuss in detail below (Section \ref{sec:centralsDef}) our procedure for identifying the central galaxies and the centers of the groups.

In order to derive an estimate for the  characteristic projected  radii, given the uncertainties in the group velocity dispersions mentioned above, we used our fiducial total group masses and derived  $\hat{R}_{200}=\left(\frac{GM_{GROUP}}{[10H(z)]^2}\right)^{1/3}$, with $H(z)=H_0\sqrt{\Omega_M(1+z)^3+\Omega_{\Lambda}}$ the Hubble constant at the given redshift.  Note that we use the measured values of $M_{GROUP}$, which are assumed to be proportional to the $M_{200}$ values; for this reason we have distinguished $\hat{R}_{200}\neq {R}_{200}$. For simplicity, however,  we will drop the $\hat{R}_{200}$ notation, but it should be kept in mind that a conversion factor is needed to rescale our size measurements to the formal  $R_{200}$ values. 

\subsubsection{A non-trivial challenge: Which is the central and which are the satellites?} \label{sec:centralsDef}

Inspecting the properties that our \emph{ZENS} groups and their central galaxies would have,
 if we used solely luminosities or nominal best-fit stellar masses for the identification of such centrals and centers, highlights some shortfalls in all those definitions. For example, \citet{Peng_et_al_2012}
 combine a requirement on luminosity with a requirement on stellar mass to determine 
 the central galaxies in their SDSS group sample, to minimize the effects of recent star formation or dust reddening in the identification of the centrals; these effects can be substantial in a luminosity-based approach, especially by introducing stellar mass dependent biases. It is clear however that the identification of  `the most massive galaxy' is affected by random and systematic errors in the derivation of galaxy stellar masses; these errors are not customarily included in the identification of the central galaxies. 

A result of these shortfalls is that the alleged central galaxies that are identified based 
on a (luminosity or even a) best-fit stellar mass criterion often lie at the projected spatial 
or kinematic outskirts of the groups of which they are supposed to be the centers. 
We have tested that this consideration holds in both the two of the currently most used clustering-based
 group catalogs of \citet[][]{Yang_et_al_2007} (for the SDSS) and \citet{Eke_et_al_2004}
  (for the 2dFGRS). For example, identifying the group center with the nominal most massive galaxies, 
  we find that roughly 50\% of the \emph{ZENS} groups  suffer from this unphysical `displacement' 
  of their own alleged centers. Other authors have addressed the issue of contamination and/or incompleteness in 
  samples of central vs.\ satellite galaxies, at least in terms of establishing a global statistical effects
  (without however applying an active correction to their final analysis). For example, Weinmann et al.\ 2009 tested 
  their grouping algorithm against SDSS mock catalogues, and found a contamination of  $\sim 30\%$  of centrals in their sample of satellites and vice versa. 
\citet{Skibba_et_al_2011} find that the brightest halo galaxy is often a satellite and not the central one; the probability that a satellite galaxy is more luminous than the central galaxy appear to increase with halo mass (reaching $\sim40\%$ at $\sim 10^{14}\Msol$).

In our study we implemented a procedure for improving the identification of the
 central galaxies, which also gives insight on the origin of central vs.\ satellite contaminations. 
 In detail, we scrutinized the properties of the nominal most massive galaxies for each group in our
  sample, and we retained them as their `centrals' only if they were  compatible, within the errors on the stellar mass estimates, with being the
   most massive galaxies in their groups, {\it and} furthermore resulted in self-consistent solutions
    in the (projected) spatial and (line-of-sight) velocity domains. 
    That is to say, for a galaxy to be a good central in a group halo, it must be its most
     massive inhabitant {\it and} must be compatible with the inferred spatial {\it and} velocity centroids of this halo. For cases in which the galaxy with the highest formal best fit stellar mass did not satisfy simultaneously these criteria, we either found an alternative galaxy within the group which provided such self-consistent solution, or flagged the groups as `unrelaxed', in order to keep in our analyses the information that, for these systems, none of their member galaxies satisfied all conditions for being genuine `central' galaxies. 

Quantitatively, we implemented these criteria by requesting that not only
 $(i)$ the central galaxy be the most massive member of the group `within the errors' of our stellar mass estimates, 
 but also that $(ii)$ its projected location lies within the inner circular area centered on the stellar-mass-weighted geometric center of the group, and enclosed within a radius  0.5$R_{200}$; and $(iii)$ its inferred line-of-sight velocity lies within one standard deviation of the median of the velocity distribution for that group.

\begin{figure*}
\begin{center}
\includegraphics[width= \textwidth,height=\textwidth]{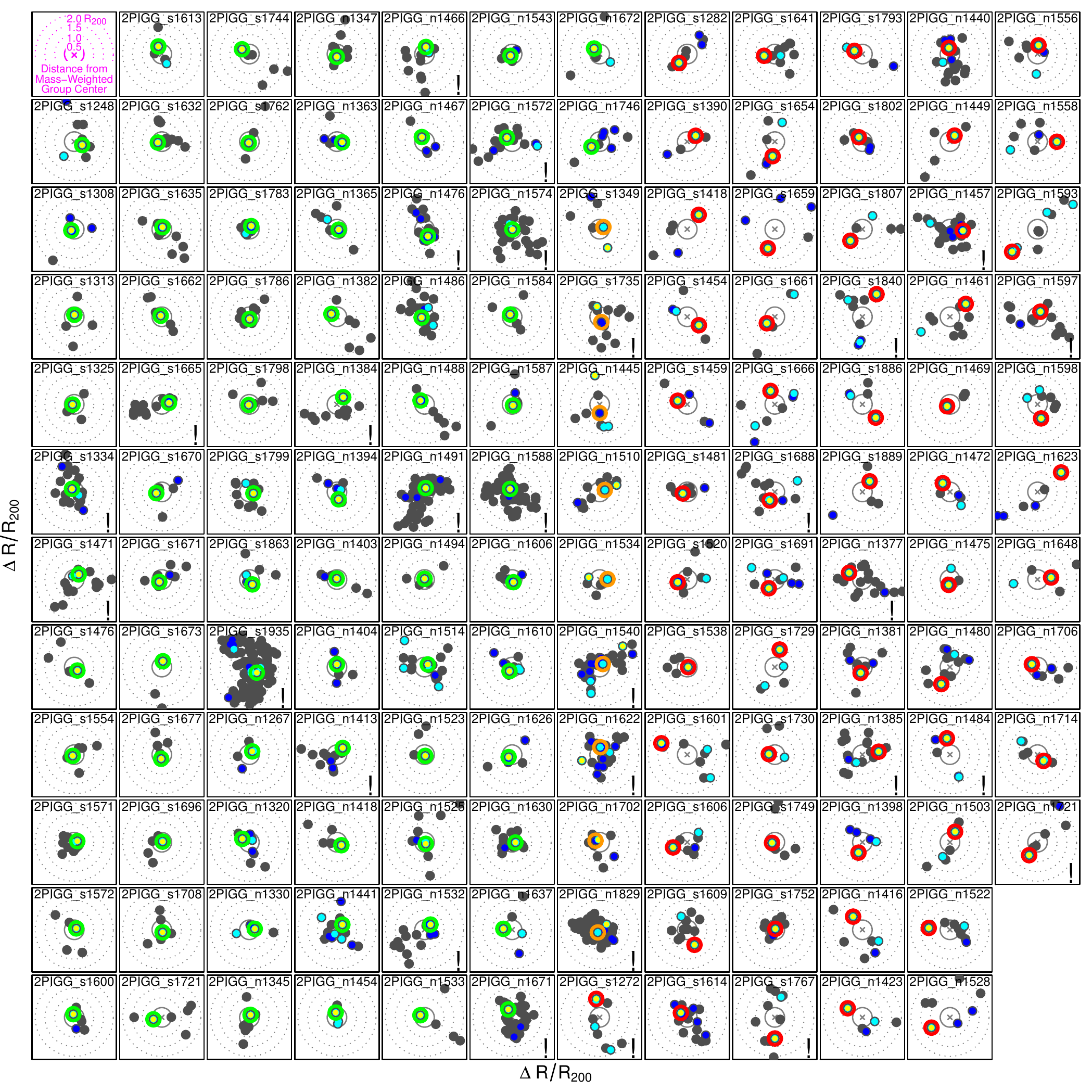}
\caption{\label{fig:cenSpatial} The spatial distributions of galaxies in each of the 141 \emph{ZENS} group. 
 Each panel presents the (projected) radial distribution around the mass-weighted geometric center in each 
  group, with the scale set by our estimate of $R_{200}$ for that halo. 
 Concentric line-circles mark galactocentric distances of 0.5, 1.0, 1.5, and 2.0$R_{200}$, respectively,
 to the mass-weighted center of the groups (marked by  grey crosses, not always visible). 
 Yellow symbols indicate the nominal most massive member of each group, 
 based on our fiducial definition of stellar mass of a galaxy, i.e.,   the best-fit stellar mass obtained by fitting a large number of synthetic templates,
   spanning a large range of star formation histories, to the observed galaxy SED.
     Light and dark blue symbols represent galaxies with nominal     stellar masses within a factor of two and four, respectively, from the nominal highest mass in the group.
      Green and orange circles around the symbols for the galaxies 
      assumed to be the centrals identify groups which we have labeled as `relaxed'; 
red circles identify `unrelaxed' groups. Green identifies groups in which, 
based on spatial and velocity considerations,  the nominal most massive galaxy in the group
 is confirmed to be the central galaxy. Orange identifies groups in which    the nominal most massive galaxy is not consistent in the spatial and/or velocity domain with being the center of the group, but another galaxy member in the group $(i)$ has an integrated probability $\geqslant10\%$ to have a stellar mass higher than the nominal most massive galaxy, and $(ii)$ also satisfies the spatial and velocity criteria described in the text. These alternative most massive galaxies, highlighted in orange, are assumed to be the central galaxies of their host groups.
The groups marked with a ``!"  are those for which the WFI pointings did not cover their entire extent; some galaxy members located at the group outskirts have been missed, but these are  low mass galaxies, unlikely to be the centrals. In general, missed galaxies do not impact the analyses of the central galaxies (see Appendix \ref{App:missedgals} for details).}
\end{center}
\end{figure*}

\begin{figure*}
\begin{center}
\includegraphics[width=\textwidth,height=\textwidth]{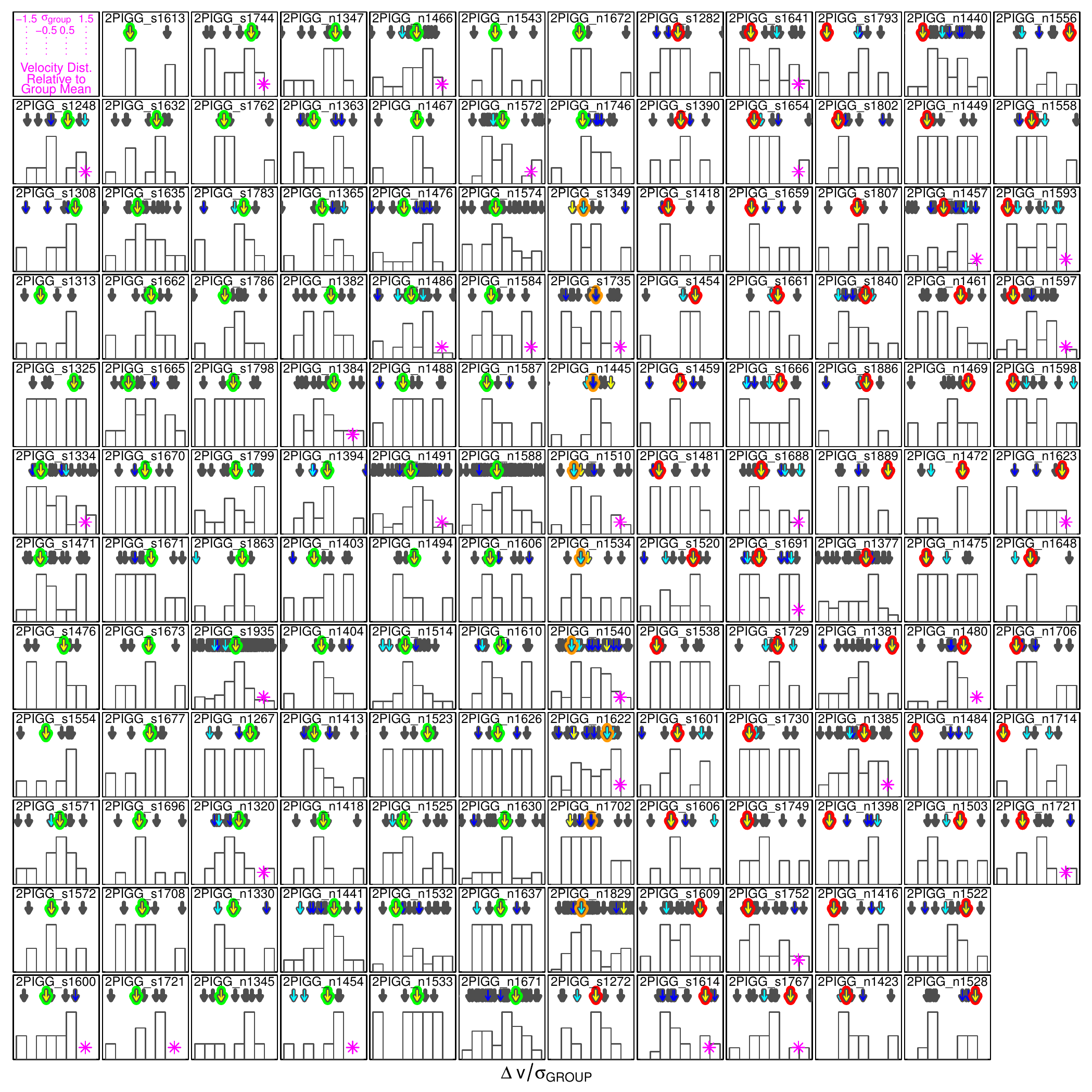}
\caption{\label{fig:cenVelocity} The line-of-sight velocity distributions of galaxies in the 141
\emph{ZENS} groups.  Each panel presents the velocity distribution relative to the median
 redshift of the group; the eight bins from left to right indicate velocities equal to $-2$, $-1.5$, $-1.0$, $-0.5$, 0, 0.5, 1, 1.5, and 2 
 the value of $\sigma_{GROUP}$, respectively (i.e., the velocity scale is normalized to the velocity dispersion of each individual group). 
Note that only galaxies with velocities between $-1.5\sigma<v<1.5\sigma$ are here plotted, and that arrows showing individual galaxies may be overlapping in some of the panels. This implies that for a few groups not all galaxy members (which is never less than five) are visible in this figure. 
  The velocities  of individual galaxies are shown with an arrow, using the color-coding as in Figure \ref{fig:cenSpatial}.
 The magenta stars identity those groups which show sub-clustering according to the criteria described in Section \ref{subclass}.}
 \end{center}
\end{figure*}

We started the   procedure highlighted above by assuming as fiducial stellar masses for the \emph{ZENS}
 galaxies  the best-fit (i.e., minimum $\chi^2$) masses  that result  from fitting, with the code \emph{ZEBRA+} \citep{Feldmann_et_al_2006, Oesch_et_al_2010},
 a large  library of synthetic models to the galaxy SEDs\footnote{The stellar masses for the \emph{ZENS} galaxies were derived by combining the $B$ and $I$ WFI photometry with the  available multi-wavelength archival photometry
 (SDSS $u$,$g$,$r$,$i$,$z$ (\citealt{Abazajian_et_al_2009}),  the 2MASS $J$, $H$, $K$, (\citealt{Skrutskie_et_al_2006}) and  the GALEX NUV and FUV magnitudes
 Ð see Paper III for details on the procedure adopted to derive and calibrate the stellar masses. Note that, in our \emph{ZENS} analyses, we adopt for the definition of galaxy stellar mass the integral of the star formation rate, i.e., we do not subtract  the mass `returned' to the gas through stellar evolution processes.}. 
 The adoption of these fiducial stellar masses leads to the identification of a nominal central galaxy, i.e.,  the 
 galaxy member in a group which has the highest fiducial best-fit stellar mass. We then checked the spatial and velocity location of these nominal centrals. For each of the 141 \emph{ZENS} groups,  
 we show in Figure \ref{fig:cenSpatial} the location of the member galaxies relative to the mass-weighted geometric group centers; in each panel, the radial scale is set by our estimate of $R_{200}$ for the given halo. 
 The nominal most massive member of each group is indicated with a yellow point. Light and dark blue points represent galaxies with masses respectively within a factor of two and four of the nominal most massive galaxy. Less massive group members are shown as black dots. For each of the groups, the corresponding velocity distributions are shown in Figure \ref{fig:cenVelocity}.  Here each panel presents the relative line-of-sight velocity distribution about the median of the distribution, with the scale set by its dispersion. The position in velocity of  each individual galaxy is indicated with an arrow, using the same color-coding as in Figure \ref{fig:cenSpatial} for the nominal most massive member of each group, for galaxies with nominal stellar masses within a factor of two and four from the nominal most massive galaxy, and for lower mass galaxies (respectively yellow, light and dark blue, and black arrows).  

In about 50$\%$ of the \emph{ZENS} groups, the nominal centrals were confirmed
 to be fully consistent, both spatially and in velocity terms,  with lying at the bottom of the potential 
 well of their host groups. We thus confirmed these galaxies to be genuine centrals, and identified with them the centers of the groups. The yellow points/curves identifying these central galaxies in Figures \ref{fig:cenSpatial} and \ref{fig:cenVelocity} 
(and \ref{fig:cenMasses}, see below) are highlighted in green for these groups. 
In the remaining $\sim50\%$ of groups, however, the nominal centrals were either at the spatial periphery of their group halos, or appeared to be `shooting away' from them in terms of relative  velocities within their host groups. We thus investigated whether this could be due to uncertainties in our galaxy stellar mass estimates  (see Paper III).  

To address this issue we capitalized on the availability of the entire  posterior probability distributions (PPDs)  for the stellar masses. 
These PPDs are presented in Figure \ref{fig:cenMasses}. Specifically, each panel in this figure shows the PPDs
 of the stellar masses for the few top-massive galaxies in each of the \emph{ZENS} groups. For each of the plotted galaxies, the PPDs are obtained by connecting 21
   sampling points spanning the 1 to 99$\%$ quantiles. In each panel, the horizontal scale is logarithmic
    in mass and covers the range between one-tenth (leftmost value) to  three times  (rightmost value)  the nominal  highest mass; the PPD for this nominal highest mass galaxy 
     is highlighted again in yellow. Also the remaining colors are as in Figures \ref{fig:cenSpatial} and \ref{fig:cenVelocity}, i.e.,
     light and dark blue curves indicate galaxies with nominal (best fit) 
     stellar masses within a factor of two and four, respectively, of the nominal highest mass for that group.  Note that in several groups there are galaxies, with nominal masses within a factor  $\sim2-4$  from the nominal highest mass, which would thus be classified as `satellites', which however have, according to their PPDs, a substantial probability that their stellar masses are actually larger than the nominal highest stellar mass of the nominal central.

For the groups in which the nominal (best fit) most massive galaxy failed to 
pass the projected-spatial criterion and/or the line-of-sight velocity criterion to be a genuine 
central galaxy, we thus searched for an alternative viable central by requiring that this 
$(i)$ satisfies both the spatial and velocity criteria; $(ii)$ has a nominal stellar mass within
 a factor of  four from the nominal highest mass for that group, and $(iii)$ has a $\geqslant $10\% probability,
 as defined by the overlapping area with the PPD of the most massive member, to exceed the minimum mass in the PPD of this dominant galaxy.
  We found 9 \emph{ZENS} groups for which  such viable alternative centrals could be identified, 
  which we adopted as the correct central galaxies in these groups. These alternative
   centrals are highlighted with orange contours to the relevant symbols in Figures \ref{fig:cenSpatial},  \ref{fig:cenVelocity} 
and \ref{fig:cenMasses}.

For the remaining 59 groups, this iterative procedure failed to identify a galaxy member which satisfied the set criteria for being a genuine central galaxy.  For these groups the nominal most massive galaxy was thus retained as the nominal central, but we flagged these groups so as to be able to estimate the impact of their inclusion in analyses that rely either on a central/satellite separation, or on
 the knowledge of the group center. These `dubious' centrals are highlighted with red contours to the  symbols in 
 Figures \ref{fig:cenSpatial}, \ref{fig:cenVelocity} and \ref{fig:cenMasses}. 
 
In about a third (45) of the \emph{ZENS} groups., the central galaxy thus identified coincides with the original central galaxy provided by the 2PIGG catalogue. 
 In the remaining 96 groups we find  however a different solution.  The 2PIGG centers were iteratively identified from the galaxies positions,  without any knowledge of the galaxy masses. Our approach should  thus provide more robust estimates for the group centers, especially for sparse and small groups. We note that a change in the identification of the central galaxy has no impact on the FOF association of the group members, and thus on the group identification
 which is done prior to  the definition of the center itself.

\begin{figure*}[htbp]
\begin{center}
\includegraphics[width=\textwidth,height=\textwidth]{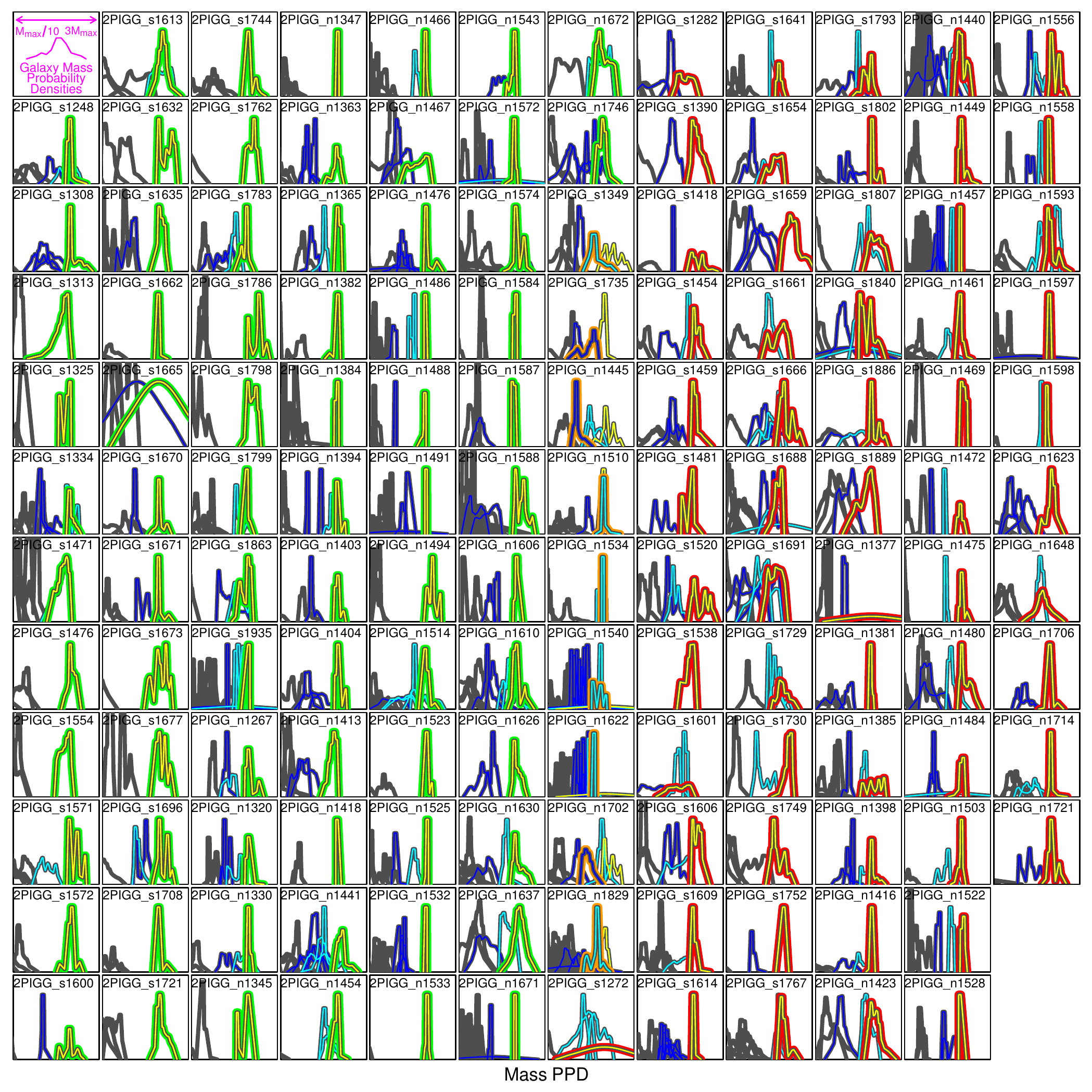}
\caption{The posterior probability distributions (PPDs) of galaxy stellar masses for the top few most massive galaxies in each \emph{ZENS} group.  The PPDs are derived by connecting 21 sample points spanning the 1 to 99\% quantiles for the stellar masses obtained by fitting a large set of synthetic templates to the observed galaxy SEDs. The horizontal scale is logarithmic in mass and ranges from one-tenth to three times the nominal (best fit)  mass of the most massive member. The color-coding is as in Figures \ref{fig:cenSpatial} and \ref{fig:cenVelocity}.
 Gaussian PPDs are assumed for galaxies which do not have WFI $B-$ and $I-$ band images (due to the limited coverage of the WFI pointings); for these, the stellar mass is inferred from the $r_F-$Mass relation (see Appendix \ref{ap:MissingGalaxies}). The widths of the Gaussian correspond to 1.5 times the observed scatter in this  relation.} 
\label{fig:cenMasses}
\end{center}
\end{figure*}

\subsubsection{Sources and effects of errors on our fiducial centrals and group centers}\label{sec:cenErrors}

The fiducial centers and centrals defined as above are correct within a certain  statistical uncertainty.
 In addition to introducing uncertainties in the estimate of the total masses of the groups, 
as discussed in Section \ref{sec:MembershipTest}, the presence of interlopers  and the absence of genuine group members  from the 2PIGG lists (which define
  the groups in our sample),  can also affect the identification of the central galaxy and thus the identification of the center of the groups.
We discuss this issue more in detail in Appendix \ref{App:missedgals}. Here we highlight that, based on our own test of comparison between the 2dFGRS catalogue and the SDSS spectroscopic and photometric catalogues \citep{Abazajian_et_al_2009}, we expect that these effects should lead to the mis-identification of the central galaxies in at most $\sim10\%$ of the cases. The dominant source of error in the identification of  centrals and satellites remains the association of galaxies with a given group through the 2PIGG algorithm. We will discuss  the impact of such an uncertainty in each individual analysis that relies on either a central-satellite split, or on the identification of the group centers. 

\subsection{Environment number four: The backbone density field of the Large Scale Structure }\label{sec:LSSEstimate}

A major achievement of  recent large spectroscopic redshift surveys, and  large multi-wavelength 
imaging surveys with accurate photometric redshifts, has been enabling the determination of a proxy 
density field produced by the large-scale structure (LSS). The projected overdensity at a position $\theta$ in celestial coordinates, and at a given redshift $z$, is defined as $\delta_{LSS}=\frac{\rho(\theta,z)-\rho_m}{\rho_m}$, with  $\rho(\theta,z)$ the comoving projected density of the tracers of the density field, and $\rho_m$ the mean projected density calculated over the global survey area at the given redshift.  Several approaches have been used to derive LSS density fields e.g., by measuring the density within a fixed volume \citep[e.g.][]{Hogg_et_al_2003,Blanton_et_al_2005,Wilman_et_a_2010}; through Voronoi-Delaunay techniques \citep[][]{Schaap_Weygaert_2000,Marinoni_et_al_2002,Gerke_et_al_2005,RomanoDiaz_Weygaert_2007,Knobel_et_al_2009,Gerke_et_al_2012}; and with adaptive approaches in which the density is calculated out the distance to a   $N^{th}$ nearest neighboring galaxy \citep[e.g.][]{Gomez_et_al_2003,Balogh_et_al_2004,Baldry_et_al_2006,Kovac_et_al_2010b}.   

 All these methods have their virtues and shortfalls, 
 as extensively discussed in the previous literature. The Nth-nearest galaxy neighbors is often
  preferred to the `fixed volume' approach because the latter washes out information on scales of order of the adopted volume; however, it has the strong disadvantage that it shifts its physical meaning from density `within a halo' to density `between halos' for galaxies which reside in groups of richness straddling across the chosen value of `N' (see also \citealt{Peng_et_al_2012} for a discussion on the correlation between N-nearest galaxy neighbor overdensity and group membership).   The Voronoi/Delaunay tessellation technique, thanks to its adaptive nature,   performs generally better then algorithms based on a fixed aperture, but can be affected by biases related to survey edge effects, redshift-space distortions and spectroscopic incompleteness \citep[see discussion in,  e.g.,][]{Cooper_et_al_2005, Kovac_et_al_2010b}.
 More generally, as discussed also by \cite{Haas_et_al_2012}, all methods contain a built-in correlation with halo mass, which hampers separating the effects on galaxies of the LSS from those of the host halos. To achieve an environmental LSS that is insensitive to halo mass, these authors construct a density field based on dimensionless galaxy luminosities/masses and distances.  
   
In our study we opt  for an alternative approach both to avoid that the estimator changes meaning (from a intra-group to an inter-group density estimator) with varying group richness, and to minimize the correlation  between halo mass and LSS density field that is introduced by construction when using a    $N^{th}$ nearest-galaxy method approach. In particular, we   adopt an $N^{th}$ nearest-neighbor estimator, but modified so as to use, 
as tracers of the LSS density field,    the groups themselves (treated as point masses
   of mass $M_{GROUP}$), rather than their member galaxies. 
Thus, $\rho(\theta,z) =\sum_i^N w_i/(\pi d_{N}^2)$ with $N$ is the chosen number of nearest (point-mass) groups, which we set to 5, $d_{N}$ the comoving distance to the   $N^{th}$ neighbor group and $w_i$ the weights, which we set equal to $M_{GROUP}$. Note that, by construction, in our analysis all galaxies belonging to a given 
 group are characterized by the same value of LSS overdensity $\delta_{LSS}$.

The entire 2PIGG group catalogue, supplemented by all remaining `ungrouped'  galaxies in the 2dFGRS (treated as groups with one galaxy member), was used to derive this Nth-nearest group-neighbors  overdensity field $\delta_{LSS}$. Note that by `ungrouped galaxies' we intend the 2dFGRS galaxies which do not belong to any 2PIGG group; this  does not necessarily imply that such galaxies are located in void regions (see also Appendix \ref{App:densityTests}).
Halo masses for ungrouped galaxies were calculated following the same procedure adopted for the groups; a correction to total luminosity was applied, which  assumes that  these ungrouped galaxies have companions fainter than the survey magnitude limit (see Section \ref{sec:GroupMasses}).
 We note that densities calculated at the edges of the 2dFGRS area are biased as a consequence of the limited area of the survey. To correct for this effect we followed the approach of \citet{Kovac_et_al_2010b}, and divided the computed density for the fraction of the area enclosed within $d_{N}$ which is covered by the 2dFGRS pointings. 
  For each group, the search for neighboring groups was restricted within a redshift interval of $dz=\pm0.01$; a minimum luminosity was set for the groups or ungrouped galaxies 
equal to the total (i.e. integrated to zero) luminosity of a single $b_j=19.1$ galaxy at redshift $z=0.07$.

While the fiducial  $\delta_{LSS}$ estimates that we use in our analyses are based on the 5th nearest-neighbor groups, we also computed 3rd- and 10th- nearest-group estimates and checked that our main results do not depend on which  of these representations of the LSS density field we use. The comparison between these three estimates is shown in Appendix \ref{App:densityTests}, were we also discuss for completeness a comparison of our fiducial $\delta_{LSS}$, which uses the groups as point-mass tracers of the LSS, with the more commonly used density field  derived  by using the Nth-nearest  individual member galaxies as tracers of the matter density along the comic filaments.

\begin{figure*}[htbp]
\begin{center}
\includegraphics[width=80mm,angle=90]{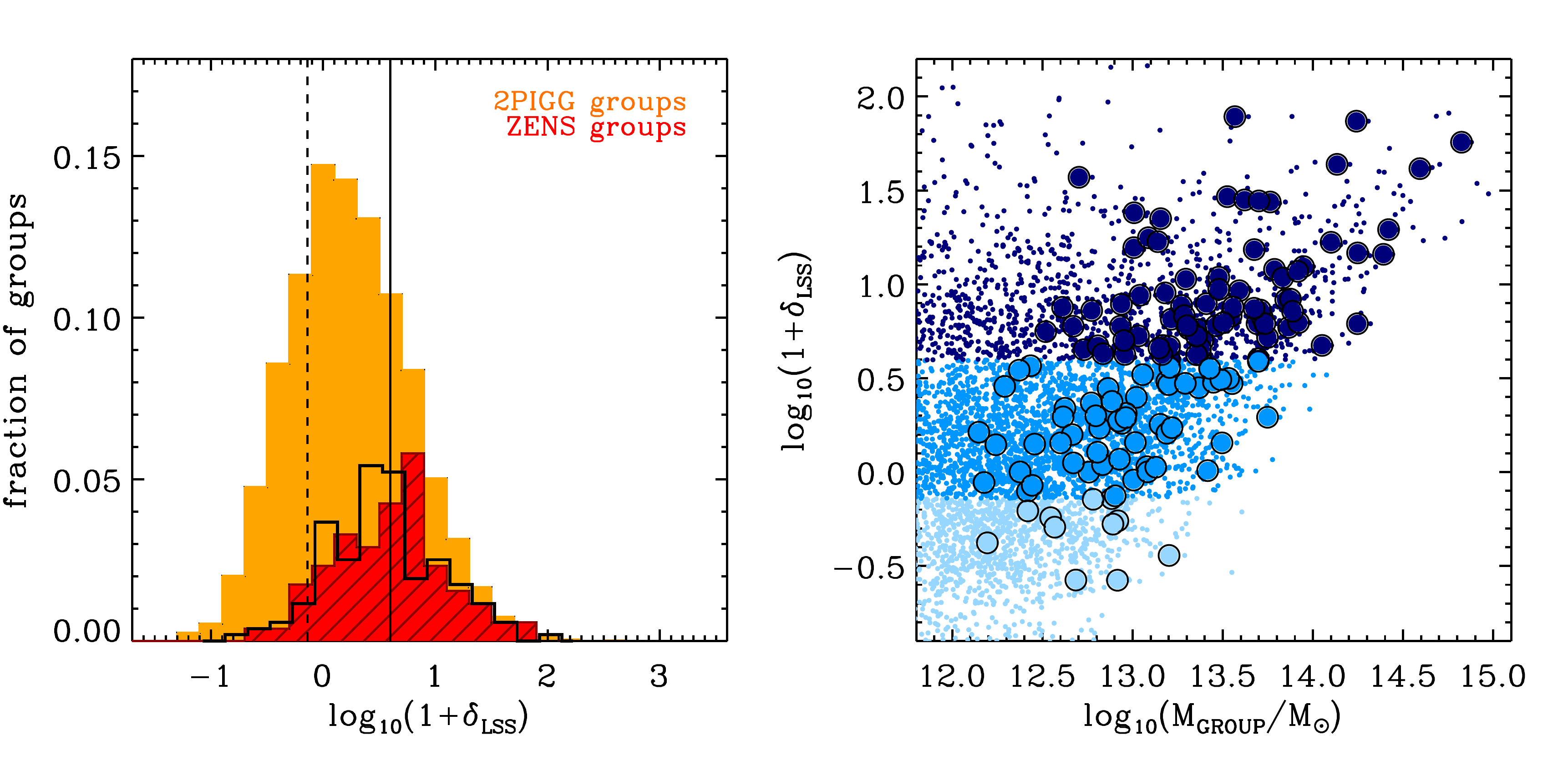}
\end{center}
\caption{\label{fig:densDist}
\emph{Left:} The distribution of the fiducial $\delta_{LSS}$ overdensities calculated with the 5-th nearest \emph{group neighbor} approach described in Section \ref{sec:LSSEstimate}.  All galaxies in a given \emph{ZENS} group have the same value of LSS overdensity, corresponding to that of their host group. The dashed histogram is for the \emph{ZENS} sample  (normalized to a 10th of the total 2PIGG groups for better visualization);   the solid histogram is for all 2PIGG groups in the redshift range $0.035<z<0.075$. 
The dashed and solid lines highlight the first and fourth quartile of the latter distribution, respectively. The black  empty histogram shows the distribution of $\delta_{LSS}$ for the \emph{ZENS} groups which is obtained by excluding the ungrouped galaxies from the sample of groups used in the calculation of $\delta_{LSS}$. The effect of including or removing ungrouped galaxies in the calculation of $\delta_{LSS}$ is minimal.
\emph{Right:} The fiducial LSS overdensity values $\delta_{LSS}$, based on the 5th nearest {\it group 
neighbor} approach described in Section \ref{sec:LSSEstimate}, as a function of group mass $M_{GROUP}$, calculated as described in Section \ref{sec:GroupMasses}. The three shades of blue highlight, from fainter to darker blue, groups in low (lowest quartile), intermediate and high (highest quartile)  LSS densities (relative to the global distribution derived for all 2PIGG groups in the $0.035<z<0.075$ redshift range, which are shown as small points with the same color scheme). High mass groups naturally live in high LSS regions; below $\sim10^{13.5}M_\odot$,  however, groups with similar masses occupy a wide range of LSS environments (as estimated by the $\delta_{LSS}$ field). At group masses below $\sim10^{13.5}M_\odot$ is thus possible to disentangle the  dependence of galaxy properties on halo mass ad LSS density.}
\end{figure*}

	The final distribution of $\delta_{LSS}$ values for the \emph{ZENS} groups (and thus galaxies within the groups) is presented in the left panel of Figure \ref{fig:densDist} as dashed histogram, and compared to the complete parent samples of 2PIGG groups in the redshift range $0.035<z<0.075$ (solid histogram). Not surprisingly the \emph{ZENS} sample is shifted towards higher density regions compared to the global distributions of 2PIGG groups, which extend down to smaller associations of two members only (and is in turn  shifted to higher densities relative to the whole 2dFGRS catalogue, which includes also ungrouped galaxies). Note that our fiducial estimate of $\delta_{LSS}$ does not produce the tail at very high overdensities that is observed when the individual galaxies are used as tracers in Nth-nearest neighbor calculations of the LSS density field (see Appendix \ref{App:densityTests}); this high-density  tail  is indeed mostly due to small inter-galaxy separations {\it within} massive halos with richness larger than the adopted `N' value.  
	
	To enable comparisons with the global galaxy population (in addition to relative comparisons within the \emph{ZENS} sample), we split the distribution of $\delta_{LSS}$ sampled by the entire 2PIGG catalogue in four quartiles (one to four in order of increasing density), and label the \emph{ZENS} groups as residing in \emph{low} LSS environments those groups whose local overdensity is not higher  than the value characterizing the first quartile of the global 2PIGG distribution  (dashed vertical line in the left panel of Figure \ref{fig:densDist}).  Similarly, we label as residing in \emph{high} LSS density regions those \emph{ZENS} groups with a local overdensity larger than the threshold defining the fourth quartile (solid line in the left panel of Fig. \ref{fig:densDist}).  The remaining groups are labeled to reside in regions of the LSS of \emph{intermediate} density. As indicated in Table \ref{tab:ZENs_prop}, applying these criteria results in 8$\%$ , 37$\%$ and 55$\%$ of \emph{ZENS}  groups being located respectively in low, intermediate and high density environments. 
 
	Note that, as shown in the right panel of Figure \ref{fig:densDist}, the $\delta_{LSS}$ values that we have adopted to describe the underlying density of the cosmic web do correlate, as expected, with the mass of the groups. 
	Given the approach that we have used to compute $\delta_{LSS}$, this is however mostly a reflection of the physical  fact that the more massive groups inhabit, by definition, high density regions of the Universe (which tend to be highly clustered). 
	However, groups with masses below M$\sim10^{13.5} \Msol$ are found over a very wide range of LSS environments (as sampled by our $\delta_{LSS}$ measurements). At these masses, we can therefore compare the properties of groups of similar halo masses which live  in different LSS environments, and thus to identify trends induced by the LSS environment separately from those induced by the group halo mass.

 \subsubsection{Sources and effects of errors on our LSS (over)density estimates}
 
 \paragraph{Inclusion or exclusion of ungrouped galaxies in the 2dFGRS }
 
In the calculation of the fiducial $\delta_{LSS}$ values that we adopt in our analysis we included all 2dFGRS galaxies which are not associated with any of the 2PIGG groups (i.e., also  the `ungrouped' galaxies in the 2dFGRS catalogue). We  checked however whether the LSS density field that we measure at the location of the \emph{ZENS} groups depends on whether these ungrouped galaxies are included or excluded in the computation of the (Nth-nearest-group-neighbor-based) LSS density field.    Figure \ref{fig:densDist} also shows the distribution of $\delta_{LSS}$ which is obtained when excluding the ungrouped galaxies (black line). We further discuss in Appendix \ref{App:densityTests} that the use of one or another of these two alternative realizations of the LSS densities shifts a group at most 
 to an adjacent density quartile of the global distribution of densities in $>90\%$ of the cases, with no major impact on our {\it comparative} analyses between different environments.
 
\paragraph{The choice of  N}

We also investigated the impact of the value of `N' in the Nth-nearest group-neighbor algorithm  adopted to filter the distribution of the density tracers. In  Appendix \ref{App:densityTests} we show that, in contrast with Nth-nearest neighbor computations which use the galaxies as the tracers of the LSS density field, our adoption of the groups themselves as tracers of the filamentary density distribution results in much weaker differences with the use of N=3, 5 or 10.

 \section{The \emph{ZENS}  catalogue: Environmental, structural and photometric measurements}\label{sec:ZENS_catalogue}

For the 1484 galaxies in the 141 \emph{ZENS} groups, we have measured a number of structural (Paper II) and photometric (Paper III) diagnostics.  

In  particular, we have quantified galaxy structure  both non-parametrically, through 
measurements of concentration, Gini coefficient, M20, smoothness (as done in Scarlata et al.\ 2007), as well as parametrically, through 
single-Sersic and double-component (Sersic bulge plus exponential disk) analytical fits to the two-dimensional surface brightness distributions. We have also used an isophotal analysis to quantify the strength of  bars.  All structural measurements, including bulge and disk parameters,  have been corrected in order to eliminate biases that depend not only on seeing/PSF, but also on magnitude, size, concentration and axis ratio.   We have furthermore employed  the  corrected structural measurements, including the bulge-to-total ratios, to define a {\it quantitative}  morphological classiÞcation, also validated by visual inspection of each galaxy 
in the sample,  into elliptical, early-, intermediate- and late-type disk, and irregulars. 

The photometric measurements for the galaxies as a whole include colors (total, and at various galactocentric distances); radial color gradients from analytical fits to the galaxy surface brightness proÞles, and the scatter around these gradients; total stellar masses and star-formation rates (and dust reddening), through fitting synthetic stellar population models to the near-UV to near-IR galaxy SEDs. Furthermore, through inspection of the 2dFGRS spectra, coupled with $(NUV-I)-(B-I)$ and 
$(FUV-NUV)-(NUV-B)$ color-color diagrams, we have disentangled dust-reddened galaxies from red,  quenched galaxies, and used this additional information to robustly classify galaxies into strongly star-forming or `moderately' star-forming, or quenched systems. 
We have also derived estimates for stellar masses separately for the disk and bulge components of galaxies, from the $B-I$ colors of these sub-galactic components derived from  the two-component surface brightness fits.
 
We publish electronically the \emph{ZENS}  catalogue\footnote{The \emph{ZENS} catalogue is also downloadable from http://www.astro.ethz.ch/research/Projects/ZENS.}  containing all structural and spectrophotometric \emph{ZENS} measurements for these 1484 galaxies, together with the environmental diagnostics discussed above and listed in Table  \ref{tab:ZENs_prop}. The $readme$ file  is given for convenience in Appendix \ref{App:readmecat}. 

\section{Groups with or without a central galaxy: Definition of `relaxed' and `unrelaxed' groups}\label{sec:relaxedUnrelaxed}

In Section \ref{sec:centralsDef} we saw   that  a total of  82   \emph{ZENS} groups, whose centrals are highlighted in either green (73)  or orange (9)  in Figures \ref{fig:cenSpatial}, \ref{fig:cenVelocity} 
and \ref{fig:cenMasses}, host a galaxy member which satisfies simultaneously the three criteria that we require in order to be a genuine central, i.e., having highest stellar mass within the errors, and being consistent with being the center of the group both in the spatial and velocity domains.  
The fact that in these groups the most massive galaxies have been able to establish their rank within their group potentials suggests a state of  dynamical relaxation for their host groups. Dynamically `relaxed' systems show indeed a well-defined center for the potential, and are  a golden sample to extend to low (i.e., smaller than cluster) mass scales studies of galaxy properties as a function of group-centric distance.

\begin{figure}[htbp]
\epsscale{1}
\plotone{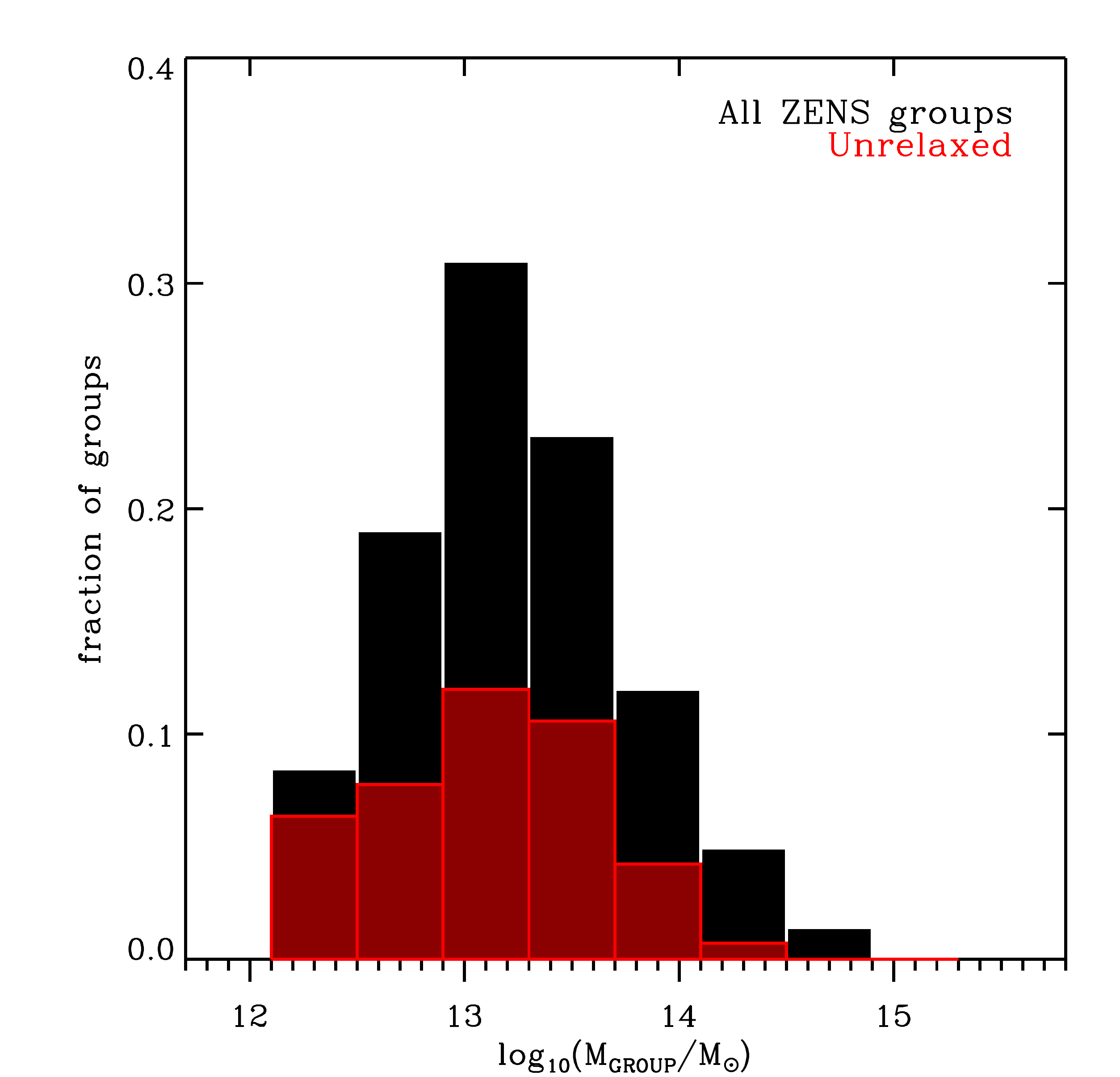}
\caption{The red histogram shows the distribution of $M_{GROUP}$ for the ZENS  groups  which are classified as `unrelaxed' according to the criteria described in Sections \ref{sec:centralsDef} and \ref{sec:relaxedUnrelaxed}. The black histogram is distribution for the total sample of  \emph{ZENS}  groups.
  Both curves are normalized to the total number of  (141) \emph{ZENS} groups.
\label{fig:UnrelaxedMass} }
\end{figure}

In the remaining 59 \emph{ZENS} groups  no galaxy member in the nominal 2PIGG group associations  satisfies simultaneously the three criteria above to be a genuine central. For these  groups we highlight in red in Figures \ref{fig:cenSpatial}, \ref{fig:cenVelocity} 
and \ref{fig:cenMasses}, the symbols for their nominal centrals, as a reminder that these, adopted as such on the basis of their nominal highest stellar masses, show a `displacement' from the groupsÕ spatial and/or velocity centers. We label these groups as `unrelaxed', to contrast them to the well-behaved, relaxed groups discussed above. Figure \ref{fig:UnrelaxedMass} shows the distribution of total group masses 
for these unrelaxed groups, comparatively with the distribution of group masses for the entire \emph{ZENS} sample; this shows that unrelaxed groups span a large range of masses, from low to relatively high ones.

We expect a physical origin to contribute to our inability to
 identify a bulletproof central galaxy in the `unrelaxed' groups in the  \emph{ZENS} sample.
Non-physical factors may also however contribute to preventing us from identifying the real central galaxy in some of these groups. The main sources of error are  again related to interlopers in group membership, incompleteness
 in the parent 2dFGRS database and/or the inherent limitations of the 2PIGG group-finding algorithm (see Section \ref{sec:2dFGRScompl},  Section \ref{sec:cenErrors}, and Appendix \ref{App:missedgals}). 
 Based on the tests that we have conducted to understand the impact of interlopers and missing galaxies in the identification of central galaxies (\ref{sec:cenErrors}), we expect that $\sim 20-25\%$  of  groups may appear as `unrelaxed' due to these catalogue failures.  The fraction that we observed is however substantially higher, of order $\sim40\%$. From this we  estimate  that, in at least $\sim10-15\%$ of groups in the \emph{ZENS} sample,  the displaced centrals are a genuine  smoking gun for an unsettled dynamical state. This may result from the accretion of  individual galaxies by the group potential and/or of group-group merger events. This rough estimate   for the fraction of genuinely unrelaxed groups in the \emph{ZENS} sample is consistent with the estimate derived from the comparison with the Yang et al.\ group sample (Section \ref{sec:YangComp}), and with a sub-clustering analysis which we describe below.

\subsection{Testing  the dynamical state of groups with a sub-clustering analysis}\label{subclass}

As a complementary method for testing the dynamical state of the \emph{ZENS} groups we  searched for substructures in  position and velocity space, following the approach described in \citet{Dressler_Shectman_1988}. In the original test, for each group, a local mean velocity ($\bar{v}_{local}$) and velocity dispersion ($\sigma_{local}$) around each member is calculated by using the $N$th nearest neighbors galaxies in the group, plus the galaxy on which the search is centered.
The quantity $\delta^2=\frac{(N+1)}{\sigma^2} \left[(\bar{v}_{local}-\bar{v} )^2+(\sigma_{local}-\sigma)^2\right]$  parameterizes the deviation of this subset of galaxies  from the group global velocity and dispersion, with $\bar{v}$ and $\sigma$  the group mean velocity and total dispersion.
Under Gaussian assumption and in the absence of substructures within the groups, the sum of the $\delta$ parameters of all galaxies in a group ($\Delta_{tot}$) will be close to the number of its members. 
As discussed in \citet{Dressler_Shectman_1988} a non-Gaussian distribution of galaxies velocities  can bias the result also in the absence of real substructures. For these reasons, the test is repeated for a number of Monte Carlo realizations, in which the position of the galaxies are held fixed, but the velocities are randomly redistributed between the group members.
Any intrinsic correlation among velocities will be thus erased and these Monte Carlo samples can be used to quantify the probability that a value of $\Delta_{tot}$ larger than the one observed can originate from a random distribution.

To optimize the test for the \emph{ZENS} groups, which have typically much lower richness than the clusters for which the test was devised, we applied the following modification to the original formulation: we chose an $N$ which depends to the group richness to calculate $\bar{v}_{local}$ and $\sigma_{local}$; specifically, we adopted  $N=0.4\times N_{members}$. This accounts for the fact that the \emph{ZENS} groups span a wide richness range, from  $N_{members}=5$ to  $N_{members}\simeq100$.
Following the above prescription, we generated 500 Monte Carlo simulations  for each \emph{ZENS} group and 
identified groups having significant sub-clustering as those in which less than $20\%$ of the Monte Carlo simulations result in a  $\Delta_{tot}$ larger than the  measured value for that given group.

More than $80\%$ of   groups  that we classify as  `relaxed' according to the criteria described in section \ref{sec:centralsDef} also show no hint for substructure in this clustering analysis; a fraction of  about $20\%$ of  `unrelaxed'   groups  show distinct  substructure  in the  $\Delta$ statistics analysis. There is therefore a good global agreement between the two approaches  in establishing that a group is a relaxed system, and in hinting at an absolute fraction of \emph{ZENS} groups that are genuinely dynamically young of order $\sim 10-15\%$.

We briefly investigate below whether and how the central and satellite  galaxy populations in `relaxed' and `unrelaxed' groups display differences that can help understanding the co-evolution of galaxies and their host group halo potentials. 

\subsection{A quick exploration of centrals and satellites  properties in relaxed and unrelaxed groups}

 To compare the distribution of galactic properties of central or satellite galaxies in relaxed groups  with those of similarly-ranked galaxies in unrelaxed groups, we limit the sample to groups with $M_{GROUP}<10^{13.5}\Msol$. Up to this halo mass, there is a fair mix of unrelaxed and relaxed groups in our sample; in contrast, the sample at higher group masses is tilted towards relaxed systems. This cut therefore helps avoiding attributing to the dynamical state of the groups differences in the galaxy populations which stem instead from a different halo mass distribution for the two categories of groups  (shown in Figure \ref{fig:UnrelaxedMass}).

We consider two bins of stellar mass, i.e., $10^{9.3} M_\odot < M < 10^{10} M_\odot$ and  $10^{10} M_\odot < M < 10^{10.7} M_\odot$. Only satellites populate the `low-mass' bin in our sample; in the `high-mass' bin both satellites and centrals are fairly  represented.  We use the measurements published in \citealt{Cibinel_el_al_2013a,Cibinel_el_al_2013b} to search for differences in galaxy half-light radii\footnote{Note that we use here the global galactic half-light radii obtained through our double-component, bulge plus disk fits to the two-dimensional galaxy surface brightness distributions; see Paper II for details. Furthermore we employ semi-major axis measurements for all galaxies, except for elliptical galaxies for which we use circularized  half-light radii} $r_{1/2}$, specific star formation rates (sSFRs), surface densities of star formation rate ($\Sigma_{SFR}$) and $(B-I)$ colors. As discussed in aPaper III, $SFR$ and $sSFR$ for galaxies in which the best-fit template result in very low $SFR<10^{-4}\Msol yr^{-1}$ are set to $SFR=10^{-4}\Msol yr^{-1}$ and $sSFR=10^{-14}\Msol yr^{-1}$, respectively. 

\begin{figure*}[htbp]
\epsscale{1}
\includegraphics[width=180mm]{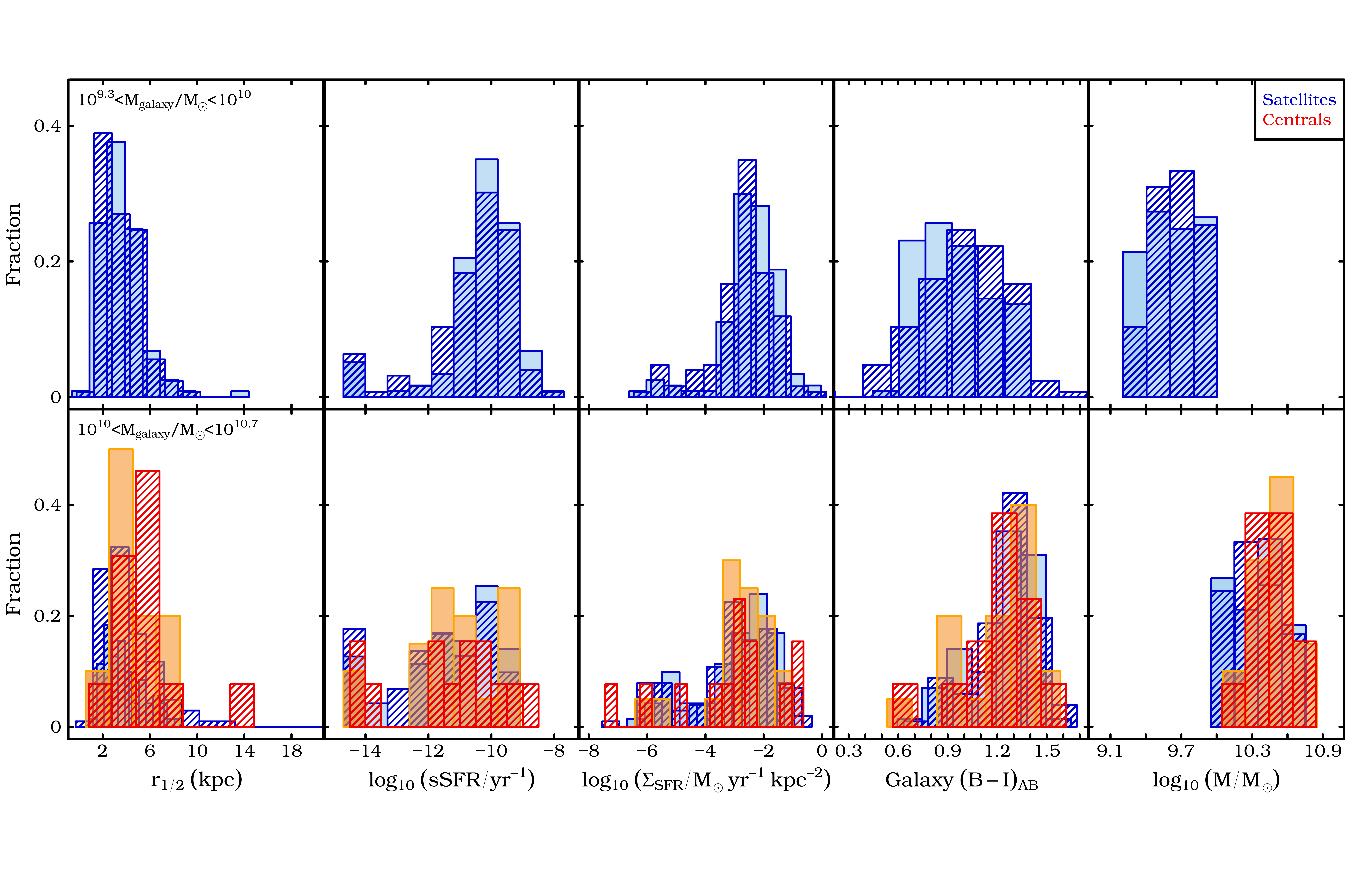}
\caption{From left to right, the figure shows the distribution of galaxy half-light radii, specific star-formation rates, star-formation rate surface densities, $(B-I)$ colors and stellar masses for central (red/orange) and satellite (dark-/light-blue) galaxies in two bins of stellar mass (top and bottom rows, as indicated in the figure). Dark hatched histograms are for galaxies in relaxed groups, light filled histogram are for galaxies in  unrelaxed groups. To match the distribution of halo  masses of the unrelaxed sample and avoid spurious effects in the comparison between relaxed and unrelaxed groups due to differences in the distributions of group masses between these two families (see Figure \ref{fig:UnrelaxedMass}), only groups with $M_{GROUP}<10^{13.5}\Msol$ are  used in this figure.}
\label{fig:Fig14} 
\end{figure*}

The results are presented in Figure \ref{fig:Fig14}.
This shows, from left to right, the distribution of $r_{1/2}$, $sSFR$, $\Sigma_{SFR}$, $(B-I)$  and stellar mass for central (red/orange) and satellite (dark/light blue) galaxies in our low (top) and high (bottom) bins of stellar mass. 

We find a global similarity between color and star formation  properties of central galaxies in relaxed and unrelaxed groups in our high-mass bin $10^{10} M_\odot < M < 10^{10.7} M_\odot$ (see Appendix \ref{newtable}).
The median half-light radius of central galaxies in relaxed groups is however   larger than for (alleged) centrals of similar mass in the non-relaxed groups (5.30$_{-0.32}^{+0.40}$   kpc and 3.77$_{-0.25}^{+0.69}$ kpc, respectively, with a KS-test probability for the  size distributions of relaxed and unrelaxed groups to be different of about 90\%). The fact that no statistically significant effect is seen in either the sSFR or the $\Sigma_{SFR}$ distributions between the two samples of centrals  is possibly a reflection of relatively large errors on these quantities (since the median stellar mass within the mass bin for the relaxed and unrelaxed groups is virtually identical, i.e.,   respectively  10.50$_{-0.06}^{+0.04} M_\odot$     versus   10.50$_{-0.04}^{+0.02} M_\odot$).  

The median size of central galaxies in the high mass bin well matches the median size of satellite galaxies of similar masses (both in relaxed and unrelaxed groups, respectively equal to  3.27$_{-0.11}^{+0.31}$  kpc and      3.41$_{-0.13}^{+0.31}$      kpc). There is a small shift in galaxy mass between centrals and satellites within this mass bin,  i.e.,   10.30$_{-0.02}^{+0.03} M_\odot$                              for satellites in relaxed groups, to be compared with the corresponding value for centrals given above; this is however not sufficient to explain the difference in median size. Once again  this may well be evidence that, in at least some of the unrelaxed groups,  the nominal central galaxy is not a real `central', but a satellite galaxy which  has been mistaken for a central due to survey incompleteness issues. 
It may also partly indicate, however, that central galaxies in relaxed group potentials further grow in size  relative to (pseudo)centrals in young/merging groups; this may happen thanks to accretion of low mass satellites in virialized group potential, as shown by numerical simulations (e.g., Hopkins et al.\ 2008; Feldmann et al.\ 2010).

Comparing amongst themselves,   at constant stellar mass, satellite galaxies,   these   have virtually identical properties independent of whether they inhabit relaxed or  unrelaxed groups, with only a hint  in our data for   low mass, $10^{9.3} M_\odot < M < 10^{10} M_\odot$ satellites in unrelaxed groups to be  on average $0.05$ magnitudes    bluer  than galaxies of similar rank and mass in relaxed groups (at the $\sim90\%$ probability level).  We note that we do not detect a similar effect in the sSFR (or $\Sigma_{SFR}$)  diagrams; this again we interpret as possibly  due to  dilution of signal due to intrinsic uncertainties in   the SFR values derived from SED fits. All median values for the histograms of Figure \ref{fig:Fig14} are listed in Appendix \ref{newtable}.

\subsection{The stellar content of relaxed and unrelaxed groups}

The fraction of group mass that is in the form of stars is an indication of how efficiently star formation has progressed in a given halo. We thus ask whether there are detectable differences in the relation between total stellar mass  and halo mass between relaxed and unrelaxed ZENS groups.

Following a similar approach as the one described in section \ref{sec:GroupMasses} to calculate the group total luminosity, we derive the total stellar mass that is locked in galaxies  within a given halo as follows. We first sum  the incompleteness-weighted mass of all member galaxies above the completeness limits, for star-forming and quenched galaxies separately, i.e, $M_{OBS,SF}= \sum_{i,M>M_{lim,SF}}w_{i,SF}M_{i,SF}$ and $M_{OBS,Q}= \sum_{i,M>M_{lim,Q}}w_{i,Q}M_{i,Q}$.
We use $M_{lim,SF}=10^{9.2}\Msol$ and $M_{lim,Q}=10^{10}\Msol$  for star-forming and quenched galaxies, respectively, as derived in Paper III. These estimates need to be  corrected for the stellar mass in galaxies falling below $M_{lim}$.

The correction  is done by integrating separately the mass functions of star-forming and quenched galaxies, for which we adopt  the estimates of the Schechter function parameters respectively for blue and red galaxies  provided in Table 3(a) of \citealt{Peng_et_al_2010}.  Although  \citealt{Peng_et_al_2010} provide the mass function parameters split  in quartiles of  high and low environmental density, 
we utilize the parameters obtained for the global populations, since the one-to-one matching between our group environments and their density definition is not straightforward. In analogy with the computations outlined in section \ref{sec:GroupMasses}, the correction factor,  
by which we divide $M_{OBS}$ is    $\Gamma(\alpha_{*}+2,M_{lim,SF}/M_{*})/\Gamma(\alpha_{*}+2)$ for the star-forming population (characterized  by a single Schechter function, Peng et al.\ 2010). The correction for the quenched population (described by two Schechter functions)   
 is:  
  
  \begin{equation*}
  \frac{\Phi_{*,1}\Gamma(\alpha_{*}+2,M_{lim,Q}/M_{*})+\Phi_{*,2}\Gamma(\alpha_{*,2}+2,M_{lim,Q}/M_{*})}{\Phi_{*,1}\Gamma(\alpha_{*}+2)+\Phi_{*,2}\Gamma(\alpha_{*,2}+2)} \, .
  \end{equation*}

The resulting contribution from galaxies below the completeness limits  to the total stellar mass is of order  6\% and 15\%  for quenched and star-forming galaxies, respectively. The total mass in galaxies $M_{tot, galaxies}$ is finally obtained as the sum of the corrected masses for the star-forming and quenched populations, $M_{tot, galaxies}=M_{SF}+M_{Q}$. 

 \begin{figure}[htbp]
\includegraphics[width=90mm,angle=90]{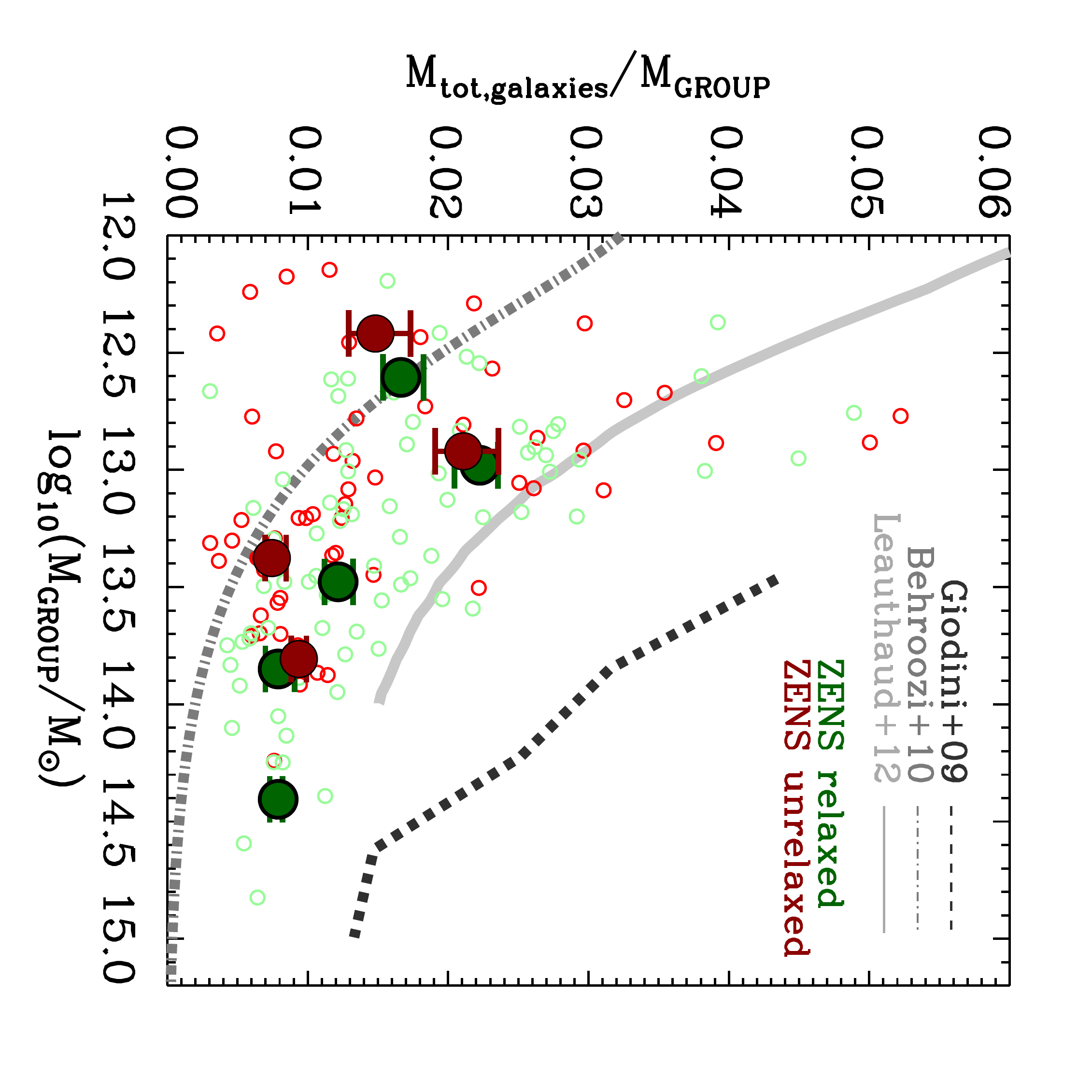}
\caption{\label{fig:FigMstellar} Fraction of group halo mass which is converted into stars within galaxies as a function of   group halo mass. Small empty points are the values for the individual ZENS groups: green   show  relaxed groups  and   red  shows unrelaxed groups. Large  symbols with error-bars of corresponding colors are  median values for the relaxed and unrelaxed ZENS groups.  For a qualitative comparison, estimates   from the literature are shown: 
the \citealt{Giodini_et_al_2009} relation for  a sample of X-ray selected groups at $0.1\leqslant z \leqslant 1$ (dashed line), 
the \citealt{Behroozi_et_al_2010} relation derived from halo abundance matching at z=0.1 (dash-dotted line) and the  halo occupation-based estimate of   \citealt{Leauthaud_et_al_2012}   at redshift z=0.37 (solid line).}
\end{figure}

 The relation between   $M_{GROUP}$ and $M_{tot, galaxies}$  for relaxed and unrelaxed groups is plotted in Figure \ref{fig:FigMstellar}, where we also show the results of  other literature studies for groups and clusters. 
A quantitative comparison between the different samples  is made difficult by a number of factors, including differences in the groups selection criteria, in redshift  and in the assumptions and methodologies used for calculating  the total stellar masses (and halo masses). This is at least in part the cause for the scatter in the relation resulting from the direct comparison of the different studies.
Nevertheless,  a number of similarities can be highlighted. The ZENS groups show a clear dependence of the stellar mass fraction on the halo mass itself, as also found in the other analyses. Specifically,
 the fraction of halo mass which is in the form of stars in the ZENS  groups generally amounts to about 1-2\%,  and it is a factor of order two higher at group masses $<10^{13}\Msol$ ($\sim$2\%) than at  group masses $>10^{13.5}\Msol$ ($\sim0.8\%$). This increase in efficiency of star formation in Milky-Way sized halos, i.e. at the low-end of the mass distribution of the ZENS groups, has indeed been previously highlighted as a quite fundamental (and redshift independent) fact of nature \citep{Behroozi_et_al_2013}. Interestingly, there is no evidence for relaxed and unrelaxed groups having significant different stellar mass fractions or dependence on halo mass. Thus, at face value, the efficiency of conversion of gas into stars is not affected by the  dynamical state of the group.

\section{Summary and Concluding Remarks}\label{sec:endfinally}

Motivated by the picture that both the mass of a galaxy, and its immediate and distant environment, may impact how the galaxy evolves  and  its redshift zero properties, and by the uncertainty on which mass and which environment are the relevant ones to galactic life, we undertake the \emph{ZENS} project, which uses new and archival multi-wavelength  data for  a statistically complete sample of 1627 galaxies brighter than $b_J=19.45$ which are members of 141 $\sim10^{12.5-14.5} M_\odot$, $0.05<z<0.0585$  groups. The emphasis of \emph{ZENS}  is to explore the dependence of key galactic populations diagnostics on the large-scale environment, on the mass of the host group halo, on the location of galaxies within their group halos, and on the central/satellite rank of a galaxy within its host group halo. The \emph{ZENS} sample is extracted from the 2PIGG catalogue of the  2dFGRS. 
We publish the \emph{ZENS}  catalogue which combines the environmental diagnostics computed in this article with the structural and spectrophotometric galactic measurements described in \citealt{Cibinel_el_al_2013a,Cibinel_el_al_2013b} .

In this first paper, introducing the project, we have described improved algorithm adopted to define the group centers, to rank galaxies as centrals or satellites in their host groups, and to separate the effects on galaxies of  groups mass and LSS density. Specifically:

 $(i)$ We have introduced a three-faceted self-consistency criterion for identifying central galaxies. These must, simultaneously, be the most massive galaxies in the group within the errorbars estimated for the galaxy stellar masses, and  must be consistent with being the spatial  and dynamical centers of the host groups.
 
 $(ii)$ We have adopted a  Nth-nearest {\it group}-neighbors computation  to  estimate  the  LSS density underlying the groups which, especially at group masses $M_{GROUP}<10^{\sim13.5} M_\odot$,  and in contrast with the commonly used Nth-nearest {\it galaxy}-neighbors approach,  is independent of group mass/richness and enables us to study separately the effects of these two distinct environments on galaxy properties.
 
Furthermore, we have used simulations, also based on semi-analytic models of galaxy evolution, to quantify the intrinsic uncertainties   in the trends of galaxy properties with the environmental parameters, that are propagated from the random and systematic errors in these parameters.
 
We have  found that at least $\sim$60$\%$ of groups are dynamically-relaxed systems with a well-identifiable central galaxy that satisfies the stringent criterion above -- and thus a well-defined center of the group. These groups enable a robust investigation  of galaxy properties with group-centric distance down to the smallest group masses sampled in \emph{ZENS}. In  the remaining $\sim40\%$ of groups there is no galaxy which  satisfies the required criteria to be a central galaxy -- and thus the center of the group potential well. We estimate that a non-negligible fraction of  these -- up to of order $\sim10-15\%$ of groups in the total \emph{ZENS} sample,  are likely  genuinely dynamically young, possibly merging groups. 

At constant stellar mass, central galaxies in relaxed and unrelaxed groups have similar  color and star formation properties, although they show  larger sizes in relaxed versus unrelaxed groups. Centrals in unrelaxed groups have sizes comparable to satellite galaxies of similar masses. These results may partly arise from the misclassification of satellite galaxies as central galaxies in groups, in our analysis labelled as unrelaxed, for which however the identification of their dynamical state and of the central galaxy might be hampered by observational errors. We estimate  that  in about  two-thirds of nominally `unrelaxed' groups, the lack of identification of a self-consistent  central galaxy has its roots in the incomplete spectroscopic and photometric coverage of the 2PIGG and 2dFGRS surveys, respectively. 
Therefore, our use of the term 'unrelaxed' should be 
read as highlighting the  important  fact that the alleged central galaxies, and thus the centers in these 
groups, should be handled with care.  

Partly however the lack of dependence   central galaxies properties on nominal dynamical state of the group may   be evidence   that the properties of central galaxies are shaped by their own mass content and not by their group environment, with the exception of a growth in size  in dynamically-relaxed halos due to secular  accretion of smaller satellites. 

Over the whole galaxy mass range of our study, satellites have indistinguishable physical properties (in terms of sizes, optical colors, sSFRs and $\Sigma_{SFR}$) independent of whether they are hosted by relaxed or unrelaxed groups. 

Furthermore, relaxed and unrelated groups appear to have similar   gas-to-star conversion efficiencies, which, as found in other studies, peak  around the$10^{12.5} M_\odot$ halo mass; this  suggests that the efficiency of conversion of gas into stars within halos may be largely independent of the  dynamical state of the group. A more detailed investigation of this important issue is postponed to a future dedicated paper. 

The only possible difference between relaxed and unrelaxed potentials is a very modest shift towards redder $(B-I)$ colors for  $<10^{10} M_\odot$ satellites  in relaxed relative to unrelaxed groups.  A possible explanation  is that, at the higher masses, satellites   are either unaffected by the group environment, or they   reach their final state as they first enter the potential of a relatively small group, with subsequent group-group  mergers   having no  further  impact on their properties  (see also \cite{DeLucia_et_al_2012} for theoretical support to this scenario). 

The marginally redder color of  low-mass satellites in relaxed relative to unrelaxed groups may be due to satellites orbiting since longer times within the former relative to the latter, or to quenching of star formation in these systems for processes that are active or at least most efficient  in relaxed  group potentials. Independent studies  also point at the low-mass satellite-quenching by physical processes   acting within virialized halos.

In future \emph{ZENS}  analyses we will investigate whether including or excluding the unrelaxed groups from  any given specific diagnostic will impact our main conclusions, and how, and will explicitly comment on this when  it will.

\section*{acknowledgments}
A.C., E.C. and C.R. acknowledge support from the Swiss National Science Foundation.
This publication makes use of data from ESO Large Program 177.A-0680, and data products from the Two Micron All Sky Survey, which is a joint project of the University of Massachusetts and the Infrared Processing and Analysis Center/California Institute of Technology, funded by the National Aeronautics and Space Administration and the National Science Foundation. GALEX (\emph{Galaxy Evolution Explorer}) is a NASA Small Explorer mission. The Millennium Simulation databases used in this paper and the web
application providing online access to them were constructed as part of
the activities of the German Astrophysical Virtual Observatory.

\clearpage

\LongTables      %comment this in aastex
\begin{deluxetable}{lcccccccccccc}
\tablewidth{0pt}
\tabletypesize{\tiny}
\tablewidth{0pt}
\tablecaption{Coordinates, properties and LSS environment of the \emph{ZENS} groups. \label{tab:ZENs_prop}}
\tablehead{
\colhead{name} & \colhead{RA} & \colhead{DEC} & \colhead{$z$} & \colhead{$r_{rms}$} & \colhead{$\hat{R}_{200}$} & \colhead{$\sigma$} & \colhead{L$_{GROUP}$} & \colhead{N$_m$} & \colhead{M$_{GROUP}$} & \colhead{$\log(1+\delta_{LSS})$} & \colhead{Env.} \\
& \colhead{(B1950)}  & \colhead{(B1950)} & &   \colhead{(Mpc)} & \colhead{(Mpc)} & \colhead{($$km s$^{-1}$)} & \colhead{($10^{10}\Lsol$)} &  & \colhead{($10^{12}\Msol$)} &  & 
}
\startdata
 $^2$ 2PIGG-s1248        &  23:41:34.53  &  -26:44:18.8  &  0.05188  & 0.362 &    0.340    &  600.7  &    4.990    &    8  &    4.683    &     0.78  &  4  \\   
 $^2$ 2PIGG-s1272        &  23:37:55.83  &  -30:07:02.9  &  0.05079  & 0.784 &    0.465    &  247.6  &    8.602    &    9  &    11.98    &     0.03  &  2  \\   
 $^2$ 2PIGG-s1282        &  23:43:44.24  &  -27:40:01.7  &  0.05038  & 0.446 &    0.356    &  148.8  &    5.414    &    8  &    5.363    &     0.65  &  4  \\   
 $^2$ 2PIGG-s1308        &  23:59:16.70  &  -35:45:24.3  &  0.05143  & 0.338 &    0.328    &  287.4  &    4.678    &    5  &    4.211    &     0.34  &  3  \\   
 $^2$ 2PIGG-s1313         &  23:43:59.64  &  -28:17:42.9  &  0.05020  & 0.454 &    0.445    &   57.2  &    7.976    &    6  &    10.49    &     0.40  &  3  \\   
 $^5$ 2PIGG-s1325         &  22:28:47.25  &  -27:04:44.6  &  0.05066  & 0.222 &    0.363    &   42.0  &    5.624    &    5  &    5.719    &    -0.00  &  2  \\   
 $^5$ 2PIGG-s1334         &  23:31:22.74  &  -30:08:48.2  &  0.05157  & 0.785 &    0.840    &  301.9  &    26.53    &   23  &    70.70    &     0.92  &  4  \\   
 $^2$ 2PIGG-s1349        &  23:53:45.32  &  -25:48:25.1  &  0.05122  & 0.272 &    0.236    &  132.0  &    2.378    &    6  &    1.564    &    -0.38  &  1  \\
\enddata
\end{deluxetable}

%%%%%%%%%%%%%%%%%%%%%%%%%%%%%%%%%%%%%%%%%
%%%%%%%%%%%%%%  APPENDIX   %%%%%%%%%%%%%%%%%%%%
%%%%%%%%%%%%%%%%%%%%%%%%%%%%%%%%%%%%%%%%%

\appendix 

\section{The impact on \emph{ZENS} of the original 2dFGRS magnitude limits } \label{App:2dFGRS_limits}

The 2dFGRS team made available three maps which specify for a given position
 in the sky $\theta$: a) the extinction-corrected magnitude limit of the survey $b_{j,lim}(\theta)$;
  b) the redshift completeness $R(\theta)$, - the number of galaxies with measured redshift 
  relative to the parent APM survey  catalog, which is the photometric basis of the 2dFGRS; 
  and c) the parameter $\mu(\theta)$ that enters  the expression for the magnitude-dependent 
  redshift completeness, $c_z(b_j,\mu(\theta))=\gamma\left[1- \exp(b_j-\mu(\theta))\right]$, with $\gamma=0.99$ \citep{Colless_et_al_2001}. The overall redshift completeness around a given set of celestial coordinates is given by: 
 $C(\theta,b_j)=R(\theta)c_z(b_j,\mu(\theta))/\bar{c}_z(\mu(\theta))$.
  The factor $\bar{c}_z(\mu(\theta)$ is a normalization constant derived from the average of $c_z(b_j,\mu)$ over the expected apparent magnitude distribution of the survey galaxies \citep{Colless_et_al_2001, Norberg_et_al_2002, Cole_et_al_2005} and can be calculated using equation 7 of \citet{Colless_et_al_2001}. 
  
\begin{figure}[hbtp]
\begin{center}
\includegraphics[width=80mm,angle=90]{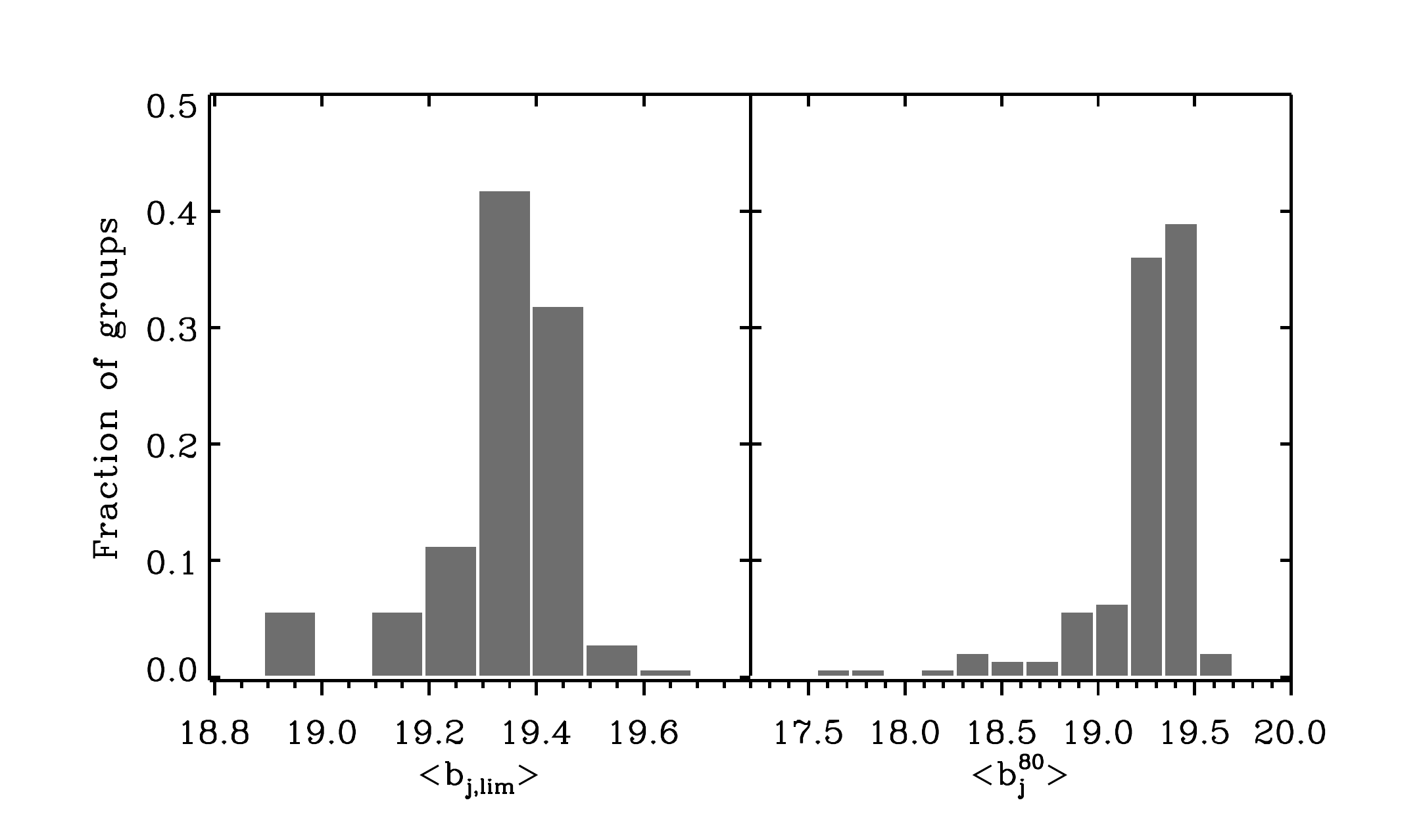}
\end{center}
\caption{\label{fig:surveyCompletness} Completeness of the 2dFGRS over the targeted \emph{ZENS} fields. {Left:} Average value of the limiting magnitude in the 2dFGRS catalog, without constraints on completeness. \emph{Right:} Mean limiting  magnitude for the \emph{ZENS} fields, imposing a 80$\%$ completeness level  in the 2dFGRS images.}
\end{figure}

	The 2PIGG catalogue is constructed only from those fields and sectors of the 2dFGRS which have a high number of measured redshifts and during the selection of the \emph{ZENS} groups we furthermore restricted the sample to the most complete ones (i.e. those which have galaxy weights from the 2PIGG catalogue $<1.6$). This ensures that the average completeness $R(\theta)$ in the group, defined as the mean of all values  at the positions in the sky of the member galaxies,  is typically $\sim90\%$.
	We thus compute the limiting faint magnitude at which the survey is complete at the $80\%$ level ($<b_{j}^{0.80}>$) from mean estimates of the limiting magnitude without constraints on completeness ($<b_{j,lim} >$) and by inverting the expression for  $C(\theta,b_j)$ given above.  In calculating the factor $\bar{c}_z(\mu(\theta)$ we use a bright and faint magnitude limit of $b_j=14$ and $b_{j,lim}(\theta)$, respectively. Figure \ref{fig:surveyCompletness} shows the derived distribution of $<b_{j,lim} >$ and $<b_{j}^{0.80}>$.  There are small variations, amongst the \emph{ZENS} groups, in the faintest magnitude reached by the original  2dFGRS data. As shown in the Figure, the effect is however small, with only a handful of groups having $<b_j^{80}> $ brighter than 19. Most of the \emph{ZENS} groups are complete down to the  ($<b_{j,lim} >$) limit. We have checked in all cases that none of our results are affected by this modest  field-to-field scatter in completeness in the \emph{ZENS} fields.
  
We finally applied corrections for spectroscopic completeness. As done in the 2dFGRS studies, these are obtained by assigning to each galaxy a weight $w$  defined as $w=1/C(\theta,b_j)$, such that the complete number of galaxies $N$ (total, or of a given type)  is $N=\sum_i 1/w_i$.

\section{Impact of `missed' galaxies on our analyses}\label{App:missedgals}

\subsection{Searching in the SDSS for galaxies missed by the 2dFGRS} \label{App:SDSSMatch}

The ZENS fields lie in regions of the 2dFGRS that have  an average redshift completeness of 87\%, with  some variations: 128 groups have a completeness of at least 80\% while 13 groups have a lower completeness between 72\% and 80\%. Four of these latter groups are in the ZENS ÔunrelaxedÕ class.  Comparing the positions of the ZENS groups with respect to the 2dFGRS survey boundaries, we found that 9 unrelaxed groups may be close enough to the survey edges/gaps (within 1 Mpc) to have been only partially covered by the 2dFGRS observations.

To understand these and other    biases in the \emph{ZENS} sample introduced by galaxies `missed' by the 2dFGRS\footnote{This approach is similar to the one we adopted in Section \ref{sec:MembershipTest} to search for suitable galaxies in the 2dFGRS which had not been associated with a given group by the 2PIGG algorithm.},
we studied the SDSS DR7 spectroscopic catalogue of \citet{Abazajian_et_al_2009}.
 About a quarter (43 of 141) \emph{ZENS} groups are located in fields that overlap with the  SDSS. For each of these 43 groups, the  search for missed galaxies was performed
on circular projected areas of radius equal to 1.5 times the r.m.s. radius of the group, centered on  the nominal most massive galaxy. 

To set an operational definition, we considered as plausible missed galaxies in each of these groups galaxies  with coordinates within these circular
   areas, and with redshifts between $z_{min}-\delta<z<z_{max}+\delta$. Here  $z_{min}$ and $z_{max}$ are
    the minimum and maximum redshift of the galaxies in the given 2PIGG group,
     a $\delta$ value from  $10\%$ up to 30$\%$ of the redshift interval spanned by the nominal 
     2PIGG galaxy members of that group was explored.

With  $\delta=30\%$  we found a total of 56 `extra' galaxies in the SDSS which satisfied these criteria 
     in 19 of the total 43 \emph{ZENS} groups with  SDSS pointings, to be compared with a total of 267 nominal 2PIGG members in these groups. 
A summary of   the fields with these `extra' galaxies is given in Table \ref{tab:SDSSextra}.
 As indicated in  this Table, only a small fraction of these galaxies have magnitudes below the 
  nominal selection limit of the 2dFGRS (for 26 galaxies we could not find the
   information on the $b_j$ magnitude; for these 
 we used the relation between $b_j$ and SDSS $g$ 
  magnitudes for those galaxies in which both are available). An analysis of the images shows that
   fiber collisions should not be a main reason for the absence of these galaxies
  from the 2dFGRS catalog. Although ultra-compact galaxies could
   be missed due to a star/galaxy misclassification,  generally  these galaxies seem  simply casualties of the 2dFGRS statistical sampling.
  
The statistics above suggests that of order $\sim40-50\%$ of the \emph{ZENS} groups Ð 
  and in general of the 2PIGG groups Ð are potentially missing some member galaxies
   above the 2dFGRS magnitude limit, due to their absence from  the parent 2dFGRS catalog. 
    We use this information to assess an order of magnitude estimate for the impact of the plausible extra members on our analyses, including group mass estimates as well as the definition of centrals and satellites.  

 In Table \ref{tab:SDSSextra} we show the nominal group masses for these 19 groups; 
 even assuming that all missed SDSS galaxies are additional members of the relevant \emph{ZENS}
  groups changes the group masses by less than $30\%$ in  85\% of them.
 In two groups the change in mass would be   $\sim60\%$, and in one the mass would change by 
 a factor of two. 
 
We then ran on these 19 groups the algorithm described in Section \ref{sec:centralsDef} for the 
identification of central and satellite galaxies, this time also including the 56 extra galaxies
 (using the total stellar masses provided in the MPA/JHU value added catalog
  for  the masses of the extra SDSS galaxies; in Paper III we use a sample in common
   to show that there are no severe systematics between our estimates for galaxy stellar masses and those of this catalog).
    Since for the extra SDSS galaxies we do not have information on the full PPDs for their stellar mass, we generated artificial Gaussian PPDs,     centered on the galaxy MPA/JHU stellar mass, and with a standard deviation of 0.3 dex.  
     Only in four of these 19  groups (2PIGG-n1363, 2PIGG-n1457, 2PIGG-n1469 and 2PIGG-n1540) the inclusion of the extra SDSS galaxies results in a possible change
      in the identification of the central galaxy.
       Three of these four potential `SDSS centrals' have structural, morphological stellar mass and star formation
        properties very similar to those of the nominal \emph{ZENS} central. In the remaining case,  
        the `SDSS central' is a  quenched E/S0 galaxy, 
        in contrast with the nominal \emph{ZENS} central,  which had an intermediate 
        disk morphology and an intermediate SFR (see Paper III for our
         definitions of  quenched, moderately star-forming and strongly star-forming). 
While in principle such situation may lead to uncertainties in  the analysis of the central and satellite galaxy populations,
 the global statistics are comforting. 

We estimate the incompleteness relative to the SDSS as follows.  We first assumed that all 
missed SDSS galaxies are physically associated with the 19 groups in question, 
and that the true central galaxies in the four aforementioned groups, for which 
the inclusion of the SDSS extra sample leads to a change in the identification 
of the central, are indeed the newly added SDSS galaxies rather than the 
nominal \emph{ZENS} centrals. Considering all 43 groups for which we 
know whether they are (or not) missing  SDSS galaxies, we then defined $(i)$ 
the number of centrals that we should have observed,  $n_{centrals}=43$;  
$(ii)$ the number of centrals that we have correctly identified, 
$n_{centrals,obs}=39$; $(iii)$ the  number of satellites that we should have
 observed, $n_{sats}=505$ (i.e., the total sample of 492 2PIGG members
  of the 43 groups in question, plus the 56 extra SDSS galaxies found in 
  total for this sample, minus 43, the number of their centrals); 
  $(iv)$  the number of satellites which are misclassified as centrals, 
  $n_{false-cen}=4$, and, finally,  $(v)$ the number of correctly identified satellites 
  $n_{sats,obs}=449$ (i.e., the total sample of 492 2PIGG members, minus 43 centrals). 
  We then estimate  the level of incompleteness due to 
  missing SDSS galaxies in the 2dFGRS sample as, for the central galaxies,
   $1-n_{centrals,obs}/n_{central}\sim10\%$, and, for the satellites, $1-n_{sats,obs}/n_{sats}\sim10\%$.
 This implies a level of contamination of satellites incorrectly 
 identified as centrals of $n_{false-cen}/n_{centrals}\sim10\%$. All these are upper limits to the fraction of misidentifications, since not all `extra' galaxies identified as described above will be missed group members. 
 We therefore conclude that this specific source of uncertainty in the identification of the central (and thus satellite) galaxy populations  is not a dominant one. Such identification remains mostly affected by other factors such as  the global impact of the friends-of-friends clustering algorithm
  used for the identification of bound galaxy groups.
 
 For the four groups with a candidate `missed' central galaxy, we had to decide to which galaxy to assign the rank of  central. We maintained the identification of the central galaxies in these  groups with the original centrals found amongst the nominal 2PIGG group members, and checked that this choice does not affect any of our conclusions.

The above checks  imply virtually  no effect  of these missed potential galaxy group members on  any of our  studies of the group environment based on our group mass estimates, galaxy membership and central/satellite ranking.

\subsection{ZENS galaxies missed by ZENS pointings}\label{ap:MissingGalaxies}

Another possible source of error in the estimate of group mass and identification of centrals and satellites are galaxies missed in the WFI observations. For 28 of the 141  groups, the WFI pointings did not  cover their entire
extent, resulting in a total of 172 members for which no $B-$ and $I-$band imaging is available.
 These groups are indicated with an exclamation mark in Figure \ref{fig:cenSpatial}. 
For 20 groups the fraction of missing members is $<20\%$ of the original 2PIGG group richness, for other six is between $20-30\%$ and only for two massive groups (2PIGG-s1935 and 2PIGG-n1377) is as high as $40-45\%$.
We include in the \emph{ZENS} catalogue these galaxies out-of-WFI-field,  setting to null entries all quantities which rely on the WFI photometry except for the galaxy mass.  A mass estimate for these galaxies was in fact  obtained from the linear relation between the SuperCOSMOS Survey $r_F$ magnitude  (provided in the 2dFGRS data release and corrected for galactic extinction) and the SED inferred galaxy mass, as derived for the \emph{ZENS} galaxies with available $B-$ and $I-$band observations.
We assumed for the mass probability distribution of these galaxies a Gaussian centered on the mass predicted by the $r_F-M$ relation, and having a deviation equal to 1.5 times the observed scatter of this relation.

\begin{deluxetable*}{lccccc} 
 \tablewidth{0pt}
\tabletypesize{\scriptsize}
\tablewidth{0pt}
\tablecaption{ZENS groups with extra candidate galaxy members in the SDSS \label{tab:SDSSextra}}
\tablehead{
 \colhead{Name} & \colhead{Nominal members}  & \colhead{SDSS candidates}  & \colhead{Below $b_{j,lim}$} &  \colhead{$M_{GROUP}$} 
 }
\startdata
2PIGG$\_$m1363               &     8            & 2        & -     & 7.781$\times 10^{12}$ \\
2PIGG$\_$m1377               &   23            & 1        &  -    & 7.506$\times 10^{13}$   \\
2PIGG$\_$m1381                &  10             & 3        &  -    & 1.223$\times 10^{13}$   \\
2PIGG$\_$m1384                &  11              & 1        &  1  & 5.796$\times 10^{13}$  \\
2PIGG$\_$m1418                &  5                & 1        & -    & 8.660$\times 10^{12}$  \\
2PIGG$\_$m1457                &  30              & 9        & 1   & 1.741$\times 10^{14}$   \\
2PIGG$\_$m1469                &   6               & 1        & -    & 2.051$\times 10^{13}$  \\
2PIGG$\_$m1472                &    5              & 1        & -    &  2.446$\times 10^{13}$   \\
2PIGG$\_$m1486                &  23             & 4         & -    & 8.869$\times 10^{13}$ \\
2PIGG$\_$m1522$^*$                & 10              & 2         & -    & 3.189$\times 10^{13}$ \\
2PIGG$\_$m1523                &  5               & 1         & -    & 1.035$\times 10^{13}$  \\
2PIGG$\_$m1525                &   11            & 2         & -    &  1.381$\times 10^{13}$  \\
2PIGG$\_$m1532                &  15             & 1         & -    & 7.710$\times 10^{13}$   \\
2PIGG$\_$m1540                &  32             & 10      & 2    &  1.766$\times 10^{14}$ \\
2PIGG$\_$m1543               &   6              & 1         & -     & 8.449$\times 10^{12}$ \\
2PIGG$\_$m1572               &  19             & 1         & -     & 4.717$\times 10^{13}$  \\
2PIGG$\_$m1584                &   6              & 1         & -     & 8.936$\times 10^{12}$  \\
2PIGG$\_$m1597               &  16             & 4         & -     &  8.237$\times 10^{13}$   \\
2PIGG$\_$m1598               &    9             & 2         & 1    &  2.671$\times 10^{13}$  \\
2PIGG$\_$m1622               &  27             & 10       & -     &  1.261$\times 10^{14}$    \\
\hline \\ 
\enddata
 \tablecomments{The \emph{ZENS} groups for which candidate `extra galaxy members' were found in the SDSS DR7 spectroscopic sample, according to the criterion described in Appendix \ref{App:SDSSMatch}. For each groups we specify the number of original 2PIGG members (column 2), the number of SDSS galaxies which are not in 2dFGRS (column 3), and, amongst these, the number of galaxies whose magnitudes lie below the 2dFGRS selection limits (columns 4). In column 5 we list the fiducial group masses  based on the extrapolation of the luminosity function (in units of $\Msol$), as sampled by the 2PIGG galaxies (see Section \ref{sec:GroupMasses}). \\ $*$ The entry for m1522 refers to the group center fixed on the nominal most massive galaxy, which was rejected to be the central by our test described in Section \ref{sec:centralsDef}; centering the search for extra galaxies on the newly-assigned central galaxy results in no extra galaxies potentially associated with this group according to the search criteria in the SDSS catalogue that are detailed in Appendix \ref{App:SDSSMatch}.}
\end{deluxetable*}
 
We again asked whether these `missed' galaxies could be the true central galaxies in groups for which we failed to find a self-consistent solution in Section \ref{sec:centralsDef}. In the majority of  cases, such galaxies that lie beyond the WFI pointings are small satellites in the outskirts of the groups ($R>0.5R_{200}$).
In fact, 85$\%$ of the galaxies that fall outside our WFI pointings are substantially  more than a factor of four less massive than the most massive group member;  about 60$\%$ of this `missed' galaxies have masses below the mass completeness limit of  quenched galaxies in our study  ($10^{10}\Msol$). 
Only for three groups, 2PIGG-s1272,  2PIGG-s1665 and  2PIGG-n1377, our scheme for the definition of the group center identified one of the galaxies with no $B-$ and $I-$band WFI  imaging as a possible candidate central galaxy. From a statistical perspective, this is again a negligible contributor to the misidentification of central and satellite galaxies in our sample.

For these three groups we adopted as central the  `missed' galaxy which suitably satisfied all criteria to have such rank. However, given that the uncertainties on the masses for these `missed' galaxies are substantially larger than for the rest of the sample, we flag these groups in the \emph{ZENS}  catalog; this gives us the chance to  check whether any of our results  change when these groups are included/excluded from our analyses, and/or when we adopt as the central galaxy the galaxy which satisfies this criteria within the members with WFI  $B-$ and $I-$ observations.
 As expected, given that only three groups are involved, in none of our studies so far these groups affect  any of our  conclusions.

\subsection{Search in the 2dFGRS for potential missed  members of the \emph{ZENS}-2PIGG groups}\label{App:ExtraCandidatesProp}

As mentioned in Section \ref{sec:MembershipTest}, the search for 2dFGRS galaxies not included in a given \emph{ZENS}
 (i.e. 2PIGG) group but with magnitudes, coordinates and redshifts within ranges that could possibly make them
  members of this group (according to the criteria listed in Section \ref{sec:MembershipTest}),
   resulted in a total of 52 galaxies distributed over 24 of the \emph{ZENS} groups. 
Figure \ref{fig:Extra2dFGRScandidates} shows the spatial (left panels) and velocity (right panels) distributions for these possible candidate members, in relation to the galaxies which compose the \emph{ZENS} group extracted from the 2PIGG catalog. Note that often they  cluster both spatially and in velocities. While statistically their identification with independent groups is validated by the comparisons with mock catalogues, it is clear that, on a group-to-group basis, it is not possible to exclude that, at least some of these galaxies, may be missed members of the 2PIGG groups that we study in \emph{ZENS}. The velocity dispersions and masses of these groups would however  not change substantially if the potential extra candidate members were added to them, as these groups have already relatively high total masses, as shown in  Figure \ref{fig:MassGrpExtra} (see also Section \ref{sec:MembershipTest}).  Furthermore, none of these extra galaxies would qualify as being the central galaxies in the \emph{ZENS} groups of which they could potentially be extra group members. 
With respect to the satellite population, the 2dFGRS extra galaxies with masses above the  passive (`quenched') mass completeness limit of  $>10^{10}\Msol$ (22 in total), not included in the 2PIGG catalog, would add a contribution of only $4\%$ to the  \emph{ZENS} satellite sample with similar properties. A similar fraction of order 4\% applies for star-forming galaxies above  the mass completeness threshold of $10^{9.2}\Msol$.
We therefore do not consider these extra galaxies in any of our   analyses, for which we adopt the nominal galaxy membership in the \emph{ZENS} groups of the  2PIGG catalogue.

\begin{figure}[htbp]
\begin{center}
\includegraphics[width=185mm]{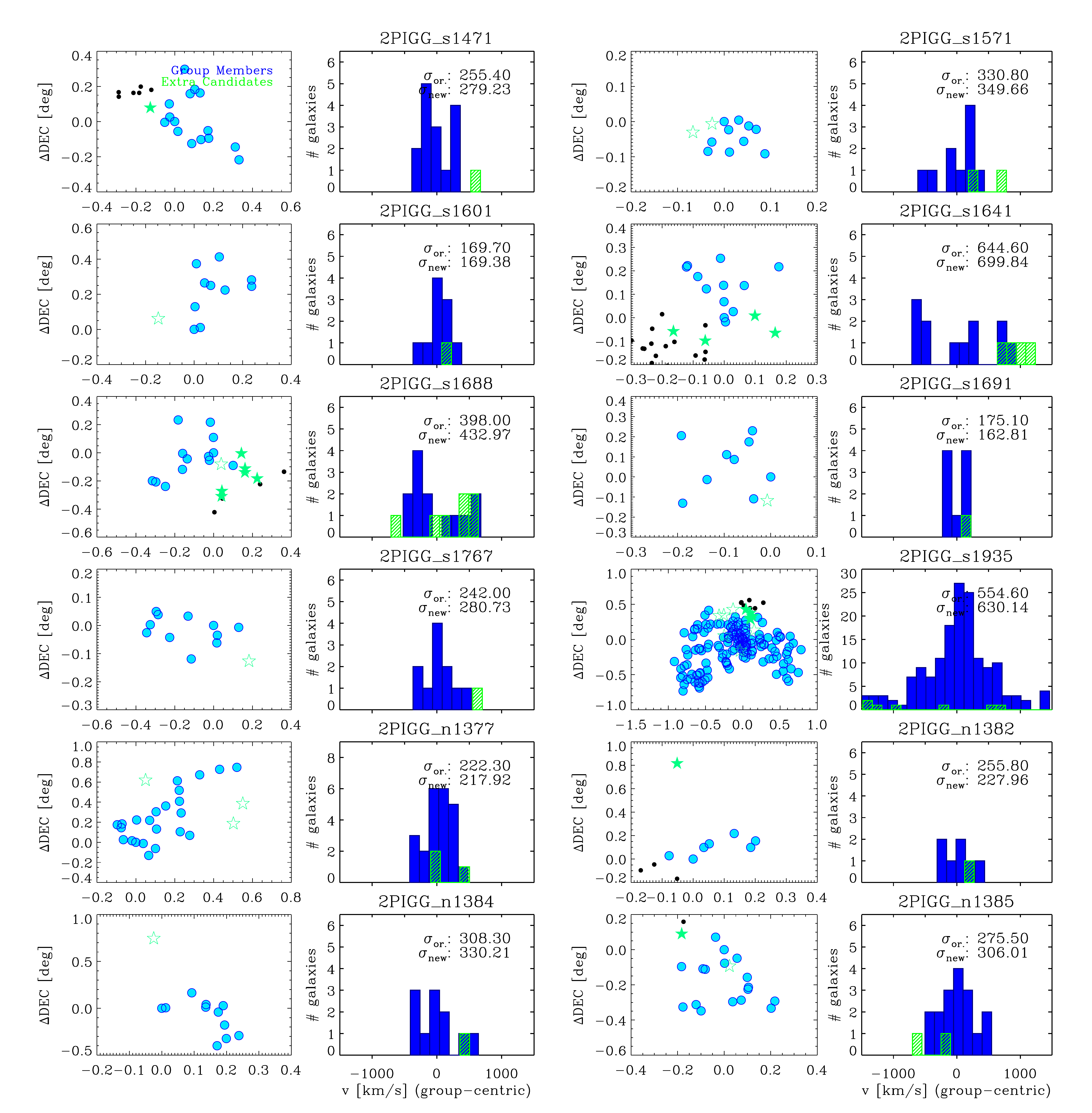}  
\end{center}
\caption{
{\it Left:} The projected spatial distribution, relative to the group central galaxy as defined in Section \ref{sec:centralsDef} (placed at the (0,0) position) of potential group member galaxies in the 2dFGRS catalog, which are not listed as members of the 2PIGG groups that we use in \emph{ZENS}. The nominal 2PIGG member galaxies are shown as filled circles, the potential extra group members as stars. Amongst these potential `extra' candidate members for a given 2PIGG group, we identify with empty stars galaxies which are not associated with any other 2PIGG group, and with filled stars galaxies which are associated with a different 2PIGG group. We also show as black dots the remaining galaxies members of these other 2PIGG groups to which the filled-star galaxies belong, although these dot-galaxies do not qualify to be potential extra members of our \emph{ZENS} groups, 
according to the definition discussed in Section \ref{sec:MembershipTest}. {\it Right:} The corresponding distribution of relative velocities of galaxies with respect to the mean redshift of the group. Solid histograms show the velocities for the original 2PIGG group members, and  dashed histograms give the relative velocities of the potential `extra' candidate members. The values of the velocity dispersion of the groups, computed before and after the inclusion of these potential extra members, are given in the top-right corners of the plots. These velocity dispersions are calculated with the gapper estimator as in \citet{Eke_et_al_2004}.\label{fig:Extra2dFGRScandidates}}
\end{figure}

\begin{figure}[htbp]
\begin{center}
\includegraphics[width=185mm]{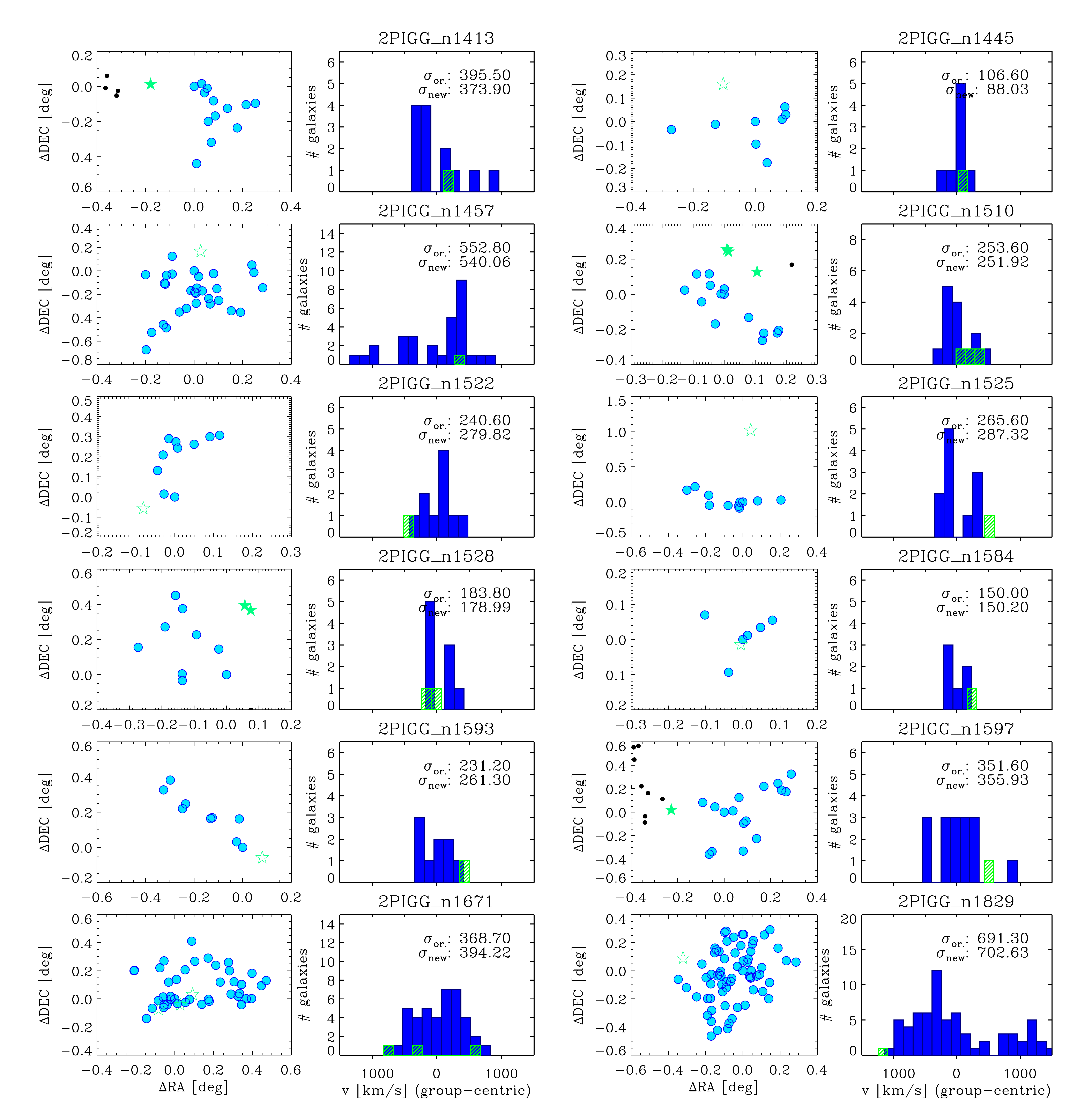}  
\end{center}
\caption{Continued.}
\end{figure}

\begin{figure}
\epsscale{1.1}
\begin{center}
\includegraphics[width=80mm]{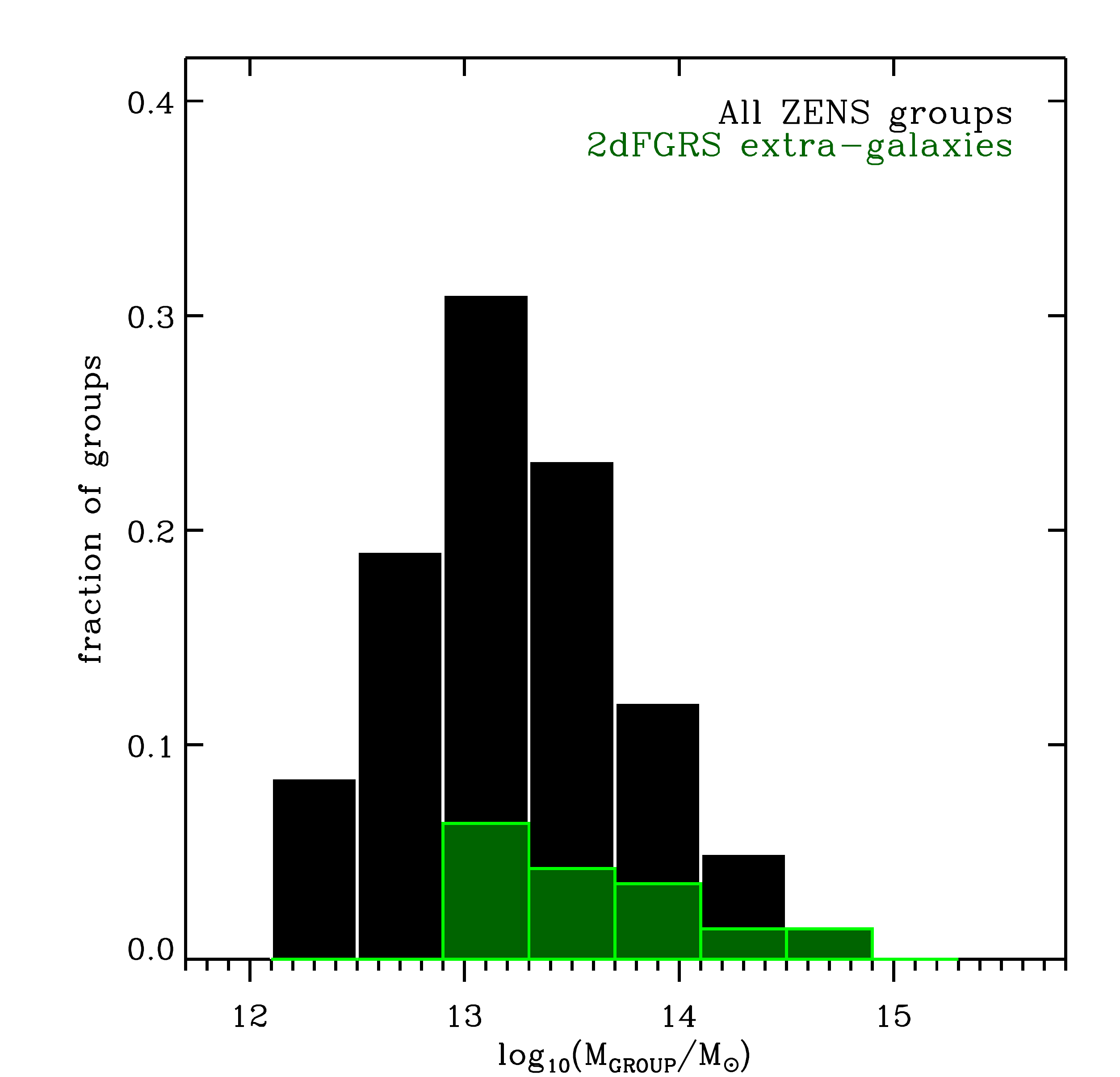}
\end{center}
\caption{\label{fig:MassGrpExtra}In green we show the distribution of fiducial group masses for the 
24 \emph{ZENS} groups for which we have found, in the 2dFGRS catalog, galaxies 
which are consistent with being additional group members (according to the definition given in Section \ref{sec:MembershipTest}). 
For comparison, the black histogram shows the distribution of fiducial group masses for the entire \emph{ZENS} sample. 
The green histogram is normalized to the total number of groups in the \emph{ZENS} sample.} 
\end{figure}

\section{Test on the Robustness of the Fiducial LSS Density Estimates}\label{App:densityTests}

\begin{figure*} [htbp]
\begin{center}
\includegraphics[width=60mm,angle=90]{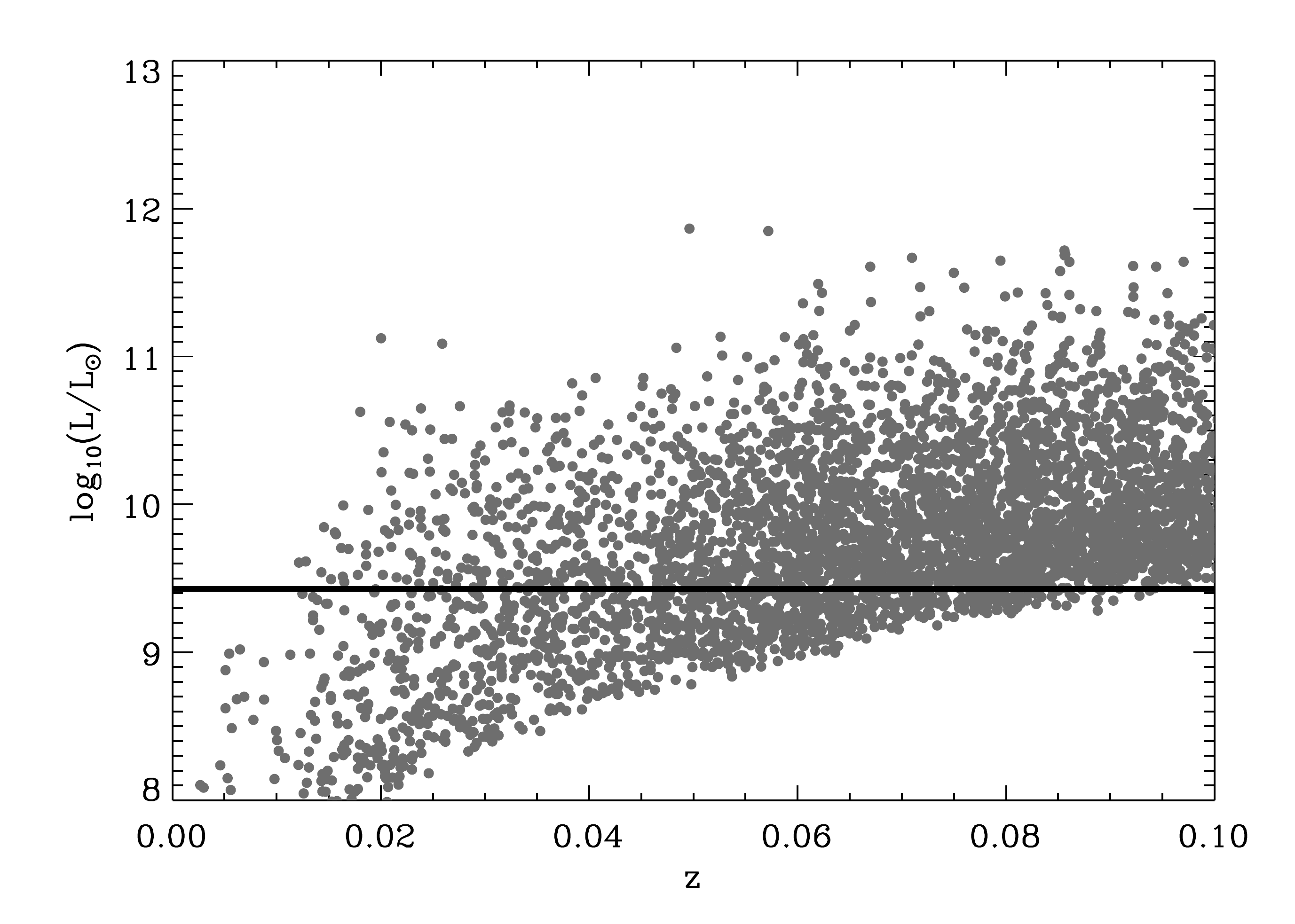}
\end{center}
\caption{\label{Fig:LSSGroupSelection}Luminosity of the 2PIGG groups and ungrouped galaxies 
 used to derive the $Nth-group-neighbor$  LSS density field (see Section \ref{sec:LSSEstimate}). 
  The solid line shows the minimum group luminosity considered in the computation, 
  i.e.,  $L=10^{9.43} \Lsol$; this corresponds to the total (i.e. integrated to zero) luminosity of a $b_j=19.1$
   individual galaxies at $z=0.07$.  Densities at each group location are calculated considering
    all other groups in the 2PIGG catalog, plus the remaining ungrouped  galaxies in the 2dFGRS, within a redshift range of $\Delta z=\pm0.1$ from the given group. Only one every ten points is plotted for clarity.}
\end{figure*}

As discussed in Section \ref{sec:LSSEstimate}, we adopted a Nth-nearest {\it group} neighbor 
algorithm to compute our fiducial LSS densities at the location of the \emph{ZENS} fields.
The  volume-limited sample of 2PIGG  \emph{groups} used in the construction of the density
 field and the imposed minimum luminosity are plotted in Figure \ref{Fig:LSSGroupSelection}.     
 
 \subsection{The role of  'ungrouped'  galaxies}
 
\begin{figure}[htbp]
\begin{center}
\includegraphics[width=100mm,angle=90]{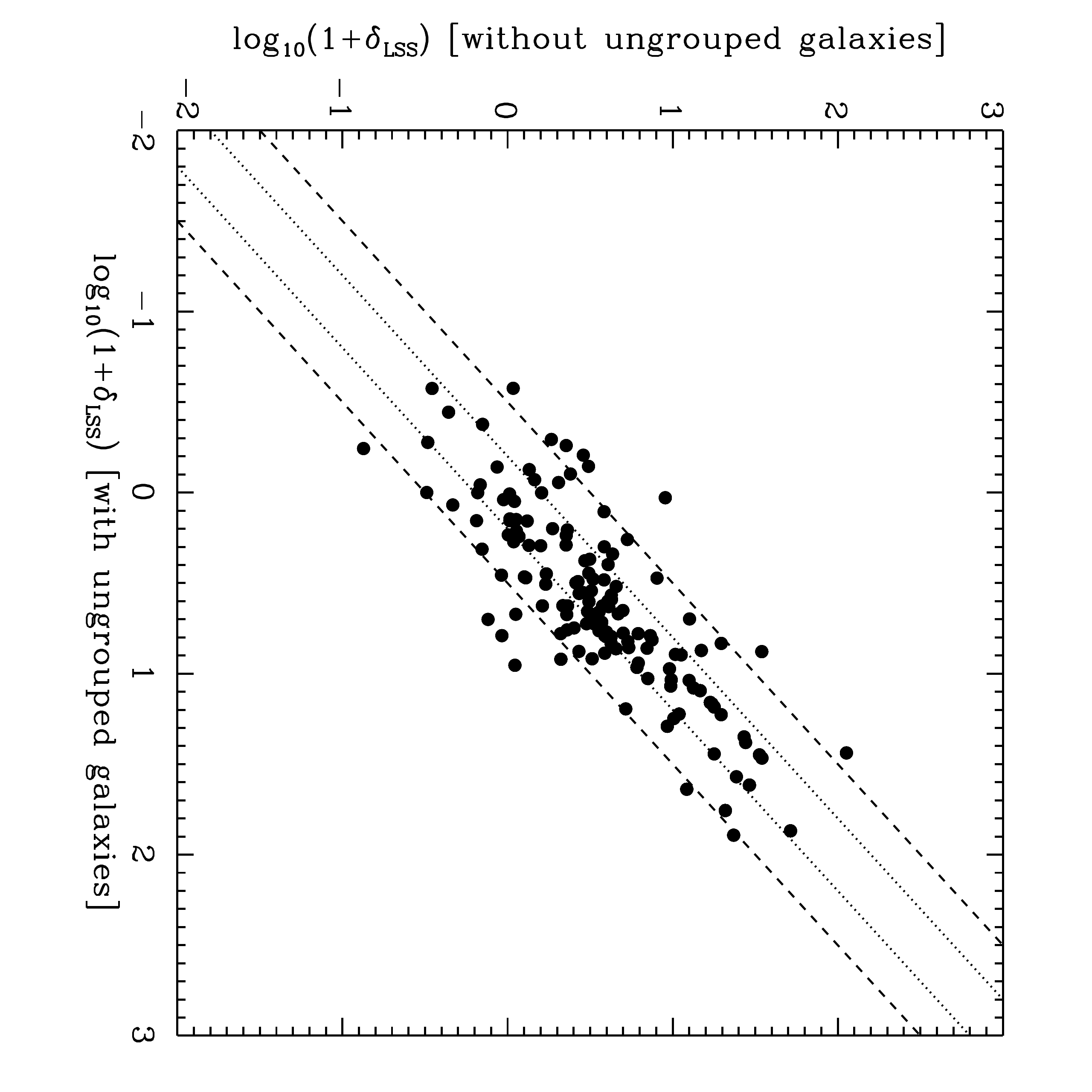}
\end{center}
\caption{\label{fig:NoIso}Comparison between the fiducial $Nth-group-neighbor$  LSS overdensities 
for the \emph{ZENS} groups, and those obtained excluding the ungrouped galaxies
 in the 2dFGRS. Dotted and dashed lines highlight differences of 0.2 dex and 0.5 dex, respectively.}
\end{figure}

As explained in the main text, we investigated whether the fiducial LSS density values for
the \emph{ZENS} groups are significantly affected by the addition or removal of  galaxies  in the 2dFGRS which were not identified as members of any group in the 2PIGG catalog. 
This is shown in Figure \ref{fig:NoIso}. There is a good correlation between 
the two measurements of LSS density: in only $\sim$10$\%$ of the cases the 
difference between the overdensities derived with and without the ungrouped galaxies 
is larger than 0.5 dex. 

We  note that, as evident from the left panel of Figure \ref{fig:densDist}, the exclusion of  the 'ungrouped' galaxies moves the peak of the $\delta_{LSS}$ distribution towards slightly lower values. This is a consequence of a $\sim40\%$ increase of the typical distance to the 5th-nearest-neighbor when the `ungrouped' galaxies are excluded. As also emphasized  in Section \ref{sec:LSSEstimate},  this  is  evidence that  the `ungrouped' galaxies may not be isolated systems in voids regions. For our purposes, the key point is that  including or excluding  these `ungrouped' galaxies does not alter significantly our LSS density measurements, and thus the trends with such density that we investigate in our study.

 \subsection{A comparison with standard Nth nearest galaxy neighbors density estimates}
    
Many studies in the past several years have adopted a Nth-nearest {\it galaxy} neighbor approach to derive 
an estimate for the LSS density field. In our case, we opted instead for the use of the groups as the 
density tracers, rather than the galaxies, to avoid the drawback of switching from a density within 
the groups, for groups with richness $>N$, to a density outside of the groups, for groups with 
richness $<N$. We highlight below this shortfall of the Nth-nearest galaxy neighbor density field, 
which we also computed (but never used in our analyses, for the reason above). 

Similarly to what is customarily done \citep[e.g.][]{Gomez_et_al_2003,
Balogh_et_al_2004, Baldry_et_al_2006,Kovac_et_al_2010b}, we computed the 
Nth-nearest {\it galaxy} field using a volume-limited sample of galaxies, in our case, 
with $M_{b_j}<-18.3-z$ in the Vega system. This brightness limit corresponds 
to the absolute magnitude of a galaxy having a $b_j=19.1$ at the maximum redshift 
of the \emph{ZENS} sample. This was chosen such as to have a uniform depth/completeness
 over the bulk of the \emph{ZENS} groups (see Figure \ref{fig:surveyCompletness}), 
 and yet to provide an adequate number of tracers. Neighbor galaxies were searched 
 within a velocity range of $\pm1000$ km s$^{-1}$ centered at the given galaxy redshift; 
 galaxies were weighted for spectroscopic incompleteness during the computation.

\begin{figure}[htbp]
\begin{center}
\includegraphics[width=80mm,angle=90]{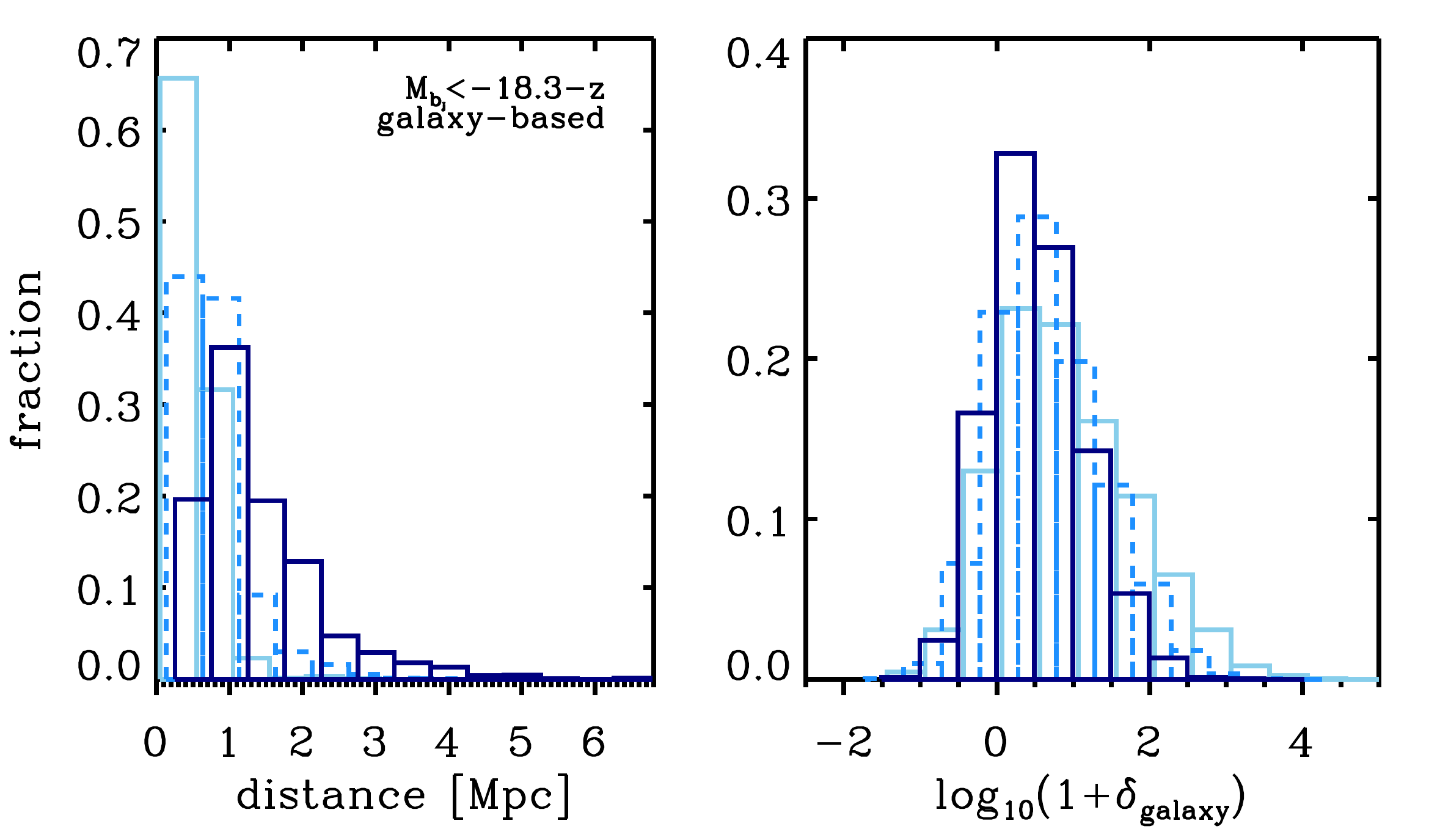}
\end{center}
\caption{\label{fig:galaxyDensities}The distribution of distances (left) and densities (right) resulting from a  
3rd (solid, light blue), 5th (dashed, blue) and 10th (solid, dark blue) nearest {\it galaxy} neighbor computation
 of the LSS field. We stress that we do not use these densities values in our analysis, since we prefer the adoption of our fiducial LSS density estimates that are based on using the groups Ð instead of  the group member  galaxies Ð as tracers of the LSS density field. }
\end{figure}  
 
The distribution of typical distances to the   $Nth$ nearest galaxy neighbor, 
 with N=3, 5 or 10, is shown in the left panel of Figure \ref{fig:galaxyDensities}. 
 The distances to the 3rd and 5th nearest galaxy neighbors peak at $\sim 0.5-1$Mpc, 
 a separation which is comparable to the typical radius of many of the \emph{ZENS} groups. 
 This is not surprising given that the \emph{ZENS} groups have at least 5 members; 
 at these distance scales, the $Nth$ nearest galaxy neighbor density estimates
  mostly probe the variation of density within the group themselves. 
  Also the 10th nearest galaxy neighbor densities, at high richness values, 
  will probe the environment inside massive groups rather than be a genuine proxy for
   the LSS density field. We note that, given the luminosity limit discussed above to
    ensure a uniform completeness and depth, the density field that we 
   calculate using the $Nth$ nearest galaxy neighbor density field uses a sub-sample
    of the 2PIGG galaxies (and hence of the galaxies used for the definition of the \emph{ZENS} groups). 
    Thus, also for the \emph{ZENS} groups with five members, 
    a partial contamination from interlopers  is in principle possible 
    when using the 5th nearest galaxy neighbor approach. From the number
    of galaxies in the \emph{ZENS} sample which are below the limit of $M_{b_j}<-18.3-z$, 
     we estimate this contamination to be about $20\%-25\%$.  
The corresponding distributions of overdensities for the   $Nth$ nearest galaxy neighbor 
     realizations  with N=3, 5 or 10, are shown in the right panel of Figure \ref{fig:galaxyDensities}.
      As commented in the main text, and as a consequence of galaxy-galaxy `clustering' 
      within the groups, a tail at high densities is observed, which is not  present in our fiducial $5th$ nearest {\it group} neighbor computation of the LSS density field.

\begin{figure}[htbp]
\begin{center}
\includegraphics[width=110mm,angle=90]{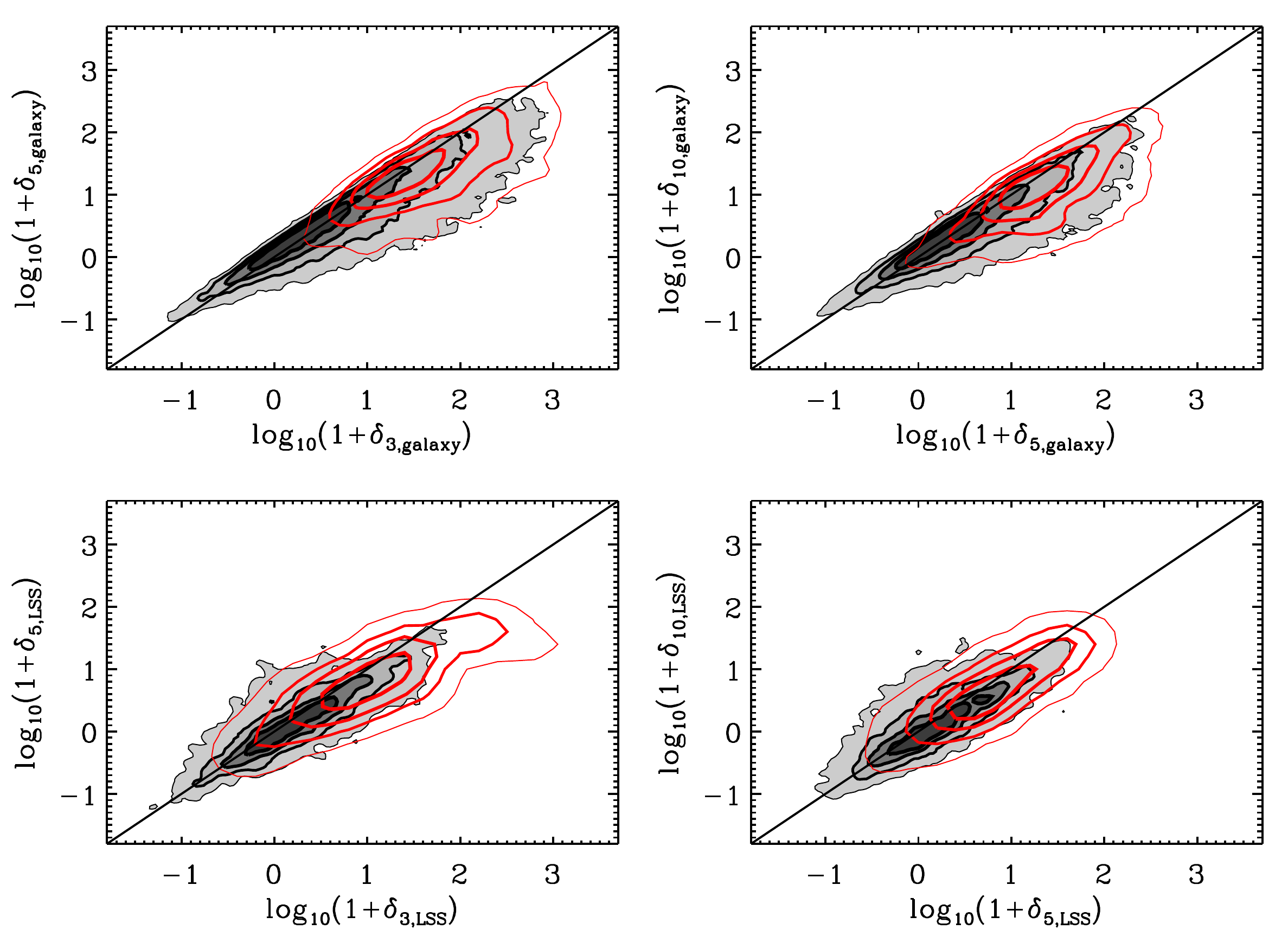}
\end{center}
\caption{\label{fig:densityComp}\emph{Top}: Comparison between the overdensities  calculated using the distance to the third ($\delta_3$), fifth ($\delta_5$)  or tenth ($\delta_{10}$) nearest {\it group} neighbor. The upper panels are for the density calculated using as tracers the volume limited sample of galaxies from the 2dFGRS with $M_{b_j}<-18.3-z$; the lower panels show the densities obtained using the groups in the 2PIGG catalogue with $L>10^{9.43}\Lsol$ as tracers. The grey areas are the values for the all galaxies or groups in the 2dFGRS with $0.035<z<0.075$; the red contours are for the \emph{ZENS} sample.}
\end{figure}

\subsection{The negligible impact of the choice for N when using the groups as density tracers}     
  
In Figure \ref{fig:densityComp}   we show the comparison between the
LSS (over)densities  calculated using the distances to  Nth-nearest  galaxy or group neighbors. In particular we compare the cases of N=3,5 and 10.  Upper and lower panels show respectively 
the densities calculated using as tracers the volume limited sample of $M_{b_j}<-18.3-z$ 
 {\it galaxies}  in the entire 2dFGRS catalogue, and the densities obtained using as tracers the {\it groups} in the 
2PIGG catalogue with $L>10^{9.43}\Lsol$. The grey areas are the values for the all galaxies 
or groups in the 2dFGRS with $0.035<z<0.075$; the red contours are for galaxies or groups in the \emph{ZENS} sample. 
The \emph{ZENS} galaxies and groups are slightly shifted towards higher density, 
reflecting the selection of our \emph{ZENS} sample. The figure shows (again) that the density field traced by the {\it galaxies} shows an extended tail below the identity 
 line at $\log(1+\delta)\sim2$, which is the signature that galaxy-based densities obtained with small apertures
 tend to be biased by local density peaks within group halos. The Figure also shows that
 our adopted LSS density estimates, that use the groups themselves as tracers of the  LSS density field, are less sensitive to the choice of  ``N" than estimates based on the Nth-nearest individual galaxies.    
   
\clearpage
\section{Median properties of centrals and satellites in relaxed and unrelaxed groups}\label{newtable}

\begin{deluxetable}{lccc}
\tablecaption{\label{tab:newparams}}\tablehead{
\colhead{Quantity}   & \colhead{Relaxed}  & \colhead{Unrelaxed}  }
\startdata
\cline{1-3}\\    
\multicolumn{3}{c}{$9.3<\log_{10}(M/\Msol)<10 $}  \\
\cline{1-3}\\    
\multicolumn{3}{c}{Satellites}  \\
 \cline{1-4}\\         
 $r_{1/2}$[kpc]                                                                                      &  3.28$_{-0.12}^{+0.17}$       &     3.29$_{-0.12}^{+0.15}$             \\                                                                                                             &                                                  &                                                        &                                      \\
 $\log_{10}(sSFR/yr^{-1})$                                                                 &  -10.25$_{-0.11}^{+0.07}$    &     -10.10$_{-0.08}^{+0.06}$          \\                                                                                                             &                                                   &                                                       &                                       \\
$\log_{10}(\Sigma_{SFR}/\Msol yr^{-1} kpc^{-2})$               & -2.59$_{-0.08}^{+0.07}$   & -2.39   $_{-0.06}^{+0.07}$        \\ 
                                                                                                              &                                                         &                                                           &                 \\
(B-I)  	                                                                                   &  0.99$_{-0.02}^{+0.02}$ &   $0.93_{-0.02}^{+0.03}$         \\  
                                                                                                              &                                                         &                                                           &                 \\
                                                                                                                                                                                       &                                                       &                                                             &                 \\
\cline{1-3}\\    
\cline{1-3}\\     
\multicolumn{3}{c}{$10<\log_{10}(M/\Msol)<10.7 $}  \\
\cline{1-3}\\    
\cline{1-3}\\    
\multicolumn{3}{c}{Satellites}  \\
 \cline{1-3}\\         
 $r_{1/2}$[kpc]                                                                                         &  3.27$_{-0.11}^{+0.31}$  &     3.41$_{-0.13}^{+0.31}$                 \\                                                                                                                 &                                            &                                                        &                     \\
 $\log_{10}(sSFR/yr^{-1})$                                                                     &  -11.38$_{-0.26}^{+0.13}$  &     -11.01$_{-0.29}^{+0.12}$         \\                                                                                                                   &                                                &                                                      &                   \\               
 $\log_{10}(\Sigma_{SFR}/\Msol yr^{-1} kpc^{-2})$                            &  -3.00$_{-0.22}^{+0.11}$  &  -2.63   $_{-0.26}^{+0.12}$           \\                                                                                                                  &                                              &                                                       &                    \\                                                                           
(B-I)   	                                                                                                  &  1.31$_{-0.02}^{+0.01}$    &   $1.30_{-0.03}^{+0.01}$              \\                                                                                                                &                                                &                                                      &                  \\
  \cline{1-3}\\    
  \multicolumn{3}{c}{Centrals}  \\
  \cline{1-3}\\    
$r_{1/2}$[kpc]                                                                                   &  5.30$_{-0.32}^{+0.40}$  &    3.77$_{-0.25}^{+0.69}$          \\
                                                                                                                    &                                                       &                                                             &                            \\
 $\log_{10}(sSFR/yr^{-1})$                                                                        &  -10.92$_{-0.83}^{+0.35}$    &     -11.16$_{-0.23}^{+0.32}$                       \\                                                                                                                     &                                                  &                                                                &                            \\             
 $\log_{10}(\Sigma_{SFR}/\Msol yr^{-1} kpc^{-2})$                               &  -2.70$_{-0.59}^{+0.38}$      &  -2.66   $_{-0.14}^{+0.20}$                          \\                                                                                                                     &                                                  &                                                                 &                            \\                                                                             
(B-I)   	                                                                                                   &  1.24$_{-0.03}^{+0.04}$        &   $1.28_{-0.07}^{+0.03}$                              \\                                                                                                                    &                                                   \enddata
\tablecomments{The table  lists the median values derived from the histograms of Figure \ref{fig:Fig14}, for central and satellite galaxies in two bins of stellar mass, for relaxed and unrelaxed groups with $M_{GROUP} < 10^{13.5} M_\odot$.
}
\end{deluxetable}

\newpage

\section{The readme file of the enclosed \emph{ZENS} catalog}\label{App:readmecat}
 
We finally list  below,  for each galaxy in the sample, the structural and photometric measurements presented in Paper II and III, and the environmental diagnostics discussed in this paper. Table \ref{tab:ZENScatalog}  matches the $readme$ file that accompanies  the \emph{ZENS} catalogue which we publish in this paper.

\LongTables 
\begin{longtable*}{lccll} 
 \tablewidth{0pt}
\tabletypesize{\scriptsize}
\tablecaption{The readme file for the ZENS catalogue of structural, photometric and environmental properties for the sample galaxies \label{tab:ZENScatalog}}
\tablehead{
\colhead{Column} & \colhead{Format} & \colhead{Units}  & \colhead{Label}  & \colhead{Comments} 
 }
\startdata
\hline \\ 
1 & a11 & -- & GroupID &	Group identification (1)	\\
2 & a10 & -- & 2dFID &	Galaxy 2dFGRS identification	\\
3 & a17 & -- & ZENSID &	Galaxy identification in ZENS	\\
4 & f7.5 & -- & groupz &	Group redshift	\\
5 & f11.6 & deg & RAdeg &	Right Ascension  in decimal degrees (J2000)	\\
6 & f11.6 & deg & DEdeg &	Declination in decimal degrees (J2000)	\\
7 & f7.5 & -- & galz &	Galaxy 2dFGRS heliocentric redshift	\\
8 & f5.3 & -- & wComp &	Galaxy completeness weight (2)	\\
9 & f5.3 & Mpc & RmsRad &	Group r.m.s. radius from 2PIGG catalog	\\
10 & f5.3 & Mpc & R200 &	Group $R_{200}$ radius (3)	\\
11 & f6.1 & km/s & Sigma &	Group velocity dispersion from 2PIGG catalog	\\
12 & e10.4 & \Lsol & Lgroup &	Group luminosity as in Eke et al. 2004	\\
13 & i3 & -- & Nmemb &	Group richness from 2PIGG catalog	\\
14 & e10.4 & \Msol & Mgroup &	Group mass $M_{GROUP}$ \\
15 & f6.3 & -- & DeltaLSS &	Large scale overdensity  $\delta_{LSS}$ (4)	\\
16 & i1 & -- & qLSS &	Quartile large scale overdensity (5)	\\
17 & i1 & -- & relaxFlag &	Group dynamical status (6)	\\
18 & i1 & -- & cenFlag &	Flag identifying central galaxy and group center (7)	\\
19 & f6.3 & R$_{200}$ & dR200 &	Galaxy distance from group center in units of R200	\\
20 & e11.4 & \Msol & MassBest &	Galaxy mass from ZEBRA+ best-fit template (8)	\\
21 & e11.4 & \Msol & erMassBest &	Lower limit on MassBest (9)	\\
22 & e11.4 & \Msol & ErMassBest &	Upper limit on MassBest (9)	\\
23 & e11.4 & \Msol & MassMedian &	ZEBRA+ median likelihood galaxy mass	\\
24 & e11.4 & \Msol & MassP16 &	The 16th percent. of  galaxy stellar mass likelihood distribution	\\
25 & e11.4 & \Msol & MassP84 &	The 84th percent. of  galaxy stellar mass likelihood distribution	\\
26 & e11.4 & \Msol & maxLMass &	ZEBRA+ maximum likelihood galaxy mass	\\
27 & i3 & -- & MType &	Morphological type (10)	\\
28 & i3 & -- & MergerFlag &	Merger flag (11)	\\
29 & f6.2 & -- & nI &	Galaxy GIM2D I-band raw S\'ersic index from single S\'ersic fit	\\
30 & f6.2 & -- & ernI &	GIM2D formal 99$\%$ confidence lower error on nI	\\
31 & f6.2 & -- & ErnI &	GIM2D formal 99$\%$ confidence upper error on nI	\\
32 & f6.2 & -- & nB &	Galaxy GIM2D B-band raw S\'ersic index from single S\'ersic fit		\\
33 & f6.2 & -- & ernB &	GIM2D formal 99$\%$ confidence lower error on nB	\\
34 & f6.2 & -- & ErnB &	GIM2D formal 99$\%$ confidence upper error on nB	\\
35 & f6.2 & -- & nIcorr &	Galaxy GIM2D I-band corrected S\'ersic index  from single S\'ersic fit (12)	\\
36 & f6.2 & -- & nBcorr &	Galaxy GIM2D B-band corrected S\'ersic index  from single S\'ersic fit (12)	\\
37 & f7.3 & kpc & gReI &	Galaxy GIM2D I-band raw half-light radius from single S\'ersic fit; semi-major axis 		\\
38 & f7.3 & kpc & ergReI &	GIM2D formal 99$\%$ confidence lower error on gReI	\\
39 & f7.3 & kpc & ErgReI &	GIM2D formal 99$\%$ confidence upper error on gReI	\\
40 & f7.3 & kpc & Delta-gReI &	Single vs. double component scatter on I-band half-light semi-major axis (12)	\\
41 & f7.3 & kpc & gReB &	Galaxy GIM2D B-band raw half-light radius  from single S\'ersic fit; semi-major axis\\
42 & f7.3 & kpc & ergReB &	GIM2D formal 99$\%$ confidence lower error on gReB	\\
43 & f7.3 & kpc & ErgReB &	GIM2D formal 99$\%$ confidence upper error on gReB	\\
44 & f7.3 & kpc & Delta-gReB &	Single vs. double component scatter on B-band half-light semi-major axis (12)	\\
45 & f7.3 & kpc & gReIcorr &	Galaxy GIM2D I-band corrected half-light radius  from single S\'ersic fit; semi-major axis (11)	\\
46 & f7.3 & kpc & gReBcorr &	Galaxy GIM2D B-band corrected half-light radius  from single S\'ersic fit; semi-major axis (11)	\\
47 & f7.3 & -- & gEllipI &	Galaxy GIM2D I-band raw ellipticity  from single S\'ersic fit		\\
48 & f7.3 & -- & ergEllipI &	GIM2D formal 99$\%$ confidence lower error on gEllipI	\\
49 & f7.3 & -- & ErgEllipI &	GIM2D formal 99$\%$ confidence upper error on gEllipI	\\
50 & f7.3 & -- & gEllipB &	Galaxy GIM2D B-band raw ellipticity  from single S\'ersic fit		\\
51 & f7.3 & -- & ergEllipB &	GIM2D formal 99$\%$ confidence lower error on gEllipB	\\
52 & f7.3 & -- & ErgEllipB &	GIM2D formal 99$\%$ confidence upper error on gEllipB	\\
53 & f7.3 & -- & gEllipIcorr &	Galaxy GIM2D I-band corrected ellipticity  from single S\'ersic fit (12)	\\
54 & f7.3 & -- & gEllipBcorr &	Galaxy GIM2D B-band corrected ellipticity  from single S\'ersic fit (12)	\\
55 & f7.3 & kpc & diskhI-tot &	Galaxy GIM2D I-band disk scale-length from pure exponential fit; semi-major axis (14)	\\
56 & f7.3 & kpc & erdiskhI-tot &	GIM2D formal 99$\%$ confidence lower error on diskhI-tot	\\
57 & f7.3 & kpc & ErdiskhI-tot &	GIM2D formal 99$\%$ confidence upper error on diskhI-tot	\\
58 & f7.3 & kpc & diskhB-tot &	Galaxy GIM2D B-band disk scale-length from pure exponential fit; semi-major axis (14)	\\
59 & f7.3 & kpc & erdiskhB-tot &	GIM2D formal 99$\%$ confidence lower error on diskhB-tot	\\
60 & f7.3 & kpc & ErdiskhB-tot &	GIM2D formal 99$\%$ confidence upper error on diskhB-tot	\\
61 & i3 & -- & tDecompI &	GIM2D vs.\ Galfit Flag for I-band bulge+disk decomposition (15)	\\
62 & i3 & -- & tDecompB &	GIM2D vs.\ Galfit Flag for B-band  bulge+disk decomposition (15)	\\
63 & f8.4 & -- & BTI &	I-band bulge-to-total ratio from bulge+disk decomposition (16)	\\
64 & f8.4 & -- & erBTI &	Formal 99$\%$ confidence lower error on BTI	\\ 
65 & f8.4 & -- & ErBTI &	Formal 99$\%$ confidence upper error on BTI	\\
66 & f8.4 & -- & BTB &	B-band bulge-to-total ratio  from bulge+disk decomposition (16)	\\
67 & f8.4 & -- & erBTB &	Formal 99$\%$ confidence lower error on BTB	\\
68 & f8.4 & -- & ErBTB &	Formal 99$\%$ confidence upper error on BTB	\\
69 & f7.3 & kpc & bulgeRe-I &	I-band bulge half-light radius from bulge+disk decomposition; semi-major axis (16)	\\
70 & f7.3 & kpc & erbulgeRe-I &	Formal 99$\%$ confidence lower error on bulgeRe-I	\\
71 & f7.3 & kpc & ErbulgeRe-I &	Formal 99$\%$ confidence upper error on bulgeRe-I	\\
72 & f7.3 & kpc & bulgeRe-B &	B-band bulge half-light radius  from bulge+disk decomposition; semi-major axis (16)	\\
73 & f7.3 & kpc & erbulgeRe-B &	Formal 99$\%$ confidence lower error on bulgeRe-B	\\
74 & f7.3 & kpc & ErbulgeRe-B &	Formal 99$\%$ confidence upper error on bulgeRe-B	\\
75 & f7.3 & kpc & diskh-I &	I-band disk scale-length  from bulge+disk decomposition; semi-major axis (16)	\\
76 & f7.3 & kpc & erdiskh-I &	Formal 99$\%$ confidence lower error on diskh-I	\\
77 & f7.3 & kpc & Erdiskh-I &	Formal 99$\%$ confidence upper error on diskh-I	\\
78 & f7.3 & kpc & diskh-B &	B-band disk scale-length from bulge+disk decomposition; semi-major axis (16)	\\
79 & f7.3 & kpc & erdiskh-B &	Formal 99$\%$ confidence lower error on diskh-B	\\
80 & f7.3 & kpc & Erdiskh-B &	Formal 99$\%$ confidence upper error on diskh-B	\\
81 & f7.3 & -- & nBulge-I &	I-band bulge Sersic index from bulge+disk decomposition (16)	\\
82 & f7.3 & -- & ernBulge-I &	Formal 99$\%$ confidence lower error on nBulge-I	\\
83 & f7.3 & -- & ErnBulge-I &	Formal 99$\%$ confidence upper error on nBulge-I	\\
84 & f7.3 & -- & nBulge-B &	B-band bulge Sersic index from bulge+disk decomposition  (16)	\\
85 & f7.3 & -- & ernBulge-B &	Formal 99$\%$ confidence lower error on nBulge-B	\\
86 & f7.3 & -- & ErnBulge-B &	Formal 99$\%$ confidence upper error on nBulge-B	\\
87 & f7.3 & -- & ellBulge-I &	I-band bulge ellipticity from bulge+disk decomposition (16)	\\
88 & f7.3 & -- & erellBulge-I &	Formal 99$\%$ confidence lower error on ellBulge-I	\\
89 & f7.3 & -- & ErellBulge-I &	Formal 99$\%$ confidence upper error on ellBulge-I	\\
90 & f7.3 & -- & ellBulge-B &	B-band bulge ellipticity from bulge+disk decomposition (16)	\\
91 & f7.3 & -- & erellBulge-B &	Formal 99$\%$ confidence lower error on ellBulge-B	\\
92 & f7.3 & -- & ErellBulge-B &	Formal 99$\%$ confidence upper error on ellBulge-B	\\
93 & f7.3 & deg & diskInc-I &	I-band disk inclination from bulge+disk decomposition (16)	\\
94 & f7.3 & deg & erdiskInc-I &	Formal 99$\%$ confidence lower error on diskInc-I	\\
95 & f7.3 & deg & ErdiskInc-I &	Formal 99$\%$ confidence upper error on diskInc-I	\\
96 & f7.3 & deg & diskInc-B &	B-band disk inclination from bulge+disk decomposition (16)	\\
97 & f7.3 & deg & erdiskInc-B &	Formal 99$\%$ confidence lower error on diskInc-B	\\
98 & f7.3 & deg & ErdiskInc-B &	Formal 99$\%$ confidence upper error on diskInc-B	\\
99 & f7.3 & kpc & ReI-decomp &	I-band global galaxy half-light radius from bulge+disk decomposition; semi-major axis (17) \\
100 & f7.3 & kpc & ReB-decomp &	B-band global galaxy half-light radius from bulge+disk decomposition; semi-major axis (17) \\
101 &f7.3 & kpc & zReI &	ZEST+ I-band raw half-light radius; semi-major axis	\\
102 &f7.3 & kpc & zReB &	ZEST+ B-band raw half-light radius;  semi-major axis	\\
103 &f7.3 & kpc & zReIcorr &	ZEST+ I-band corrected half-light  radius; semi-major axis (12)	\\
104 &f7.3 & kpc & zReBcorr &	ZEST+ B-band corrected half-light  radius; semi-major axis (12)	\\
105 &f7.3 & -- & zEllipI &	SEXTRACTOR/ZEST+ I-band raw ellipticity	\\
106 &f7.3 & -- & zEllipB &	SEXTRACTOR/ZEST+ B-band raw ellipticity	\\
107 &f7.3 & -- & zEllipIcorr &	SEXTRACTOR/ZEST+ I-band corrected ellipticity (12)	\\
108 &f7.3 & -- & zEllipBcorr &	SEXTRACTOR/ZEST+ B-band corrected ellipticity (12)	\\
109 &f7.3 & -- & CI &	ZEST+ I-band raw Concentration index	\\
110 &f7.3 & -- & CB &	ZEST+ B-band raw Concentration index	\\
111 &f7.3 & -- & CIcorr &	ZEST+ I-band corrected Concentration index (18)	\\
112 &f7.3 & -- & CBcorr &	ZEST+ B-band corrected Concentration index (18)	\\
113 &f7.3 & -- & GiniI &	ZEST+ I-band raw Gini index	\\
114 &f7.3 & -- & GiniB &	ZEST+ B-band raw Gini index	\\
115 &f7.3 & -- & GiniIcorr &	ZEST+ I-band corrected Gini index (18)	\\
116 &f7.3 & -- & GiniBcorr &	ZEST+ B-band corrected Gini index (18)	\\
117 &f7.3 & -- & M20I &	ZEST+ I-band raw $M_{20}$ index	\\
118 &f7.3 & -- & M20B &	ZEST+ B-band raw $M_{20}$ index	\\
119 &f7.3 & -- & M20Icorr &	ZEST+ I-band corrected $M_{20}$ index (18)	\\
120 &f7.3 & -- & M20Bcorr &	ZEST+ B-band corrected $M_{20}$ index (18)	\\
121 &f8.4 & -- & AsymI &	ZEST+ I-band raw asymmetry index	\\
122 &f8.4 & -- & AsymB &	ZEST+ B-band raw asymmetry index	\\
123 &f8.4 & -- & SI &	ZEST+ I-band raw smoothness index	\\
124 &f8.4 & -- & SB &	ZEST+ B-band raw smoothness index	\\
125 & f7.3 & kpc & RpI &	SEXTRACTOR I-Band petrosian radius; semi-major axis (19)	\\
126 & f7.3 & kpc & RpB &	SEXTRACTOR B-Band petrosian radius; semi-major axis (19)	\\
127 & f7.3 & kpc & RkI &	SEXTRACTOR I-band Kron aperture; semi-major axis (20)	\\
128 & f7.3 & kpc & RkB &	SEXTRACTOR B-band Kron aperture; semi-major axis (20)	\\
129 & i3 & -- & BarType &	Bar Flag (21)	\\
130 & f7.3 & kpc & aBar &	Bar semi-major axis	\\
131 &f6.2 & -- & fBar &	Bar strength	\\
132 &i3 & -- & SpType &	Spectral type (22)	\\
133 &i3 & -- & Dust-SBFlag &	Dusty, star-forming or post starburst flag (23)	\\
134 &f8.4 & \Msol/yr & SFRBest &	Star-formation rate from ZEBRA+ best-fit template (24)	\\
135 &f8.4 & \Msol/yr & erSFRBest &	Lower limit on SFRBest (9)	\\
136 &f8.4 & \Msol/yr & ErSFRBest &	Upper limit on SFRBest (9)	\\
137 &f8.4 & \Msol/yr & SFRMedian &	ZEBRA+ median likelihood SFR	\\
138 &f8.4 & \Msol/yr & SFRP16 &	The 16th percentile of the ZEBRA+ SFR likelihood distribution	\\
139 &f8.4 & \Msol/yr & SFRP84 &	The 84th percentile of the ZEBRA+ SFR likelihood distribution	\\
140 &f8.4 & \Msol/yr & maxLSFR &	ZEBRA+ maximum likelihood SFR	\\
141 & e11.4 & yr$^{-1}$ & sSFRBest &	Specific star-formation rate from ZEBRA+ best-fit template (24)	\\
142 & e11.4 & yr$^{-1}$ & ersSFRBest &	Lower limit on sSFRBest (9)	\\
143 & e11.4 & yr$^{-1}$ & ErsSFRBest &	Upper limit on sSFRBest (9)	\\
144 & e11.4 & yr$^{-1}$ & sSFRMedian &	ZEBRA+ median likelihood specific star-formation rate	\\
145 & e11.4 & yr$^{-1}$ & sSFRP16 &	The 16th percentile of the ZEBRA+ sSFR likelihood distribution	\\
146 & e11.4 & yr$^{-1}$ & sSFRP84 &	The 84th percentile of the ZEBRA+ sSFR likelihood distribution	\\
147 & e11.4 & yr$^{-1}$ & maxLsSFR &	ZEBRA+ maximum likelihood sSFR	\\
148 & f8.4 & mag & kcorr-B &	ZEBRA+ B-band k-correction	\\
149 & f8.4 & mag & kcorr-I &	ZEBRA+ I-band k-correction	\\
150 & f8.3 & mag & oBmag &	Galaxy Petrosian apparent B-band magnitude; observer-frame (25)	\\
151 & f8.3 & mag & rBmag &	Galaxy Petrosian apparent B-band magnitude; rest-frame (25)	\\
152 & f8.3 & mag & errBmag &	Error on rBmag	\\
153 & f8.3 & mag & BMag &	Absolute galaxy Petrosian B-band magnitude	\\
154 & f8.3 & mag & oImag &	Galaxy Petrosian apparent I-band magnitude; observer-frame (25)	\\
155 & f8.3 & mag & rImag &	Galaxy Petrosian apparent I-band magnitude; rest-frame (25)	\\
156 & f8.3 & mag & errImag &	Error on rImag	\\
157 & f8.3 & mag & IMag &	Absolute galaxy Petrosian I-band magnitude	\\
158 & f8.3 & mag & BmagKron &	Galaxy Kron apparent B-band magnitude; rest-frame (26)	\\
159 & f8.3 & mag & erBmagKron &	Error on BmagKron	\\
160 & f8.3 & mag & ImagKron &	Galaxy Kron apparent I-band magnitude; rest-frame (26)	\\
161 & f8.3 & mag & erImagKron &	Error on ImagKron	\\
162 & f8.3 & mag & B-I &	Galaxy Petrosian rest-frame (B-I) color	\\
163 & f8.3 & mag & bjmag &	Galaxy 2dFGRS $b_{j}$ Vega magnitude (27)	\\
164 & f8.3 & mag & rFmag &	Galaxy 2dFGRS/SCOS r$_{F}$ Vega magnitude (27)	\\
165 & f8.3 & mag & FUVmag &	Galaxy apparent FUV magnitude; rest-frame (25,28)	\\
166 & f8.3 & mag & erFUVmag &	Error on FUVmag	\\
167 & f8.3 & mag & NUVmag &	Galaxy apparent NUV magnitude; rest-frame (25,28)	\\
168 & f8.3 & mag & erNUVmag &	Error on NUVmag	\\
169 & f8.3 & mag & NUV-I &	NUV-I color; rest-frame	\\
170 & f8.3 & mag & NUV-B &	NUV-B color; rest-frame	\\
171 & f8.3 & mag & FUV-NUV &	FUV-NUV color; rest-frame	\\
172 & f8.3 & mag & umag &	Galaxy  apparent SDSS u magnitude; rest-frame (25,28)	\\
173 & f8.3 & mag & erumag &	Error on umag	\\
174 & f8.3 & mag & gmag &	Galaxy apparent SDSS g magnitude; rest-frame (25,28)	\\
175 & f8.3 & mag & ergmag &	Error on gmag	\\
176 & f8.3 & mag & rmag &	Galaxy apparent SDSS r-magnitude; rest-frame (25,28)	\\
177 & f8.3 & mag & errmag &	Error on rmag	\\
178 & f8.3 & mag & imag &	Galaxy apparent SDSS i-magnitude; rest-frame (25,28)	\\
179 & f8.3 & mag & erimag &	Error on imag	\\
180 & f8.3 & mag & zmag &	Galaxy apparent SDSS z-magnitude; rest-frame (25,28)	\\
181 & f8.3 & mag & erzmag &	Error on zmag	\\
182 & f8.3 & mag & Jmag &	Galaxy  apparent 2MASS J-magnitude; rest-frame (25,28)	\\
183 & f8.3 & mag & erJmag &	Error on Jmag	\\
184 &  f8.3 & mag & Hmag &	Galaxy apparent 2MASS H-magnitude; rest-frame (25,28)	\\
185 & f8.3 & mag & erHmag &	Error on Hmag	\\
186 & f8.3 & mag & Kmag &	Galaxy apparent 2MASS K-magnitude; rest-frame (25,28)	\\
187 & f8.3 & mag & erKmag &	Error on Kmag	\\
188 & f7.3 & mag & BmagSer &	Total galaxy B-band magnitude from GIM2D single S\'ersic fit; rest-frame	\\
189 & f7.3 & mag & erBmagSer &	GIM2D formal 99$\%$ confidence lower error on BmagSer	\\
190 & f7.3 & mag & ErBmagSer &	GIM2D formal 99$\%$ confidence upper error on BmagSer	\\
191 & f7.3 & mag & ImagSer &	Total galaxy I-band magnitude from GIM2D  single S\'ersic fit; rest-frame	\\
192 & f7.3 & mag & erImagSer &	GIM2D formal 99$\%$ confidence lower error on ImagSer	\\
193 & f7.3 & mag & ErImagSer &	GIM2D formal 99$\%$ confidence upper error on ImagSer	\\
194 & f7.3 & mag & BmagSercorr &	Corrected total galaxy $B$ mag from single S\'ersic fit; rest-frame (12) \\
195 & f7.3 & mag & ImagSercorr &	Corrected total galaxy $I$  mag  from single S\'ersic fit; rest-frame (12)	\\
196 & f7.3 & mag &  BmagExp      &           Total galaxy B-band magnitude from GIM2D pure exponential fit; rest-frame  \\                                           
197 & f7.3 & mag &  erBmagExp    &          GIM2D formal 99\% confidence lower error on BmagExp                            \\                                               
198 & f7.3 & mag &  ErBmagExp     &         GIM2D formal 99\% confidence upper error on BmagExp                            \\                                               
199 & f7.3 & mag &  ImagExp          &       Total galaxy I-band magnitude from GIM2D  pure exponential fit; rest-frame \\                                            
200 & f7.3 & mag &  erImagExp    &          GIM2D formal 99\% confidence lower error on ImagExp                                  \\                                         
201 & f7.3 & mag &   ErImagExp   &           GIM2D formal 99\% confidence upper error on ImagExp                        \\                                                   
202 & f7.3 & mag & oBmagBulge &	Bulge B-band magnitude from bulge+disk decomposition; observer-frame	\\
203 & f7.3 & mag & eroBmagBulge &	Formal 99$\%$ confidence lower error on oBmagBulge (29)	\\
204 & f7.3 & mag & EroBmagBulge &	Formal 99$\%$ confidence upper error on oBmagBulge (29)	\\
205 & f7.3 & mag & oImagBulge &	Bulge I-band magnitude from bulge+disk decomposition; observer-frame	\\
206 & f7.3 & mag & eroImagBulge &	Formal 99$\%$ confidence lower error on oImagBulge (29)	\\
207 & f7.3 & mag & EroImagBulge &	Formal 99$\%$ confidence upper error on oImagBulge (29)	\\
208 & f7.3 & mag & rBmagBulge &	Bulge B-band magnitude from bulge+disk decomposition; rest-frame (30)	\\
209 & f7.3 & mag & rImagBulge &	Bulge I-band magnitude from bulge+disk decomposition; rest-frame (30)\\
210 & f7.3 & mag & oBmagDisk &	Disk B-band magnitude from bulge+disk decomposition; observer-frame	\\
211 & f7.3 & mag & eroBmagDisk &	Formal 99$\%$ confidence lower error on oBmagDisk (29)	\\
212 & f7.3 & mag & EroBmagDisk &	Formal 99$\%$ confidence upper error on oBmagDisk (29)	\\
213 & f7.3 & mag & oImagDisk &	Disk I-band magnitude from bulge+disk decomposition;  observer-frame	\\
214 & f7.3 & mag & eroBmagIulge &	Formal 99$\%$ confidence lower error on oImagDisk (29)	\\
215 & f7.3 & mag & EroImagDisk &	Formal 99$\%$ confidence upper error on oImagDisk (29)	\\
216 & f7.3 & mag & rBmagDisk &	Disk B-band magnitude from bulge+disk decomposition; rest-frame (30)	\\
217 & f7.3 & mag & rImagDisk &	Disk I-band magnitude from bulge+disk decomposition; rest-frame (30)	\\
218 & e11.4 & \Msol & BulgeMass  &	Bulge stellar mass (31)	\\
219 & e11.4 & \Msol & BulgeMassP16 &	The 16th percentile bulge mass (31)	\\
220 & e11.4 & \Msol & BulgeMassP84   &	The 84th percentile bulge mass (31)	\\
221 & e11.4 & \Msol & DiskMass   &	Disk stellar mass (31)	\\
222 & e11.4 & \Msol & DiskMassP16  &	The 16th percentile disk mass (31)	\\
223 & e11.4 & \Msol & DiskMassP84  &	The 84th percentile disk mass (31)	\\
224 & f7.3 & mag & d(B-I)/dlogrVor &	 The (B-I) color gradient from Voronoi tessellated color maps	\\
225 &  f7.3 & mag & erd(B-I)/dlogrVor&	Error on (B-I)gradVor	\\
226 &  f7.3 & mag & corr\_d(B-I)/dlogrVor &	Corrected (B-I) color gradient from Voronoi tessellated color maps (32)	\\
227 & f7.3 & mag & (B-I)ReVor &	The (B-I) color at half-light radius from Voronoi tessellated color maps; rest-frame	\\
228 & f7.3 & mag & er(B-I)ReVor &	Error on (B-I)ReVor (33)	\\
229 &  f7.3 & mag & corr(B-I)ReVor &	Corrected (B-I) color at $r_{1/2}$  from Voronoi  color maps; rest-frame (32)	\\
230 & f7.3 & mag & (B-I)rms &	Dispersion around best fit color profile from Voronoi  color maps	\\
231 & f7.3 & mag & d(B-I)/dlogrGim &	The (B-I) color gradient from GIM2D best fits	\\
232 & f7.3 & mag & erd(B-I)/dlogrGim &	Error on (B-I)gradGim	\\
233 & f7.3 & mag & (B-I)ReGim &	The (B-I) color at half-light radius from GIM2D best fits; rest-frame	\\
234 & f7.3 & mag & er(B-I)ReGim &	Error on (B-I)ReGim (34)	\\
235 & e11.4&  \Msol &    FlagBadSFR      &       Flag for  ZEBRA+ template limited to a star-forming  model (35) \\
236 & i3&  -- &    cFlag      &  Contamination flag (36)  \\
237&  f7.3  &   mag  &       BmagKroncorr       &    Corrected galaxy Kron B-band magnitude; rest-frame (18)\\
238 & f7.3  & mag    &       ImagKroncorr      &     Corrected galaxy Kron I-band magnitude; rest-frame (18) \\
239 & f6.3  & --    &   zCompl        &   2dFGRS redshift completeness at galaxy position (37) \\
240 & f6.3  & mag    &  dFmagLim  &  Extinction corrected magnitude limit of 2dFGRS at galaxy position (37)    \\  
241 & f6.3  & --    &      muVal     &  2dFGRS completeness $\mu$ factor at galaxy position (37)  \\
242 & i1  & --    &     PIGGCen      &   Original 2PIGG central galaxy flag (38) \\
243 & f7.3  & deg    &     diskInc-tot-I  & I-band disk inclination from pure exponential fit \\
244 & f7.3  & deg    &     erdiskInc-tot-I & Formal 99\% confidence lower error on diskInc-tot-I \\
245 & f7.3 &  deg    &     ErdiskInc-tot-I & Formal 99\% confidence upper error on diskInc-tot-I \\
246 & f7.3  & deg    &     diskInc-tot-B & B-band disk inclination from pure exponential fit \\
247 & f7.3  & deg    &     erdiskInc-tot-B & Formal 99\% confidence lower error on diskInc-tot-B\\
248 & f7.3  & deg    &     ErdiskInc-tot-B &Formal 99\% confidence upper error on diskInc-tot-B\\
249 & i3 & --  & GalfitDev & Flag indicating deviations between GIM2D and GALFIT parameters (39)\\
250 & f5.1 &  -- &          SpecClass     &      Flag indicating type of galaxy spectrum (40) \\                                                                           

\end{longtable*}
Note (1): Projected sizes are converted into physical units assuming
          the following cosmological parameters: h=0.7, $\Omega_m$=0.3, $\Omega_\Lambda$=0.7.
          Parameters which are not available are listed as:
          -99 entries for definite positive parameters, and  +99 for definite negative parameters.
          For merging galaxy pairs we list parameters for both  the primary 
          and secondary galaxy, when available. 
          Unless specified differently, all magnitudes and colors are in the AB system.
           For galaxies outside the WFI FOV all galactic parameters, except the stellar masses,
          are given as -99 entries.\\
Note (2): Galaxy weight accounting for 2dFGRS redshift incompleteness calculated
          as described in Section \ref{App:2dFGRS_limits}.\\
Note (3): $R_{200}$ radius derived from the group mass (see text)\\
Note (4): Defined as $\log(1+\delta_{LSS})$ where $\delta_{LSS}$
          is calculated to the fifth nearest 2PIGG {\it group} (see text).\\
Note (5): 1=group located in first quartile of the distribution of $\log(1+\delta_{LSS})$;
          2=second quartile;
          3=third quartile;
          4=fourth quartile.\\
Note (6): 0=relaxed, nominal best-fit most massive galaxy is identified as ``central" (and group center);
          1=relaxed, a central galaxy satisfying criteria of Section \ref{sec:centralsDef} is identified, but it is not the nominal best-fit most massive galaxy;
          2=unrelaxed, no galaxy in the group satisfies the criteria of Section  \ref{sec:centralsDef} to be a central, 
          but nevertheless the nominal best-fit most massive galaxy is labeled as ``central" and used as the group center.\\
Note (7): 0=satellite, 1=central, 2=central if considering only galaxies with WFI $B$- and $I$-band imaging.\\
Note (8): For galaxies outside the WFI FOV the masses are inferred from the
          mass vs. r$_F$ magnitude relation as described in the text.\\
Note (9): Lower and upper limits  corresponding to an increase of 50$\%$ 
          of the best-fit $\chi^2$,  derived from the distribution of $\chi^2$ for all
          templates used in the ZEBRA+ SED fitting.\\
Note (10): 0=elliptical; 1=S0; 2=bulge-dominated spiral;
          3=intermediate disk; 4=late-type disk; 5=Irregular.\\
Note (11): 0=not merging; 1=plausible merger, no spectroscopic or photo-z confirmation; 1.5= Same as flag 1, but visible tidal tails; 2=merger, spectroscopic or photo-z confirmation;      3=close pair among group members;  4=disturbed morphology.
           Close pairs are identified as those galaxies which have a velocity difference, with respect to another group member, $\Delta v<500$ km/s,
           and lie at a projected distance from the same member, $D_{max}\leqslant48.368^{\prime\prime}$
           (equal to the maximum separation between merging galaxies type=1 or type=2. This is about 50kpc at the typical ZENS redshift).\\
Note (12): The parameter is corrected for observational biases  as described in Paper II.\\
Note (13): Together with the formal GIM2D errors we also provide an additional error which is obtained
           by the half difference between the single and double component half-light radii.\\
Note (14): Disk scale length from pure exponential GIM2D fit for the entire galaxy.
           This scale length is only available for late-type disks (Mtype=4).\\
Note (15):  GIM2D failed to provide some decompositions,
          which were successfully re-computed using  GALFIT.
          For I-band: 0=GIM2D, 1=GALFIT.
          For B-band: 0=GIM2D unconstrained;
                      1=GALFIT;
                      2=average of GIM2D unconstrained and GIM2D with ellipticity/PA fixed to I-band;
                      3=GIM2D with ellipticity/PA fixed to I-band;
                      4=GIM2D with ellipticity/PA/bulge parameters fixed to I-band;
                      5=GIM2D with ellipticity/PA/bulge/disk parameters fixed to I-band.  \\         
Note (16): -99 if no reliable decomposition is available;
                   -98 if galaxy has a late-type morphology and is described by a single component S\'ersic fit with $n < 1.5$.
                   No bulge+disk decomposition is performed on galaxies classified as ellipticals,
                   however we set B/T=1 in this catalogue for this morphological type. All other bulge and disk parameters are set to -99 for elliptical galaxies.\\
Note (17): Obtained by integration to infinity of the bulge+disk surface brightness profiles \\
Note (18): Non parametric structural index corrected for PSF and 
          observational biases as described in Paper II.\\
Note (19): This is the actual Petrosian radius, not the default SEXTRACTOR. 
                  Petrosian aperture which is 2.5 the Petrosian radius.\\
Note (20): Default SEXTRACTOR Kron aperture equal to 2.5 times $R_{Kron}$.\\
Note (21): 0=not barred, 1=barred. \\
Note (22): 0= quenched, 1=moderately star-forming, 2=strongly star-forming.\\ 
Note (23): 0=the galaxy satisfies color-color and spectral criteria;
                  1=the galaxy has an actively star-forming spectrum but has red optical-UV colors;
                  2=the galaxy has a quenched spectrum but has blue optical-UV colors and strong $H_{\delta}$ absorption;
                  3=the galaxy has a quenched spectrum, has blue optical-UV colors but no strong $H_{\delta}$ absorption (see Paper III for details).\\
Note (24): The SSFR for galaxies for which the best fit SED result in a  
           SFR$<10^{-4}$ $\Msol$yr$^{-1}$ is set equal to sSFR=$10^{-14}$ yr$^{-1}$.
           Likewise SFR$<10^{-4}$ $\Msol$yr$^{-1}$ are set to SFR=$10^{-4}$$\Msol$yr$^{-1}$.\\
Note (25): Magnitude computed in an elliptical aperture equal to 2 times the largest Petrosian radius among the $B$- and $I$-band one.
           This data is used in the derivation of stellar masses.\\
Note (26): Sextractor MAG$\_$AUTO.\\
Note (27): These are the original $b_j$ and $r_F$ magnitudes, as released by the 2dFGRS team.
           The $b_j$ magnitude is corrected for galactic extinction, but is not 
           at the rest-frame. The $r_F$ in not corrected for galactic extinction and is not at the rest-frame.
           Both the $r_F$ and $b_j$ magnitudes are in the Vega system.          \\
Note (28): 999=undetected, -99=not available.\\
Note (29): Sum in quadrature of the magnitude errors deriving from the formal GIM2D
           uncertainty on the flux and on the bulge-to-total ratio.
           Bulge lower magnitude errors are set to 99 if erBT=0.
           Disk upper magnitude errors are set to 99 if ErBT=1.\\
Note (30): For the bulge and disk components the k-corrections are obtained
           from the observed colors and the relation between the k-correction and color
           as derived for the entire galaxies (see Paper III). \\
 Note (31): -99 if no reliable B+D decompositions are available in both $B$- and $I$-band
           or color cannot be reproduced by synthetic spectral library. \\                    
Note (32): color gradients and color at the half-light radius corrected
           for observational biases as described in Paper III.\\
Note (33): This error reflects the S/N ratio obtained in the Voronoi bins at the galaxy half-light radius.
          It is set to 99 if tessellated map does not reach the half-light radius.\\
Note (34): Quadratic sum of the $B-$, $I$-surface brightness errors on a single pixel
           at the galaxy half-light radius. It is set  to 99 if surface brightness
           in one of the two bands is below the r.m.s value of the sky. \\
Note (35):  For a few  galaxies classified as moderately- or strongly star-forming from their spectral  features or location on the NUV-optical color-color diagram,
           the unconstrained ZEBRA+ fits give inconsistently low SFR and sSFR values.
           For   these galaxies ZEBRA+ was re-run imposing a star-forming template model.
           The flag in this column identifies such galaxies and is set equal to the ``incorrect" galaxy stellar mass from the
           unconstrained ZEBRA+ fits for the re-fitted galaxies, and to -99 for all other galaxies (see Paper III for details).\\         
Note (36):  0=no bright star/companion within Petrosian radius;
           1=galaxy lies close to a bright star, the parameters for this galaxy may be subject to large uncertainties;
           2=companion within the galaxy Petrosian radius;
           3=bright clump/star-clusters within Petrosian radius.\\    
Note (37): These parameters are used to calculate the magnitude and position dependent 2dFGRS redshift completeness at the ZENS galaxy positions. 
           See Section 8 of \citep{Colless_et_al_2001} and Appendix \ref{App:2dFGRS_limits}. \\
Note (38): This flag is equal to 1 if the given galaxy corresponds to the original 2PIGG group center, and otherwise equal to 0.
           For merging pairs/triplets, which have a single entry in the 2PIGG catalog, the flag is set  equal  for all merger members. \\
Note (39): 0=GIM2D and GALFIT I-band parameters agree within a factor of two; 1=At least one parameter differs more than a factor of two between the GIM2D and GALFIT I-band fits; -99=if either the GALFIT or GIM2D decomposition is not reliable/available (see Paper II).     \\
Note (40): The spectroscopic flag used, together   with the color criteria described in Paper III, to classify galaxies in strongly star-forming, moderately star forming and quenched systems (see also Figure 4 of Paper III). Specifically, 1=no emission lines (in particular \Halpha and \Hbeta); 2=\Halpha and [OIII] or [OII] in emission, but no \Hbeta; 3=strong emission in \Halpha,\Hbeta, [OII] and [OIII]. The flag is negative (from -1 to -3) if the spectrum is probing only the galaxy central region (i.e., ``nuclear spectrum").

\end{document}